\documentclass[a4paper]{article}

\usepackage{graphicx}
\usepackage{amsmath}
\usepackage{amssymb}

\def\R{\mathbb{R}}

\def\Z{\mathbb{Z}}
\def\N{\mathbb{N}}

\usepackage{bm}

\usepackage{amsbsy}

\title{\bf Wavelet transforms\\ and their applications to MHD\\ and plasma turbulence: a review}
\author{Marie Farge$^{1,}\footnote{farge@lmd.ens.fr}$ and Kai Schneider$^{2,}\footnote{kschneid@cmi.univ-mrs.fr}$  \\ ~ \\
{\small $^1$LMD-CNRS, Ecole Normale Sup\'erieure} \\
{\small 24, Rue Lhomond, 75231 Paris Cedex 6, France.}  \\ 
{\small $^2$M2P2-CNRS, Aix-Marseille Universit\'e} \\
{\small 38, Rue Fr\'ed\'eric Joliot-Curie, 13451 Marseille Cedex 13, France. } 
}
\date{\today}
\begin{document}
\maketitle


\begin{abstract}
Wavelet analysis and compression tools are reviewed and different applications to
study MHD and plasma turbulence are presented.
We introduce the continuous and the orthogonal wavelet transform and detail several statistical diagnostics based on the wavelet coefficients.
We then show how to extract coherent structures out of fully developed turbulent flows using
wavelet-based denoising.
Finally some multiscale numerical simulation schemes using wavelets are described.
Several examples for analyzing, compressing and computing one, two and three dimensional turbulent MHD or plasma flows
are presented.
\end{abstract}

\tableofcontents

\section{Introduction}

Turbulence is ubiquitous and plays a critical role for the plasma stability and confinement properties of fusion devices, {\it e.g.}, in the tokamak edge region.
Turbulence is a regime of fluid, gas and plasma flows characterized by its highly nonlinear dynamics~\cite{Bisk97}.
It  exhibits a chaotic, {\it i.e.}, unpredictible behavior and rotational motion all along a wide range of dynamically active scales.
In contrast to classical dynamical systems, which are low dimensional and conservative,
a turbulent flow is a dissipative dynamical system, whose behavior is governed by a very large, even maybe infinite, number of degrees of freedom.
Each field, {\it e.g.}, velocity, vorticity, magnetic field or current density, strongly fluctuates around a mean value
and one observes that these fluctuations tend to self-organize into so-called coherent structures,  
{\it i.e.}, vortex tubes in hydrodynamics and vorticty sheets and current sheets in magnehydrodynamics (MHD). 
The presence of coherent structures results in the strong spatial and temporal flow intermittency, which is a
key feature of turbulence.
Intermittency is understood here that the fluctuations become stronger for decreasing scale and are hence more localized.
The appropriate tool to study intermittency is the wavelet representation due to its intrinsic multiscale nature.
Indeed, it yields a sparse multiscale representation of intermittent fields since wavelets are well localized functions in both physical and Fourier space.

The classical theory of homogeneous turbulence \cite{Batchelor82} assumes that turbulent flows are statistically stationary and homogeneous.
This allows to use the Fourier space representation to analyze it ({\it e.g.}, the energy spectrum is the modulus of the Fourier transform of the velocity auto-correlation), to model it ({\it e.g.}, using Large Eddy Simulation) and to compute it ({\it e.g.}, using spectral methods).
Hence, since the Fourier representation spreads information among the phases of all Fourier coefficients, the structural information ({\it i.e.}, locality  in time and in space) is lost when one considers only the modulus of the Fourier coefficients, as usually done. 
This is a major drawback of the classical theory of turbulence and the reason why we proposed in 1988  \cite{Farge88} to replace the Fourier representation by the wavelet representation to define new analysis and computational tools able to preserve time and space local information.
If the Fourier representation is well suited to study linear dynamical systems (whose behaviour either persists at the initial scale or spreads over larger ones), it is not the case for nonlinear dynamical systems for which  the superposition principle no more holds ({\it i.e.}, they cannot be decomposed into a sum of independent subsystems to be separately studied). 
Moreover, the evolution of nonlinear dynamical systems develop over a wide range of scales, since energy is spread from the initially excited scale towards smaller and smaller ones (the so-called energy cascade) until finite-time singularities develop ({\it e.g.}, shocks), unless some dissipative mechanisms damp energy and thus avoid its ultra-violet divergence. 
The art of predicting the evolution of nonlinear dynamical systems consists of disentangling their active components from their passive components, the former being deterministically computed while the latter being discarded or statistically modeled. 
One thus performs a distillation process to only retain the components essential to predict the nonlinear behaviour.
The wavelet representation is particularly appropriate for this since it allows to track the evolution in both space and scale and to only retain the degrees of freedom  in charge of the nonlinear dynamics. 
Turbulent flows are archetypes of nonlinear dynamical systems and therefore good candidates to be analyzed, modelled and computed using the wavelet representation.

If we now focus on plasma turbulence we are uneasy about the fact that we have two different descriptions, depending on which side of the Fourier transform we look from.
\begin{itemize}
\item
 On the one hand, we have a theory \cite{Batchelor82} that assumes a nonlinear cascade in Fourier space for a range of scales, the so-called `inertial range', where the flow kinetic energy is statistically ({\it i.e.}, for ensemble or time or space averages) transferred towards smaller scales until reaching Kolmogorov's scale where  molecular dissipation transforms kinetic energy into heat.
Under these hypotheses, the theory predicts a power-law behaviour for the modulus of the energy spectrum in the inertial range.
\item
 On the other hand, if we study the flow in physical space we do not have yet a predictive theory but only empirical observations (from laboratory and numerical experiments) showing the emergence and persistence of coherent structures,  {\it e.g.}, blobs and current sheets that concentrate most of the kinetic and magnetic energy, even for very high Reynolds number flows. 
\end{itemize}

The classical methods for modeling turbulent flows, {\it e.g.}, Large Eddy Simulation (LES), suppose a scale separation ({\it i.e.}, a spectral gap) and neglect the small scale motions, although their effect onto the large scale motions is statistically modelled (supposing their dynamics to be linear or slaved to them). 
Unfortunately for those methods we have strong evidence, from both laboratory and numerical experiments, that there is no spectral gap since all scales of the inertial range are coupled and nonlinearly interact.
Moreover, one observes that coherent structures play a major dynamical role and are responsible for the transport and mixing properties of turbulent flows. 
In consequence one might ask the following questions: are coherent structures the dynamical building blocks of turbulent flows and can we extract them?
If we succeed to do so, would it be possible to represent them with a reduced number of degrees of freedom and would those be sufficient to compute the flow nonlinear dynamics?

\medskip

The aim of this review is to offer a primer on wavelets for both continuous and orthogonal transforms.
We then detail different diagnostics based on wavelet coefficients to analyze and to compress
turbulent flows by extracting coherent structures.
Examples for experimental data from the tokamak Tore Supra (Cadarache, France) and numerical simulation data of resistive drift-wave and MHD turbulence,
illustrate the wavelet tools.
Wavelet-based density estimation techniques to improve particle-in-cell numerical schemes are presented, together with a particle-in-wavelet scheme that we developed
for solving the Vlasov--Poisson equations directly in wavelet space.
Coherent Vorticity and Current sheet Simulation (CVCS), that  applies wavelet filtering to the resistive non-ideal MHD equations, is proposed as a new model for turbulent MHD flows.
It allows to reduce the number of degrees of freedom necessary to compute them, while capturing the nonlinear dynamics of the flow.
This review is based on the work  and publications we have performed within the last 15 years, in collaboration with the CEA-Cadarache
and other teams in France, Japan and United States.
Almost all material presented here has already been published in some of our papers (cited in the references),
and parts have been adapted for this review.  
Let us only mention few references of wavelet techniques that have been used to analyze and quantify plasma turbulence:
{\it e.g.}, transients \cite{DVZ97}, bicoherence \cite{DK95, VHS95, VSEHBCG95, DAFSZ14},
intermittency \cite{CRMA00} and anisotropy \cite{ALM08}.
An exhaustive review is beyond the scope of our paper and we focus here exclusively on our contributions.

\medskip

The outline of this review is the following: first, in section~2 we present wavelet analysis tools, including a short primer on continuous and orthogonal wavelets. 
Statistical tools in wavelet coefficient space are also introduced.
Section~3 focusses on coherent structure extraction using wavelet-based denoising.
Wavelet-based simulation schemes are reviewed in section~4 and section~5 draws some conclusions.

\section{Wavelet analysis}

\subsection{Wavelets: a short primer}

\subsubsection{Continuous wavelet transform}

The wavelet transform \cite{GrMo84} unfolds any signal ({\it e.g.}, in time) or any field ({\it e.g.}, in  three-dimensional space) into both space (or time) and scale (or time scale), 
and possibly directions (for dimensions higher than one). 
The building block of the wavelet transform is the `mother wavelet', $\psi(x)  \in \ L^2(\R)\ $ with $x  \in \R$, that is a well-localized function  with fast decay at infinity and at least one vanishing moment ({\it i.e.}, zero mean) or more.
It is also smooth enough in order its Fourier transform, $\hat \psi (k)$, exhibits fast decay for $|k|$ tending to infinity.
From the mother wavelet one then generates a family of wavelets, translated by $b \in \R$, the position parameter, dilated (or contracted) by $a \in \R^+$, the scale parameter, and normalized in $L^2$-norm ({\it i.e.}, $|| \psi_{a,b} ||_2 = 1$) to obtain the set
\begin{equation}
\psi_{a,b} (x) = \frac{1}{\sqrt a} \, \psi \left(\frac {x - b} {a} \right) 
\end{equation}
The wavelet transform of $f \in \ L^2(\R)$ is the inner product of $f$ with the analyzing wavelets $\psi_{a,b}$, and the wavelet coefficients, that measure the fluctuations of $f$ at scale $a$ and position $b$, are
\begin{equation}
\label{continuousWT}
\widetilde f (a,b) =  \langle f, \psi_{a,b}  \rangle = \int_{\R} f(x) \psi^{\star}_{a,b}(x) dx
\end{equation}
with $^\star$ denoting the complex conjugate.
The function $f$ is reconstructed from its wavelet coefficients, as the inner product of $\widetilde f$ with the set of analyzing wavelets $\psi_{a,b}$
\begin{equation}
f(x) \, = \, \frac{1}{C_\psi} \, \int_{\R^+} \int_{\R} \, \widetilde f (a,b) \,  \psi_{a,b} (x) \, \frac{da db}{a^2}\, ,
\label{eq_icwt}
\end{equation}
where $C_\psi = \int_{\R^+} |\hat \psi(k)|^2 k^{-1} dk$ is a constant that depends on the wavelet $\psi$.
Similarly to the Fourier transform, the wavelet transform corresponds to a change of basis (from physical space to wavelet space) and, since it is an isometry, it preserves the  inner product ($\langle f, g  \rangle = \langle \widetilde f, \widetilde g  \rangle$) (Plancherel's theorem) and conserves energy (Parseval's identity), therefore 
\begin{equation}
\int_{\R} | f(x) |^2 dx   = \frac{1}{C_\psi} \, \int_{\R^+} \int_{\R} \, |\widetilde f (a,b)|^2  \, \frac{da db}{a^2}
\end{equation}
Note that the wavelet coefficients of the continuous wavelet transform are redundant and therefore correlated.
This could be illustrated by the patterns one observes within the continuous wavelet coefficients of a white noise, which correspond to the correlation between the dilated and translated wavelets (the white noise being decorrelated by construction) and visualizes the `reproducing kernel' of the continuous wavelet transform.
Due to the fact that wavelets are well localized in physical space, the behaviour of the signal at infinity does not play any role.
Therefore both wavelet analysis and wavelet synthesis can be performed locally, in contrast to the Fourier transform which is intrinsically non local (Fourier modes are spread all over space).
One can also construct peculiar wavelets on a dyadic grid $\lambda=(j,i)$ ({\it i.e.}, scale is sampled by octaves $j$ and space by positions $2^{-j}i$) that are orthogonal to each other and are used to construct wavelet orthonormal bases.
In contrast to the continuous wavelet coefficients eq.~(\ref{continuousWT}) that are redundant and correlated, the orthogonal wavelet coefficients are decorrelated and non redundant ({\it i.e.}, a signal sampled on $N$ points is perfectly represented by $N$ orthogonal wavelet coefficients only).
As for the Fourier transform, there exists a Fast Wavelet Transform (FWT) that is even faster than the Fast Fourier Transform (FFT) whose operation count for a one dimensional signal sampled on $N$ points is proportional to $N$, instead of $N \log_2 N$ for the FFT.

\subsubsection{Orthogonal wavelet transform}

\begin{figure}[htbp]
\begin{center}
\includegraphics[width=\linewidth]{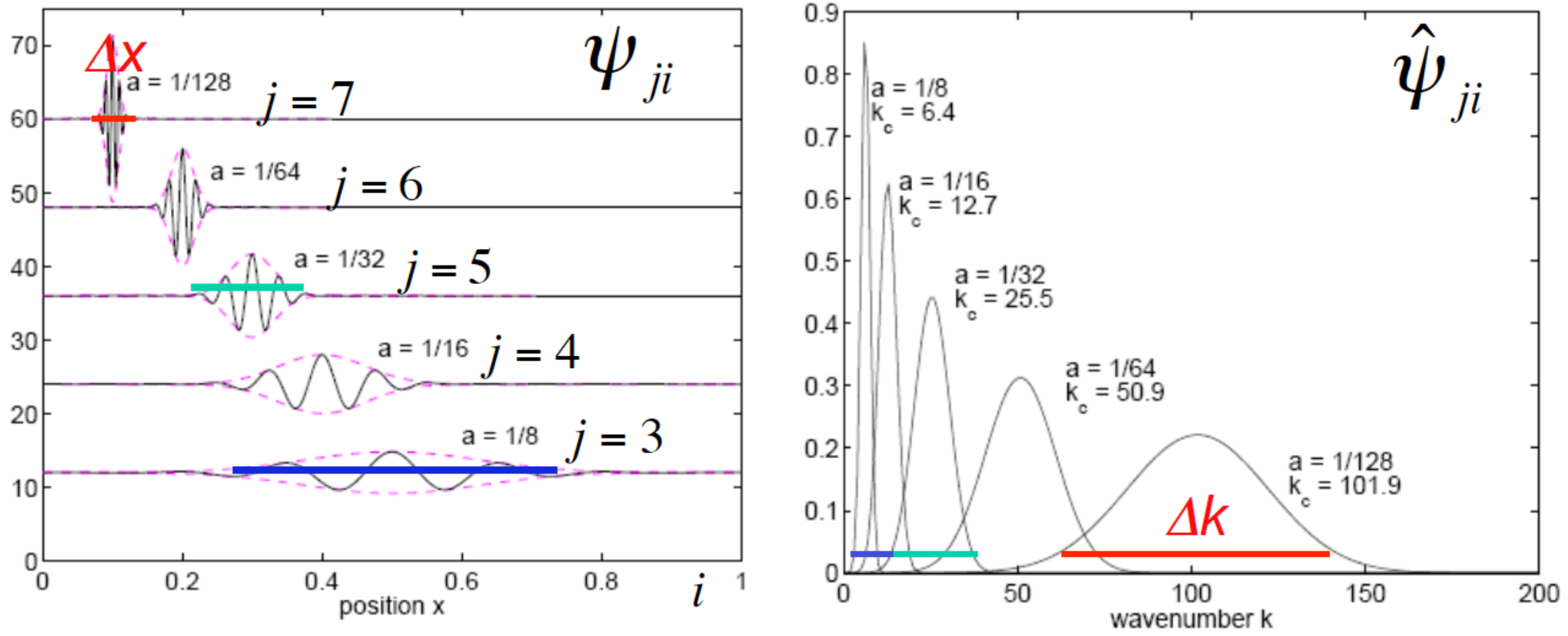}
\end{center}
\caption{Wavelet representation. Physical space (left) and spectral space (right). Note that $\Delta x \, \Delta k > C$ is due to the Fourier uncertainty principle.}
\label{fig:cwt_discrete_representation}
\end{figure}

\begin{figure}[htbp]
\begin{center}
\includegraphics[width=\linewidth]{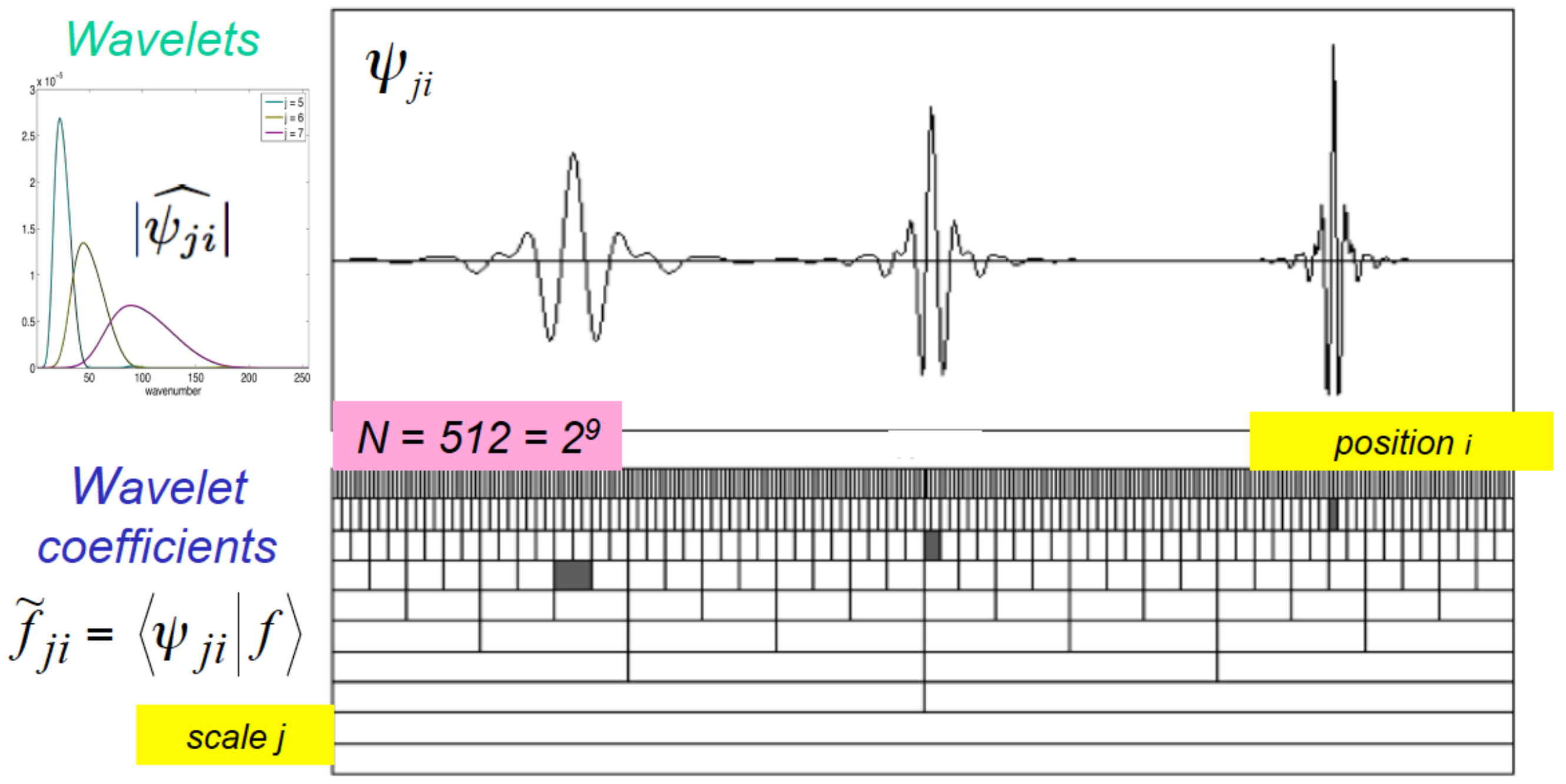}
\end{center}
\caption{Space-scale representation of an orthogonal spline wavelet at three different scales and positions, {\it i.e.}, $\psi_{6,6}, \psi_{7,32}, \psi_{8,108}$. The modulus of the Fourier transform of three corresponding wavelets is shown in the insert (top, left).}
\label{fig:owt_wl}
\end{figure}

A discrete wavelet representation is obtained by sampling dyadically the scale $a$ and the position $b$ introducing $a_j = 2^{-j}$ and $b_{ji} = i a_j$ with $i,j \in \Z$.
The resulting discrete wavelets
\begin{equation}
\psi_{ji}(x) = a_j^{-1/2} \psi \left(\frac{x - b_{ji}}{a_j} \right)  = 2^{j/2} \psi \left(2^j x - i \right)
\end{equation}
generate orthogonal bases for peculiar wavelets.
Figure~\ref{fig:cwt_discrete_representation} shows five discrete wavelets $\psi_{ji}$ for $j=3, ..., 7$ and their corresponding Fourier transforms, the modulus $|\widehat \psi_{ji}|$.
Note that the scale $2^{-j}$ is related to the wavenumber $k_j$ as 
\begin{equation}
k_j = k_\psi 2^j, 
\label{coifkj}
\end{equation} 
where $k_\psi=\int_0^{\infty}k | \hat {\psi}(k)| dk/ \int_0^{\infty} |\widehat {\psi}(k)| dk$ 
is the centroid wavenumber of the chosen wavelet. 
In Fig.~\ref{fig:cwt_discrete_representation} we observe the duality between physical and spectral space,
namely small scale wavelets are well localized in physical space and badly localized in spectral space, and vice-versa.
Denoting the support of a wavelet in physical space by $\Delta x$ and the one in spectral space by $\Delta k$ the Fourier uncertainty principle requires that
 the product $\Delta x \, \Delta k$ is bounded from below.
%
%
In this case the orthogonal wavelet coefficients of a function $f \in L^2(\R)$ are given by
\begin{equation}
\widetilde f_{ji} \; = \;  \langle f, \psi_{ji} \rangle
\end{equation}
and the corresponding orthogonal wavelet series reads
\begin{equation}
f(x) \; = \; \sum_{j,i \in \Z} \, \widetilde f_{ji} \, \psi_{ji} (x) \, .
\end{equation}
The integral in the continuous reconstruction forumla, eq.~(\ref{eq_icwt}), can thus be replaced by a discrete sum.
In practical applications the infinite sums of the wavelet series have to be truncated in both scale and position.
Limiting the analysis to the largest accessible scale of the domain $2^0 =L$
the scaling function associated to the wavelet has to be introduced and the wavelet series becomes
\begin{equation}
f(x) \; = \; \sum_{i \in \Z} \, \overline f \, \phi_{0i} (x) \; + \;  \sum_{j \ge 0,i \in \Z} \, \widetilde f_{ji} \, \psi_{ji} (x) 
\end{equation}
where $\phi$ is the scaling function and $\overline f \, = \langle f, \phi_{0i} \rangle$ the corresponding scaling coefficients.
The smallest scale $2^{-J}$ is given by the sampling rate of the function $f$ which determines the number of grid points $N = 2^J$.
The finite domain size implies that the number of positions becomes also finite and, choosing $L=1$, we obtain the range $i= 0, ..., 2^j -1$ for $j=0, ..., J-1$.
Figure~\ref{fig:owt_wl} illustrates for an orthogonal spline wavelet the discrete scale-space representation for three different scales ($j= 6, 7 , 8$)
and positions.
%
There exists a fast wavelet transform algorithm which computes the orthogonal wavelet coefficients
in  $O(N)$ operations, therefore even faster than the fast Fourier transform whose operation count is  $O(N \log_2N)$ \cite{Mallat98}. 

\begin{figure}[htbp]
\begin{center}
\includegraphics[width=\linewidth]{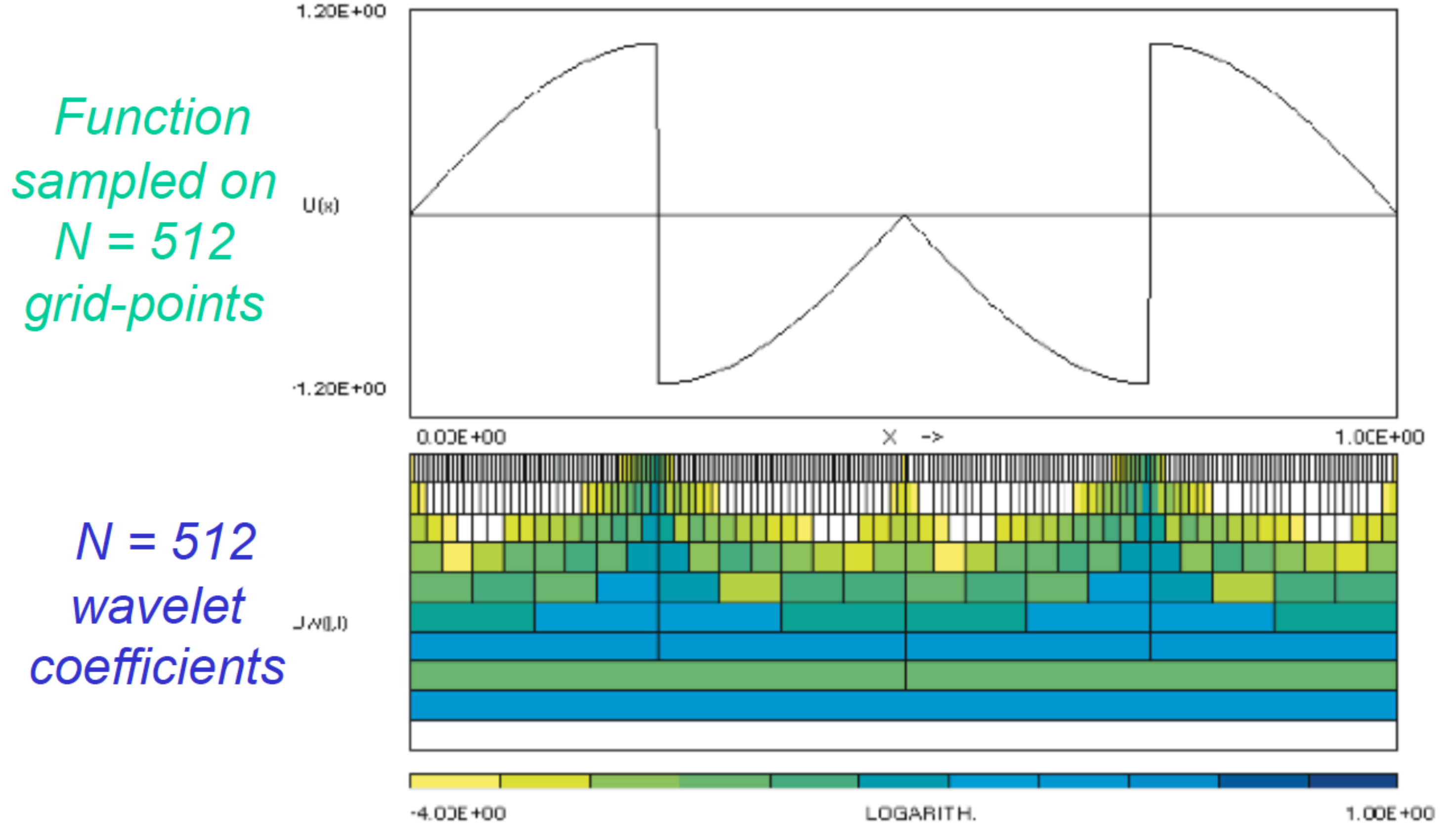}
\end{center}
\caption{Academic example: function with two discontinuities and one in its derivative (top), corresponding modulus of orthogonal wavelet coefficients in logarithmic scale using periodic spline wavelets of degree five.}
\label{fig:owt_coeff}
\end{figure}

As example we show in Fig.~\ref{fig:owt_coeff} the orthogonal wavelet coefficients of an academic function presenting discontinuities.
We observe that wavelet coefficients at small scales only have significant values in the vicinity of the discontinuities.
Hence only few coefficients are needed to represent the function after discarding the small wavelet coefficients.
%

\paragraph{Extension to higher dimensions:}

\begin{figure}[htbp]
\begin{center}
\includegraphics[width=\linewidth]{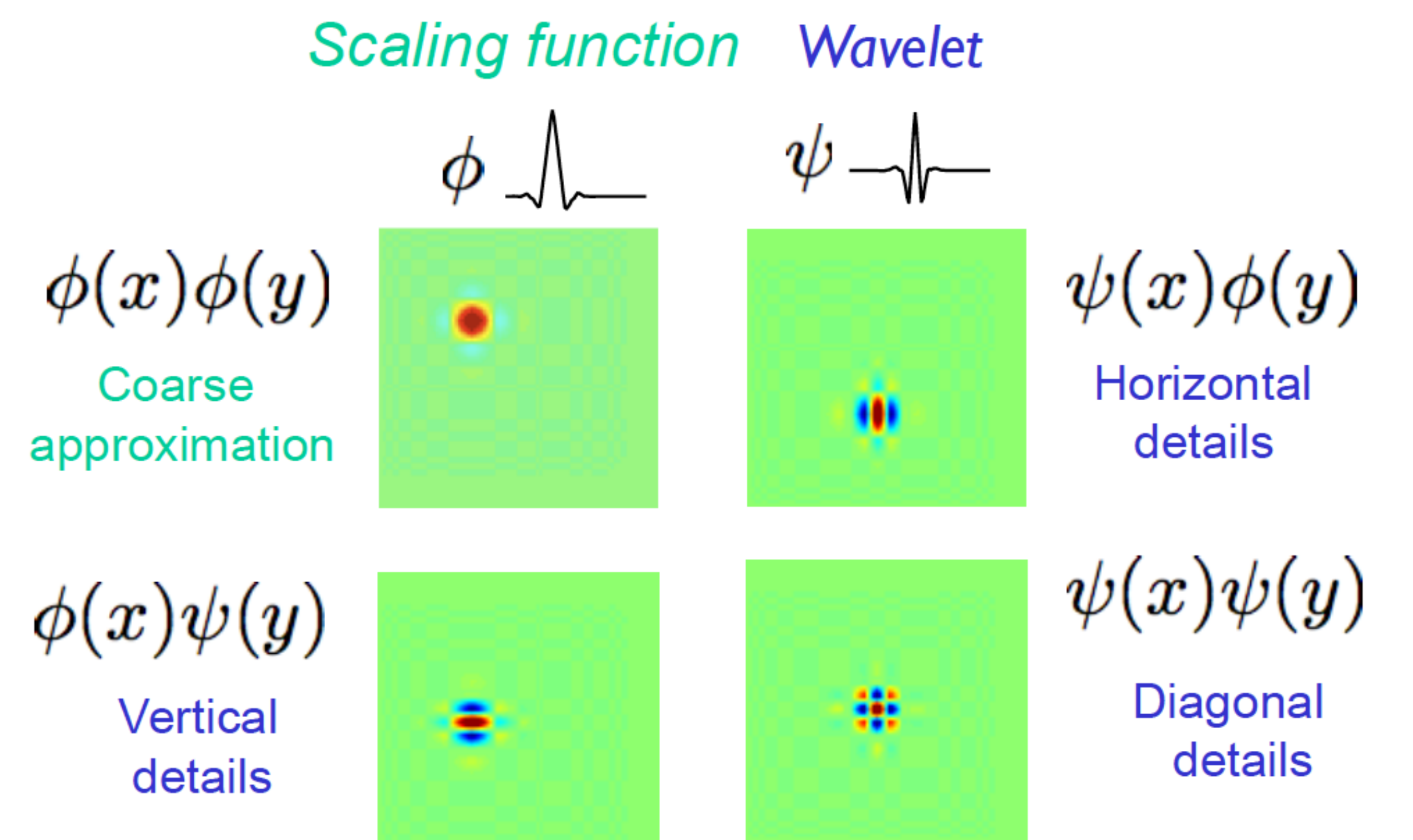}
\end{center}
\caption{Two-dimensional orthogonal wavelets. Scaling function (top, left) and the three associated directional wavelets in the horizontal (top, right), vertical (bottom, left) and diagonal (bottom, right) direction.}
\label{fig:2d_owt}
\end{figure}

The orthogonal wavelet representation can be extended to represent functions in higher space dimensions using tensor product constructions, see {\it e.g.}, \cite{Daub92, Mallat98, ScFa06}.
Figure~\ref{fig:2d_owt} shows two-dimensional orthogonal wavelets constructed by tensor products.
%

The wavelet transform can also be generalized to treat vector-valued functions ({\it e.g.}, velocity or magnetic fields) in $d$ space dimensions
by decomposing each vector component into an orthogonal wavelet series.
In the following we consider a vector field ${\bm v} = (v^{(1)}, v^{(2)}, v^{(3)} )$ for $d=3$ sampled at resolution $N = 2^{3 J}$ with periodic boundary conditions.
Its orthogonal wavelet series reads
\begin{eqnarray}
{\bm v}({\bm x}) \; = \; \sum_{j=0}^{J-1} \sum_{\mu=1}^7 
\sum_{i_1,i_2,i_3=0}^{2^j-1} {\widetilde {\bm v}}_{j,\mu,{\bm i}} \; \psi_{j,\mu,{\bm i}}({\bm x}), 
\label{OWS}
\end{eqnarray}
using 3D orthogonal wavelets $\psi_{j,\mu,{\bm i}}({\bm x})$. 
The basis functions are constructed by tensor products of a set of one-dimensional wavelets and scaling functions~\cite{Daub92, Mallat98} which have been periodized since the boundary conditions considered here are periodic.
The scale index $j$ varies from $0$ to $J-1$, the spatial index ${\bm i} = (i_1, i_2, i_3)$ has $2^{3j}$ values for each scale $2^{-j}$ and each angle indexed by $\mu = 1,\cdots,7$.
The three Cartesian directions ${\bm x} = x^{(1)}, x^{(2)}, x^{(3)}$ correspond to $\mu = 1, 2, 3$,
while $\mu=4, 5, 6, 7$ denote the remaining diagonal directions.
The wavelet coefficients measure the fluctuations of ${\bm v}$ at scale $2^{-j}$ and around position $2^{-j}{\bm i} $ for each of the seven possible directions $\mu$.
%
%
%
The contribution of the vector field ${\bm v}$ at scale $2^{-j}$ and direction $\mu$ can be reconstructed by summation of ${\widetilde {\bm v}}_{j,\mu,{\bm i}} \psi_{j,\mu,{\bm i}}({\bm x})$ over all positions ${\bm i}$:
\begin{equation}
{\bm v}_{j,\mu}({\bm x}) \; = \; \sum_{i_1,i_2,i_3=0}^{2^j-1} {\widetilde {\bm v}}_{j,\mu,{\bm i}} \, \psi_{j,\mu,{\bm i}}({\bm x}).
\end{equation}
The contribution of ${\bm v}$ at scale $2^{-j}$  is obtained by
\begin{equation}
{\bm v}_j({\bm x}) \; = \; \sum_{\mu=1}^7 {\bm v}_{j,\mu}({\bm x}).
\end{equation}
For more details on wavelets, we refer the reader to several review articles, {\it e.g.},~\cite{Farge92,FaSc06, ScFa06} 
and textbooks, {\it e.g.}, ~\cite{Daub92,Mallat98}.

\subsection{Wavelet-based statistical diagnostics}		

The physical representation gives access to both position and direction, the latter when the space dimension is larger than one.
The spectral representation gives access to both wavenumber and direction, when the space dimension is larger than one, 
but the information on position is spread among the phases of all Fourier coefficients.
The wavelet representation combines the advantages of both representations, while also giving access to scale.
For instance if we consider a three-dimensional vector-valued field, its orthogonal wavelet coefficients of each of its three components are indexed by three positions, 
seven directions and one scale.
Thus using the wavelet representation new statistical diagnostics can be designed
by computing moments of coefficients using summation, either over position, direction or scale, or any combination of them.
Second order moments correspond to energy distributions ({\it e.g.}, the energy spectrum), while higher order moments allow to compute skewness and flatness.
In the following we will present scale dependent moments, scale-dependent directional statistics and scale dependent topological statistics.
By topological statistics we mean the statistics of bilinear quantities, like
the scalar product of a vector field and its curl, {\it e.g.}, helicity.
In the following, we give a  summary of statistical diagnostics based on orthogonal wavelet analysis, here applied to a generic vector field following the lines of \cite{OYSF14}.
Decomposing a vector field into orthogonal wavelets, scale-dependent distributions of turbulent flows can be measured, including indifferent directions and also of different flow components.
For example, the energy and its spatial fluctuations can be quantified at different length scales and in different directions and hence longitudinal or transverse contributions can be determined.
In the case of an imposed magnetic field the contributions in  the directions perpendicular or parallel
to it can be distinguished.
To this end, statistical quantities based on the wavelet representation can be introduced, 
and the scale-dependent anisotropy and the corresponding intermittency of MHD turbulence can be examined.  
Here we define intermittency as a departure from Gaussianity, which is
reflected by increasing flatness when scale decreases. 
Sandborn~\cite{San59} introduced this definition in the context of boundary layer flows
and for a historical overview on intermittency we refer to \cite{ScFaKe04}. 
Alternative definitions of intermittency can be found, {\it e.g.}, in \cite{UF95}, for example a steepening of the energy spectrum proposed by Kolmogorov in 1962 \cite{K62}.
In \cite{KS01, KS00,SVCBV06} related techniques to quantify the anisotropy of the flow and its intermittency
have been proposed.
They used structure functions of either tensorial components or applied the SO3 decomposition, which is based on spherical harmonics.
Structure functions which correspond to moments of increments can be directly linked to wavelet decompositions (see, {\it e.g.}, \cite{ScFaKe04}). 
The increments are wavelet coefficients using the poor man's wavelet, {\it i.e.}, the difference of two delta distributions, which has only one vanishing moment, its mean value.
This implies that the exponent of the detectable scaling laws is limited by the order of the structure function and the scale selectivity is reduced as the frequency localization of the poor man's wavelet is rather bad.
These drawbacks can be overcome using higher order wavelets.

%
\subsubsection{Scale dependent moments}
To study the scale-dependent directional statistics we consider the component $v^{\ell}$ with $\ell=1, 2, 3$ of a generic vector field ${\bm v}$.
First we define the $q$-th order moment of 
the scale-dependent vector ${\bm v}_j({\bm x})=(v^{(1)}_j,v^{(2)}_j,v^{(3)}_j)$,
which is here either the vector field at scale $2^{-j}$ and direction $\mu$, $v_{j,\mu}^{(\ell)}$, or the vector field at scale $2^{-j}$, $v_{j}^{(\ell)}$,
\begin{equation}
M_q [ v_j^{(\ell)} ] = \langle ( v_j^{(\ell)} )^q \rangle .
\end{equation}
By construction the mean value satisfies $\langle v_j^{(\ell)} \rangle=0$ and hence the moments are automatically centered.
These scale-dependent moments are related to the $q$-th order structure functions, as shown , {\it e.g.}, in~\cite{ScFaKe04}.
In the following we consider the second order moment $M_2[v_j^{(\ell)} ]$, which is a scale-dependent quadratic mean intensity of $v_j^{(\ell)} $, and the fourth order moment $M_4[v_j^{(\ell)} ]$ which contains the scale-dependent spatial fluctuations. Both moments are related via the flatness factor.

In anisotropic turbulence typically a preferred direction can be defined, {\it e.g.}, for low magnetic Reynolds number turbulence, or rotating turbulence.
These flows have statistical symmetries, which we suppose here with respect to the $x_3$-axis.
For the remaining perpendicular components, $\ell=1,2$,  the average of these two components is taken, 
$M_q[v_j^\perp] = \{M_q[v_j^{(1)}]+M_q[v_j^{(2)}]  \} /2 $, 
and the superscript $\perp$ represents the  perpendicular contribution. 
The parallel contribution $v_j^{(3)}$ is denoted by $v_j^\parallel$.

The wavelet energy spectrum for $v_j^{(\ell)}$ is obtained
using $M_2[v_j^{(\ell)}]$ and eq. (\ref{coifkj}),
\begin{equation}
{E} [v_j^{(\ell)}]  = \frac{1}{2\Delta k_j} M_2 [ v_j^{(\ell)} ] ,
\label{wave_spe}
\end{equation}
where $\Delta k_j = (k_{j+1} - k_{j}) \ln 2$ \cite{Mene91,Addison}.
It is thus directly related to the Fourier energy spectrum and yields a smoothed version~\cite{Farge92, Mene91}.
The orthogonality of the wavelets with respect to scale and direction guarantees that the total energy is obtained by direct summation, $E=\sum_{\ell,j}{E}[v_{j}^{(\ell)}]= \sum_{\ell,j,\mu} {E}[ v_{j,\mu}^{(\ell)} ] $.

The standard deviation of the energy spectrum at a given wavenumber $k_j$ quantifies
the spatial variability 
\begin{equation}
{\sigma} [v_j^{(\ell)}] = \frac{1}{2\Delta k_j} \sqrt{ M_4 [ v_j^{(\ell)} ]  - \left( M_2 [ v_j^{(\ell)} ] \right) ^2}  \, . \label{wl_variace3d}
\end{equation}

The ratio of the fourth and second order moments defines the scale-dependent flatness factor, 
\begin{equation}
F [v_j^{(\ell)} ]=\frac{M_4 [ v_j^{(\ell)} ]}{ \left( M_2 [ v_j^{(\ell)}  ] \right)^2 } \, .\label{flatness}
\end{equation}
which quantifies the flow intermittency at scale $2^{-j}$.

The scale-dependent flatness is related to the energy spectrum (\ref{wave_spe}) and the standard deviation~(\ref{wl_variace3d}),
\begin{equation}
F[v_j^{(\ell)} ] = \left( \frac{{\sigma} [v_j^{(\ell)} ]} { {E} [v_j^{(\ell)} ] } \right)^2 \, + \, 1.\label{flatness_std}
\end{equation}
as shown in~\cite{BoLiSc07}. 
This relation illustrates that the spatial variability of the energy spectrum is directly reflected by the scale-dependent flatness.

%
\subsubsection{Scale-dependent directional statistics}
To quantify scale-dependent spatial flow anisotropy and anisotropic flow intermittency we introduce
wavelet-based measures.
Both component-wise anisotropy and directional anisotropy of the flow are considered in the following.
For the scale-dependent mean energy, $E[v_j^{(\ell)}]$, the anisotropy measure can be defined similarly  to the classical Fourier representation.
Analoguously this can be extended for its spatial fluctuations, $\sigma[v_j^{(\ell)}]$.
Using the relation between the scale-dependent flatness with the energy spectrum and its spatial fluctuations, eq. (\ref{flatness_std}), various measures of anisotropic flow intermittency can be defined.

\medskip  

\paragraph{Component-wise anisotropy:}
The scale-dependent component-wise aniso-tropy is defined by
the ratio of perpendicular to parallel energy, and its fluctuation, at a given scale scale $2^{-j}$, respectively, 
\begin{equation}
c_E(k_j) =  \frac{ E[v_j^\perp ] }{ E[v_j^\parallel ]} \; ,   \quad \quad
c_\sigma (k_j) =  \frac{ \sigma[v_j^{\perp} ]}{ \sigma[v_j^{\parallel}]} 
\label{comp_m_spec2} \; .
\end{equation}
The scale-dependent mean energy, $c_E(k_j)$ is a smoothed version of the Fourier counterpart $c(k)$.
The component-wise anisotropy of the spatial fluctuations is quantified by $c_\sigma(k_j)$. 
These measures are directly related to the component-wise flatness factors of $v_{j}^{(\ell)}$, {\it i.e.}, $F[v_j^\perp]$ and $F[v_j^\parallel]$, as shown in \cite{OYSF14}.
Combining eqs. (\ref{flatness_std}) and (\ref{comp_m_spec2}) results in
\begin{equation}
\Lambda_j^{C} \equiv  \left\{ \frac{c_\sigma (k_j)}{c_E(k_j)} \right\}^2=\frac{ F[v_j^{\perp}] -1}{ F[v_j^{\parallel}] -1}, \label{aniso}
\end{equation}
which yields a scale-dependent measure of component-wise anisotropic intermittency.

\medskip

\paragraph{Directional anisotropy:}
Scale-dependent measures for directional anisotropy can be defined using ratios of perpendicular to parallel energy and fluctuations in longitudinal or transverse directions, 
\begin{eqnarray}
d_E^L(k_j) &=&  \frac{ E [v_{j,L}^{\perp}] }{ E[v_{j,L}^{\parallel}]},  \label{comp_m_spec_L} \quad \quad
d_\sigma^L(k_j) =  \frac{ \sigma [v_{j,L}^{\perp}]  }{ \sigma [v_{j,L}^{\parallel}]}, \\
d_E^T(k_j) &=&  \frac{ E [v_{j,3}^{\perp}] }{ E [v_{j,T}^{\perp}]},  \quad \quad
d_\sigma^T(k_j) =  \frac{ \sigma [v_{j,3}^{\perp}] }{ \sigma [v_{j,T}^{\perp}]} \, . \label{comp_m_spec}
\end{eqnarray}
The {longitudinal} direction is denoted by the index $L$, {\it i.e.}, $L=\mu=\ell$.
The subscript $\mu=3$ corresponds to a {transverse} direction of the perpendicular components, 
while $T$ represents  the other {transverse} direction of the perpendicular components, 
{\it i.e.}, $T=\mu=1$ for $v_{j,\mu}^{(2)}$ or $T=\mu=2$ for $v_{j,\mu}^{(1)}$.

Three principal directions, {\it i.e.}, $\mu=1, 2$ and $3$,  out of the seven possible directions have been selected 
for the directional statistics.

The measures $d_E^L(k_j)$ and $d_E^T(k_j)$ are smoothed versions of the Fourier representation 
$2 e^{(3)}(k_3) /\{ e^{(1)}(k_1) + e^{(2)}(k_2) \}$ and $\{e^{(1)}(k_3) + e^{(2)}(k_3)\}/\{ e^{(1)}(k_2) + e^{(2)}(k_1) \}$, respectively, following the interpretation of the directional statistics proposed in \cite{BoLiSc07}.
Furthermore these quantities can be related to second order structure functions defined in physical space, and respectively we have:
\begin{equation}
\frac{ 2 D^{(3)}  (r{\hat {\bm l}}_3) } { \{ D^{(1)} (r{\hat {\bm l}}_1)+D^{(2)} (r{\hat {\bm l}}_2) \}}  \quad  {\rm and} \quad
\frac{\{D^{(1)} (r{\hat {\bm l}}_3) + D^{(2)} (r{\hat {\bm l}}_3) \} }{ \{ D^{(1)} (r{\hat {\bm l}}_2)+D^{(2)} (r{\hat {\bm l}}_1) \}} \, .
\end{equation}
Structure functions are defined as the spatial average of velocity increments, 
$D^{(\ell)} ({\bm r}) = \langle \{v^{(\ell)}({\bm x}+{\bm r}) -v^{(\ell)}({\bm x}) \}^2 \rangle $.
Here $v^{(\ell)}$ consists of contributions of $v^{(\ell)}$ to scales larger than  $2^{-j}$. 
which are obtained by low pass filtering using the three-dimensional scaling function at scale $2^{-j}$.
The unit vector of the Cartesian direction $x_\ell$ is denoted by ${\hat {\bm l}}_\ell$.


Combining eq. (\ref{flatness_std}) and eqs. (\ref{comp_m_spec_L})-(\ref{comp_m_spec}), 
yields directional anisotropy measures \cite{OYSF14}:
\begin{eqnarray}
\Lambda^{L}_j&\equiv & \left\{ \frac{d_\sigma^L (k_j)}{ d_E^L(k_j)} \right\}^2=\frac{ F[v_{j,L}^{\perp}] -1}{ F[v_{j,L}^{\parallel}]-1}, \label{measure_flat_L}\\
\Lambda^{T}_j&\equiv & \left\{ \frac{d_\sigma^T (k_j)}{ d_E^T(k_j)} \right\}^2=\frac{ F[v_{j,3}^{\perp}] -1}{ F[v_{j,T}^{\perp}] -1} \, , \label{measure_flat_T}
\end{eqnarray}
which quantify the scale-dependent anisotropic intermittency in the transverse and longitudinal directions.
They measure intermittency, not only in the plane perpendicular or in the direction parallel to for example a magnetic field ${\bm B}_0$, but also in the longitudinal or transverse directions. 
These measures are equal to one for isotropic fields, and their departure from the value one indicates the degree of flow anisotropy.

\subsubsection{Scale-dependent topological statistics}

Considering the velocity field ${\bm u}$ and the corresponding vorticity ${\bm \omega} = \nabla \times {\bm u}$
the kinetic helicity,  $H({\bm x}) = {\bm u} \cdot {\bm \omega}$, can be defined.
The helicity yields a measure of the geometrical statistics of turbulence.
Integrating the helicity over space one obtains the mean helicity $\overline H = \langle {\bm u} \cdot {\bm \omega} \rangle$.
%
The scale-dependent helicity $H_j$ was introduced in \cite{YOSKF09} 
and is defined by
\begin{equation}
H_j({\bm x}) = {\bm u}_j \cdot {\bm \omega}_j
\end{equation}
It  preserves Galilean invariance, though the kinetic helicity itself does not.
The corresponding mean helicity is obtained by summing $H_j$ over scale, $\overline H = \sum_j \, \langle H_j \rangle$
due to the orthogonality of the wavelet decomposition.

The relative helicity
\begin{equation}
h({\bm x}) = \frac{H} { |{\bm u}| \, |{\bm \omega} | }
\end{equation}
defines the cosine of the angle between the velocity and the vorticity at each spatial position.
The range of $h$ lies between $-1$ and $+1$.
The scale dependent relative helicity can be defined correspondingly
\begin{equation}
h_j({\bm x}) = \frac{H_j} { |{\bm u}_j| \, |{\bm \omega}_j | }
\end{equation}

The Euler equations of hydrodynamics conserve the mean kinetic helicity, while
in ideal MHD turbulence the mean cross helicity $\overline H^C = \langle {\bm u} \cdot {\bm b} \rangle$  and the mean magnetic
helicity $\overline H^M = \langle {\bm a} \cdot {\bm b} \rangle$ are conserved quantities. 
Here $\bm a$ is the vector potential of the magnetic field $\bm b$.
The scale dependent versions of the relative cross and magnetic helicities have been introduced in~\cite{YSOKF11}
and are defined respectively by
\begin{equation}
h^C_j({\bm x}) \; = \; \frac{H^C_j}{|{\bm u}_j| |{\bm b}_j|}
\end{equation}
with $H^C({\bm x}) =  {\bm u} \cdot {\bm b}$
and
\begin{equation}
h^M_j({\bm x}) \; = \; \frac{H^M_j}{|{\bm a}_j| |{\bm b}_j|}
\end{equation}
with $H^M({\bm x}) =  {\bm a} \cdot {\bm b}$.
These quantities define the cosine of the angle between the two vector fields.

\subsection{Application to 3D MHD turbulence}

In the following we show applications of the above scale-dependent wavelet-based measures to three-dimensional
incompressible magnetohydrodynamic turbulence. To study the anisotropy we analyze flows with uniformly imposed magnetic field considering the quasistatic approximation at moderate Reynolds numbers for different interaction parameters \cite{OYSF14}. 
For the geometrical statistics full MHD turbulence without imposed mean field is analyzed \cite{YSOKF11}.
The flows are computed by direct numerical simulation with a Fourier pseudo-spectral method at resolution $512^3$ and for further details we refer the reader to the respective publications.
%
\begin{figure}[htbp!]
\begin{center}
\includegraphics[width=0.48\linewidth]{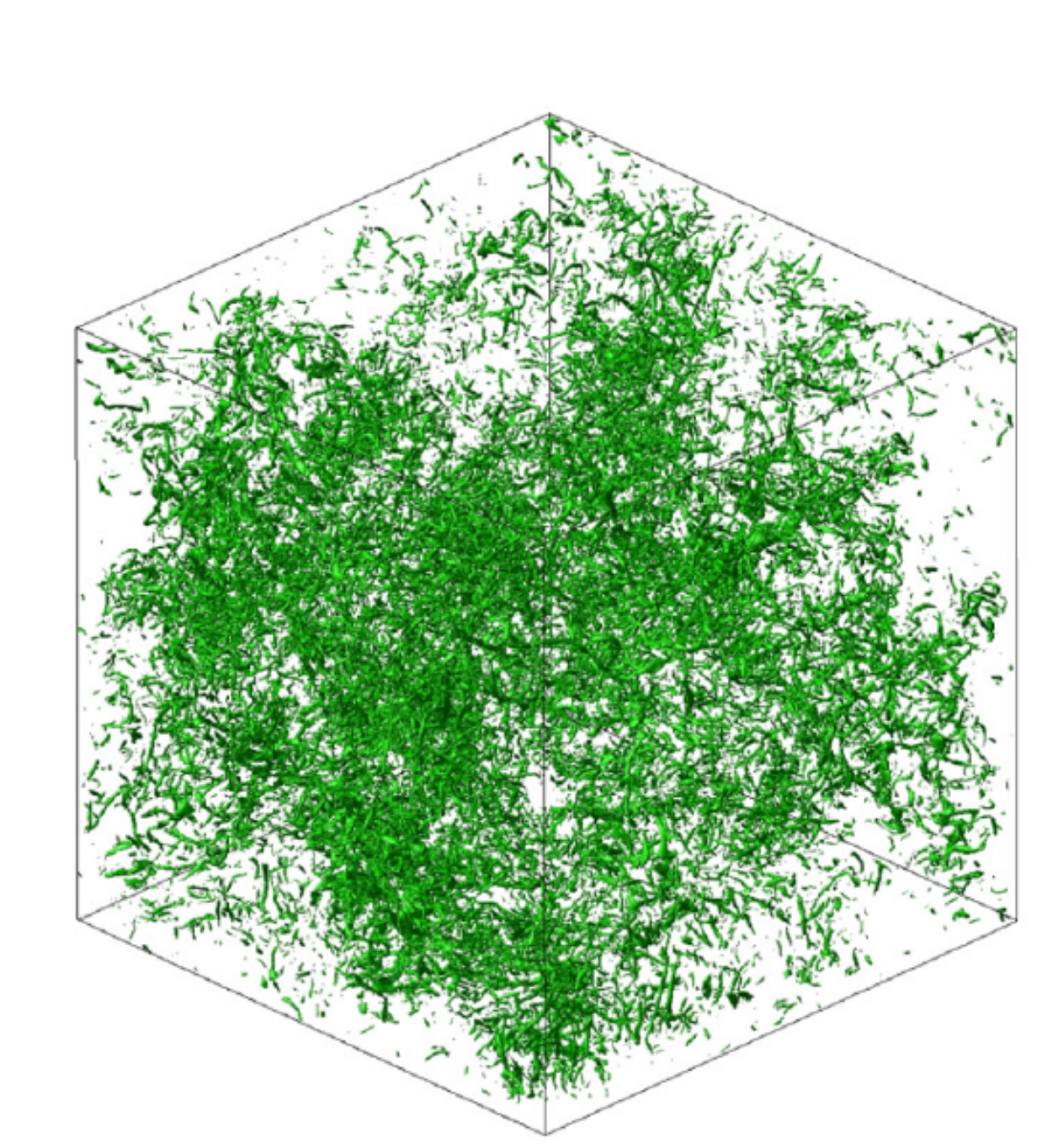}
\includegraphics[width=0.48\linewidth]{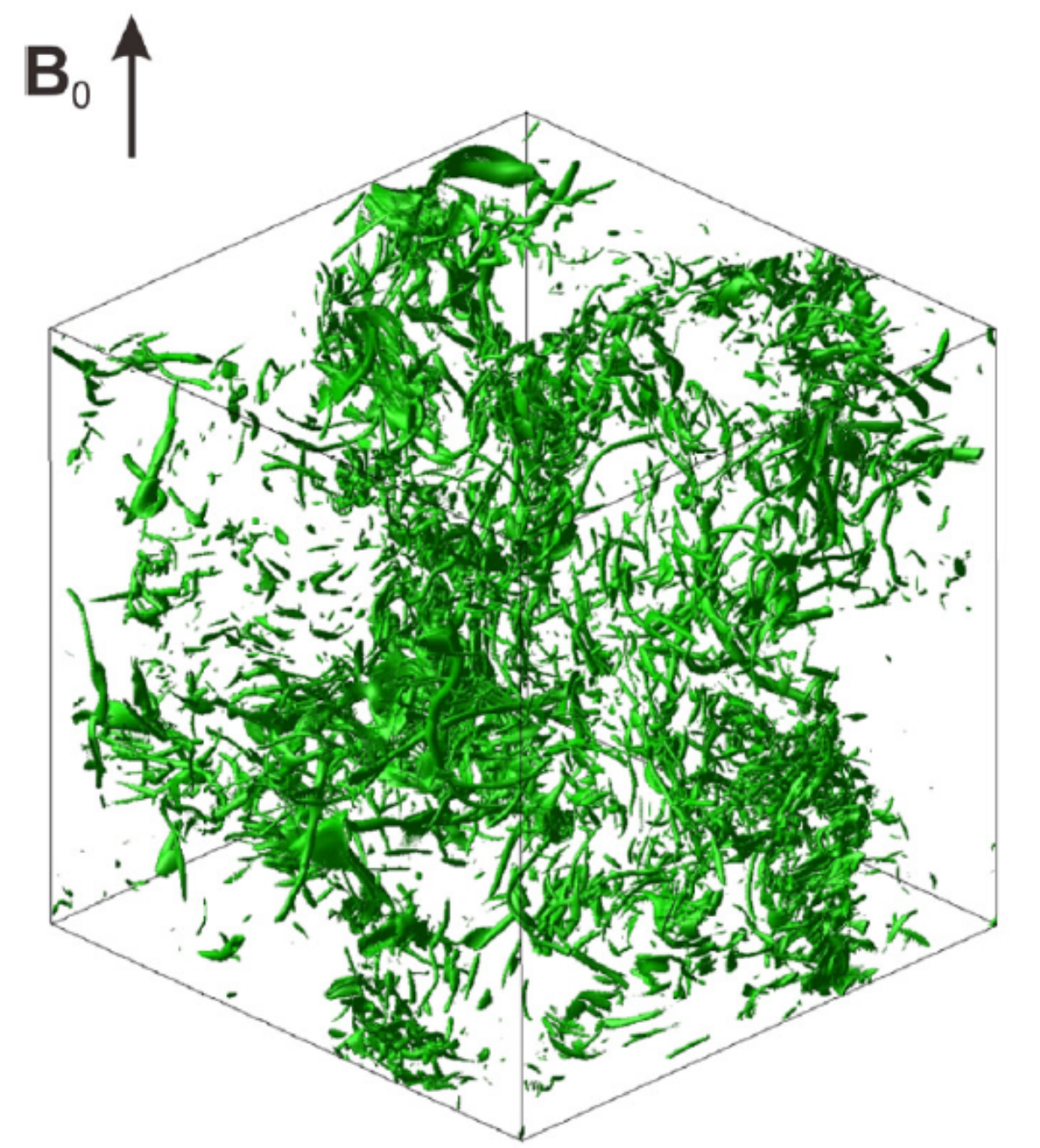}
\end{center}
\caption{QS-3D-MHD: Modulus of vorticity for quasistatic 3D MHD at $R_\lambda = 235$, with $N=0$, (left) and $N=2$ (right) computed by DNS, from~\cite{OYSF14}.}
\label{fig:visu_3dqsmhd}
\end{figure}
%
The flow structure of the quasistatic MHD turbulence is illustrated in Fig.~\ref{fig:visu_3dqsmhd}.
Shown are isosurfaces of the modulus of vorticity for two different interaction parameters $N$.
The interaction parameter characterizes the intensity of the imposed magnetic field $B_0$ (here chosen in the $z$ direction) relative to the flow nonlinearity. It is defined by $N = \frac{\sigma B_0^2 L}{\rho u'}$, where $\sigma$ is the electrical conductivity, $L$ the integral length scale, $\rho$ the density and $u'$ the rms velocity.
In the case without imposed magnetic field, {\it i.e.}, $N=0$ the flow is equivalent to isotropic hydrodynamic turbulence
and entangled vortex turbes can be observed in Fig.~\ref{fig:visu_3dqsmhd}, left.
For $N=2$ the structures are aligned parallel to the $z$ direction, {\it i.e.}, the direction of the imposed magnetic field, and
the flow is thus strongly anisotropic.

\begin{figure}[htbp]
\begin{center}
\includegraphics[width=0.49\linewidth]{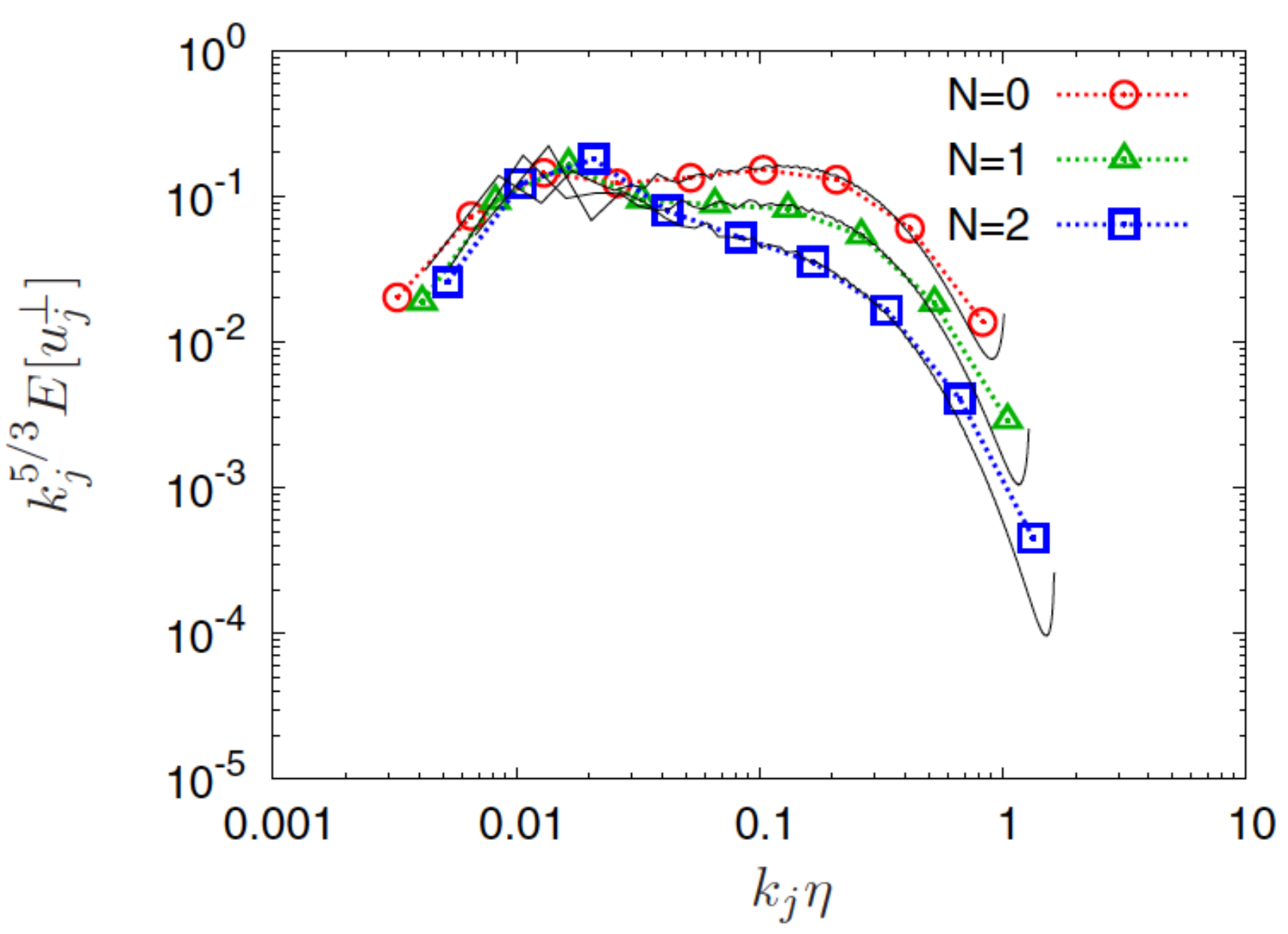}
\includegraphics[width=0.49\linewidth]{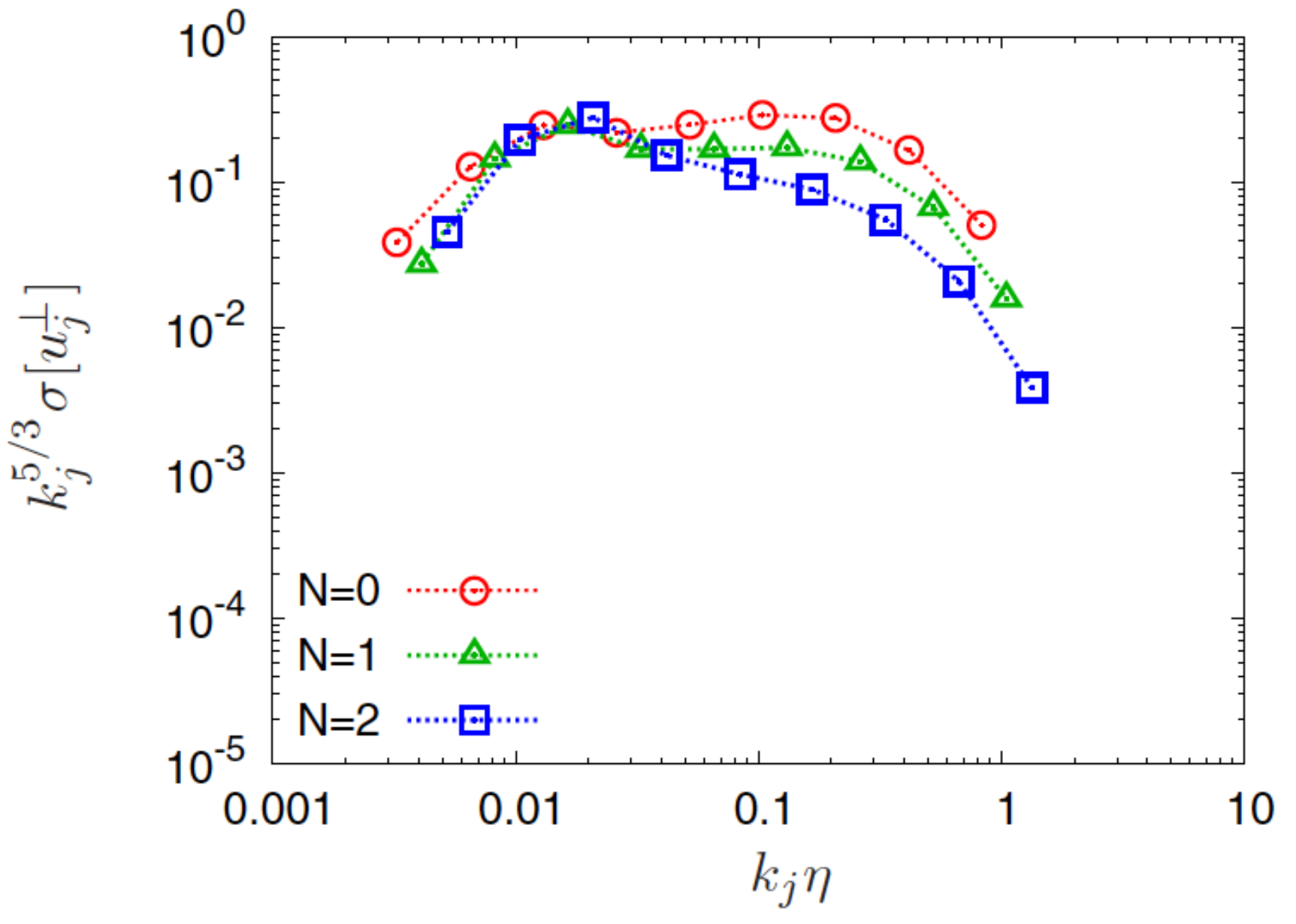}
\end{center}
\caption{QS-3D-MHD: Wavelet mean energy spectra (left) $k_j^{5/3}{ E}^{\perp} (k_j)$ together
with the Fourier energy spectra (solid lines). 
Wavelet standard deviation spectra (right) $k_j^{5/3}{ \sigma}^{\perp} (k_j)$. 
All quantites are shown for the perpendicular velocity components. 
The inset (left) shows the corresponding forcing Fourier spectra $k^{5/3} E_f (k)$. From~\cite{OYSF14}.}
\label{fig:3dqs_spectra}
\end{figure}

The wavelet energy spectra (Fig.~\ref{fig:3dqs_spectra}, left) yield information on the kinetic energy at scale $2^{-j}$ and the spatial fluctuations are quantified by the standard deviation spectra (Fig.~\ref{fig:3dqs_spectra}, right). 
All spectra have been multiplied by $k^{5/3}$ to enhance their differences at small scale.
We observe that the spectra decay with increasing normalized wavenumber $k_j \eta$ where $\eta$ is the Kolmogorov length scale.
Furthermore the wavelet spectra (dotted lines) do agree well with the corresponding Fourier spectra (solid lines).
For larger values of $N$ the spectra $E[u_j^\perp ]$ decay faster for increasing $k_j \eta$.
The standard deviation spectra of $u_j^\perp$ also decay more rapidly when $N$ becomes larger.
%
%

\begin{figure}[htbp]
\begin{center}
\includegraphics[width=0.48\linewidth]{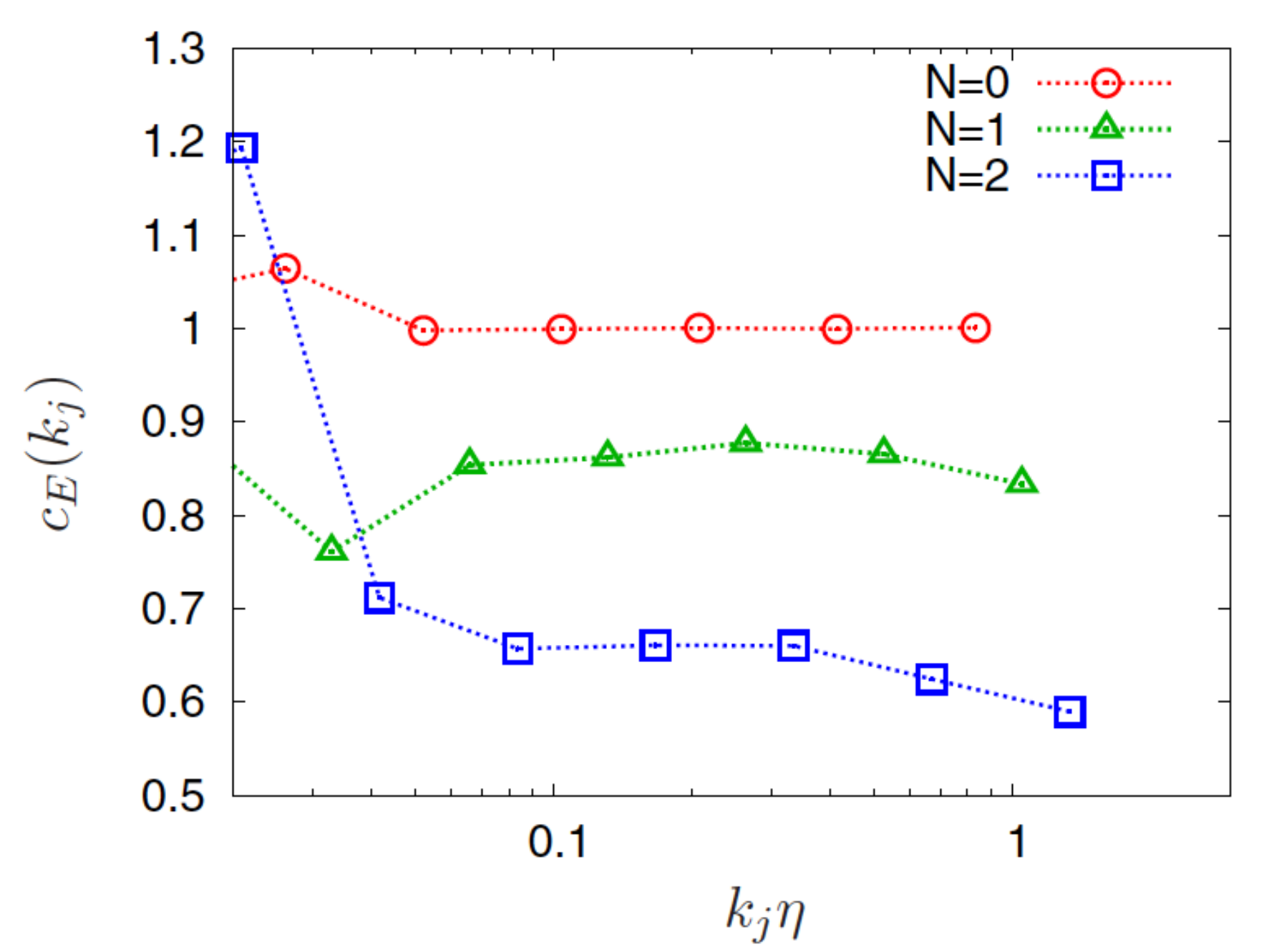}
\includegraphics[width=0.47\linewidth]{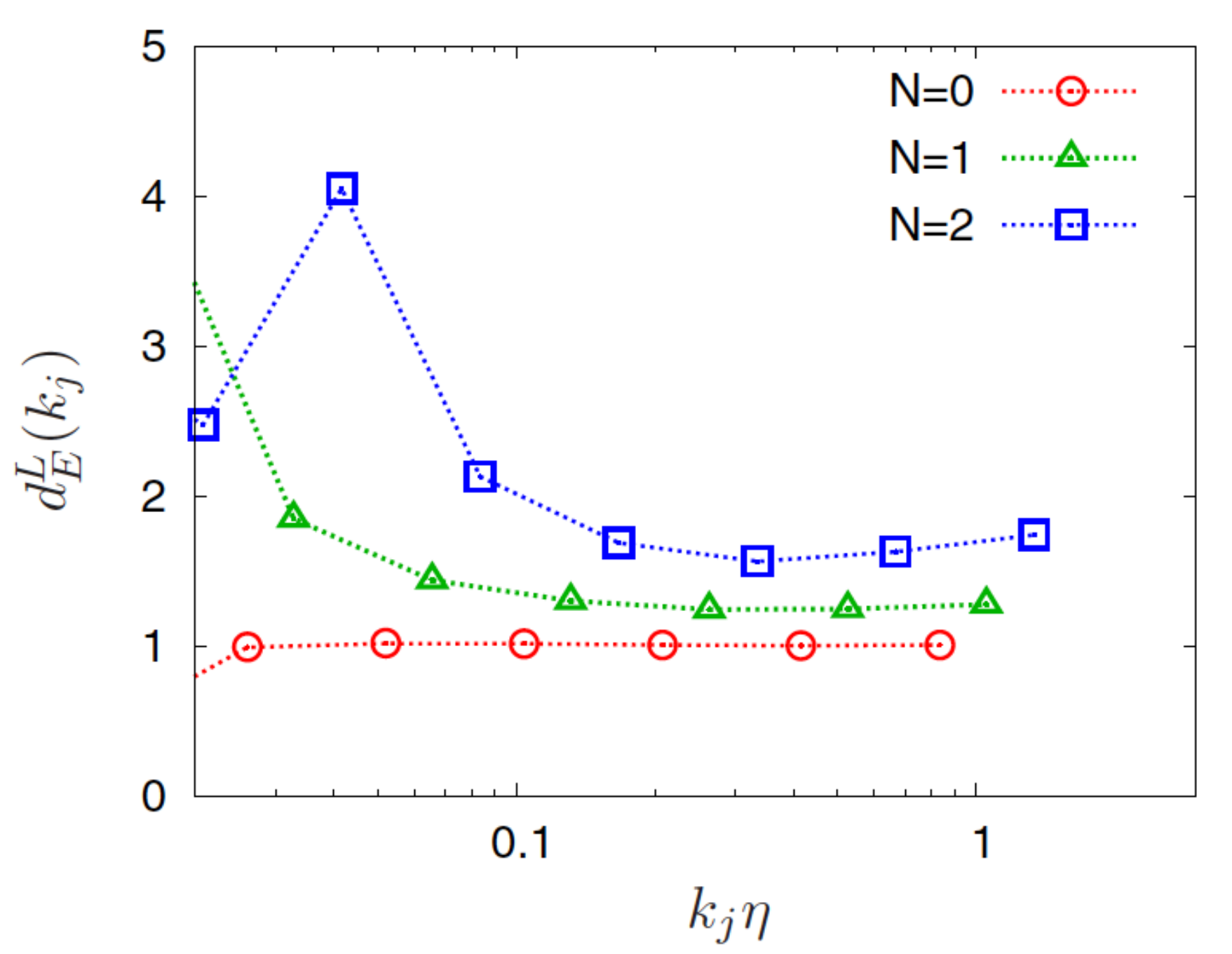}
\end{center}
\caption{QS-3D-MHD: Component-wise anisotropy measure $c_E(k_j)$ (left) and directional anisotropy measure in the longitudinal direction $d^L_E(k_j)$. From~\cite{OYSF14}.}
\label{fig:3dqs_aniso}
\end{figure}

The scale-dependent anisotropy measures allow to analyze the anisotropy at different scales.
The scale-dependent component-wise anisotropy $c_E(k_j)$ shown in Fig.~\ref{fig:3dqs_aniso}, left,
quantifies the anisotropy of the wavelet mean energy spectrum.
As expected we find for $N=0$ that $c_E(k_j) \approx 1$ as the flow is isotropic.
The departure from the value one corresponds to flow anisotropy, {\it i.e.}, for values smaller than one the energy of the parallel component is predominant of that of the perpendicular component, an obervation which holds for both cases, $N=1$ and $N=2$.
Furthermore the anisotropy is persistent at the small scales and yields smaller values for $N=2$.
Now we examine the anisotropy in different directions. 
Figure~\ref{fig:3dqs_aniso}, right, shows $d^L_E$, the flow anisotropy of the mean wavelet spectrum in the longitudinal direction.
We find that this measure yields values larger than one for $N=1$ and $2$ and values close to one for $N=0$.
For $N \ne 0$ the correlation of the velocity component parallel to the imposed magnetic field in its longitudinal direction is supposed to be stronger than the correlation of the perpendicular components.
We also see that the scale dependence gets weak for $k_j \eta > 0.1$. 
%
%

\begin{figure}[htbp]
\begin{center}
\includegraphics[width=0.48\linewidth]{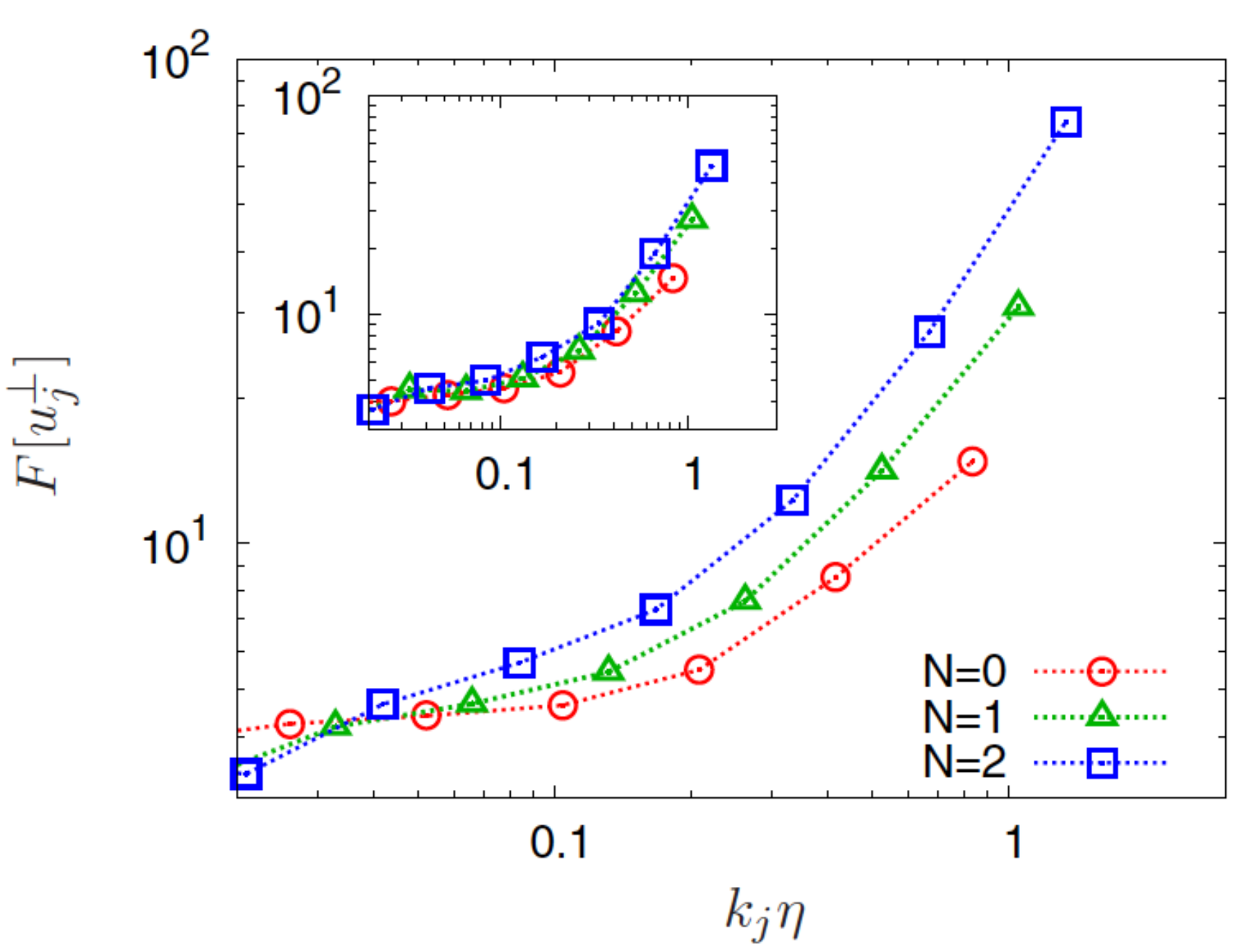}
\includegraphics[width=0.48\linewidth]{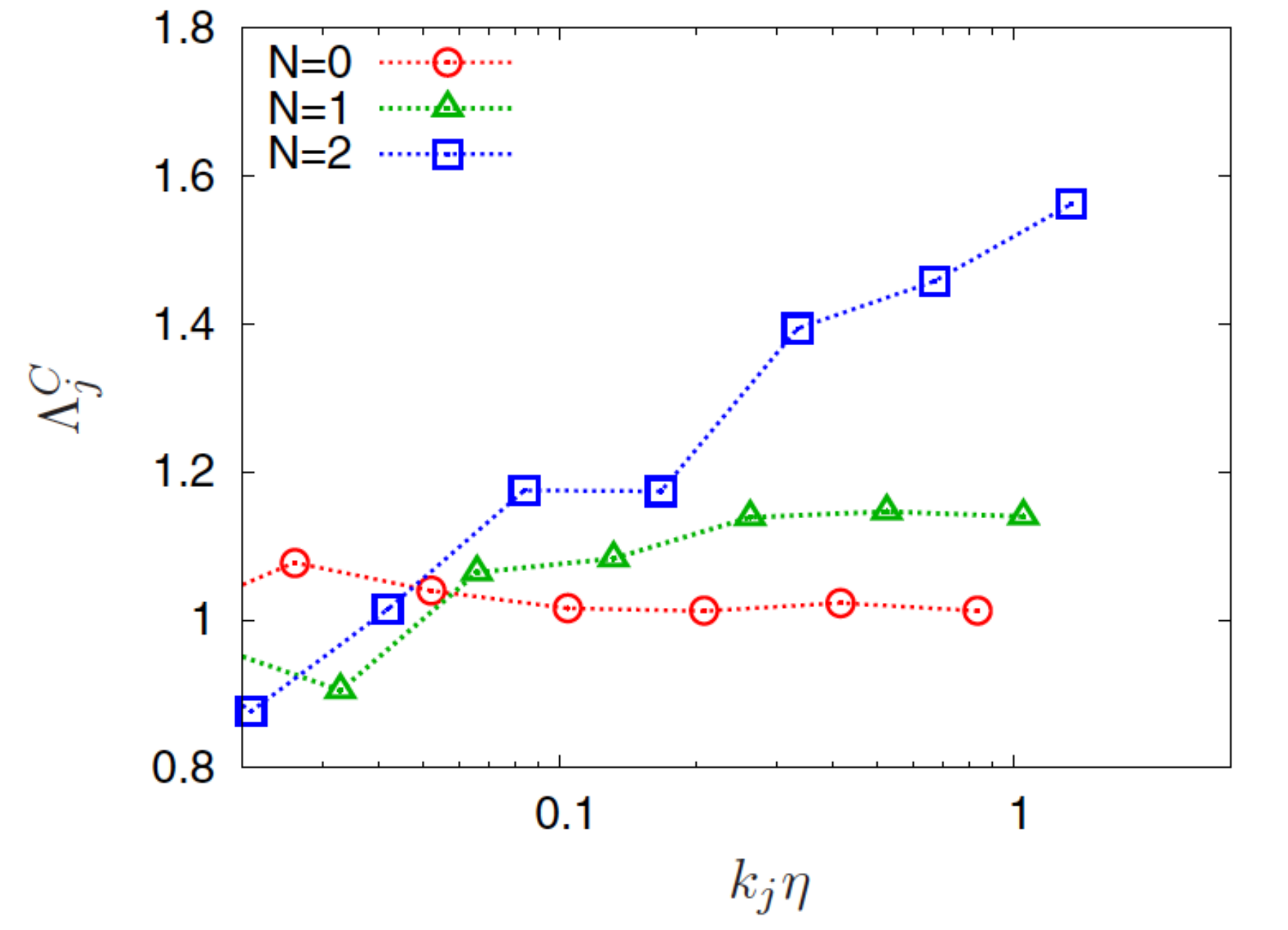}
\end{center}
\caption{QS-3D-MHD: Scale-dependent flatness of the perpendicular velocity
 $F^{\perp}_j$ with in the inset the corresponding flatness for the parallel velocity (left). Anisotropic measure of intermittency $\Lambda(k_j)$ (right). From~\cite{OYSF14}.}
\label{fig:3dqs_flat}
\end{figure}

The scale-dependent flatness of the perpendicular velocity $F[u_j^\perp]$ and of the parallel velocity $F[u_j^\parallel]$, shown in Fig.~\ref{fig:3dqs_flat}, left, quantify the intermittency of the different flow components.
In all cases we find that the flatness does indeed increase for decreasing scale.
At small scales, $k_j \eta > 1$ we also see that the flatness is larger for larger values of $N$. 
The inset shows that $F[u_j^\parallel]$  behaves similarily.

The component-wise anisotropy of the intermittency at each scale can be quantified with $\Lambda^C(k_j)$, see Fig.~\ref{fig:3dqs_flat}, right.
Again we find that for $N=0$ values close to one are found, as expected due to the isotropy of the flow.
For $N=1$ and $2$ the component-wise anisotropic intermittency $\Lambda^C(k_j)$ has values larger than one for $k_j \eta > 0.1$, which means that the perpendicular velocity becomes more intermittent than the parallel velocity at small scales. For $N=2$ this becomes even more pronounced.
%
%

\begin{figure}[htbp]
\begin{center}
\includegraphics[width=0.48\linewidth]{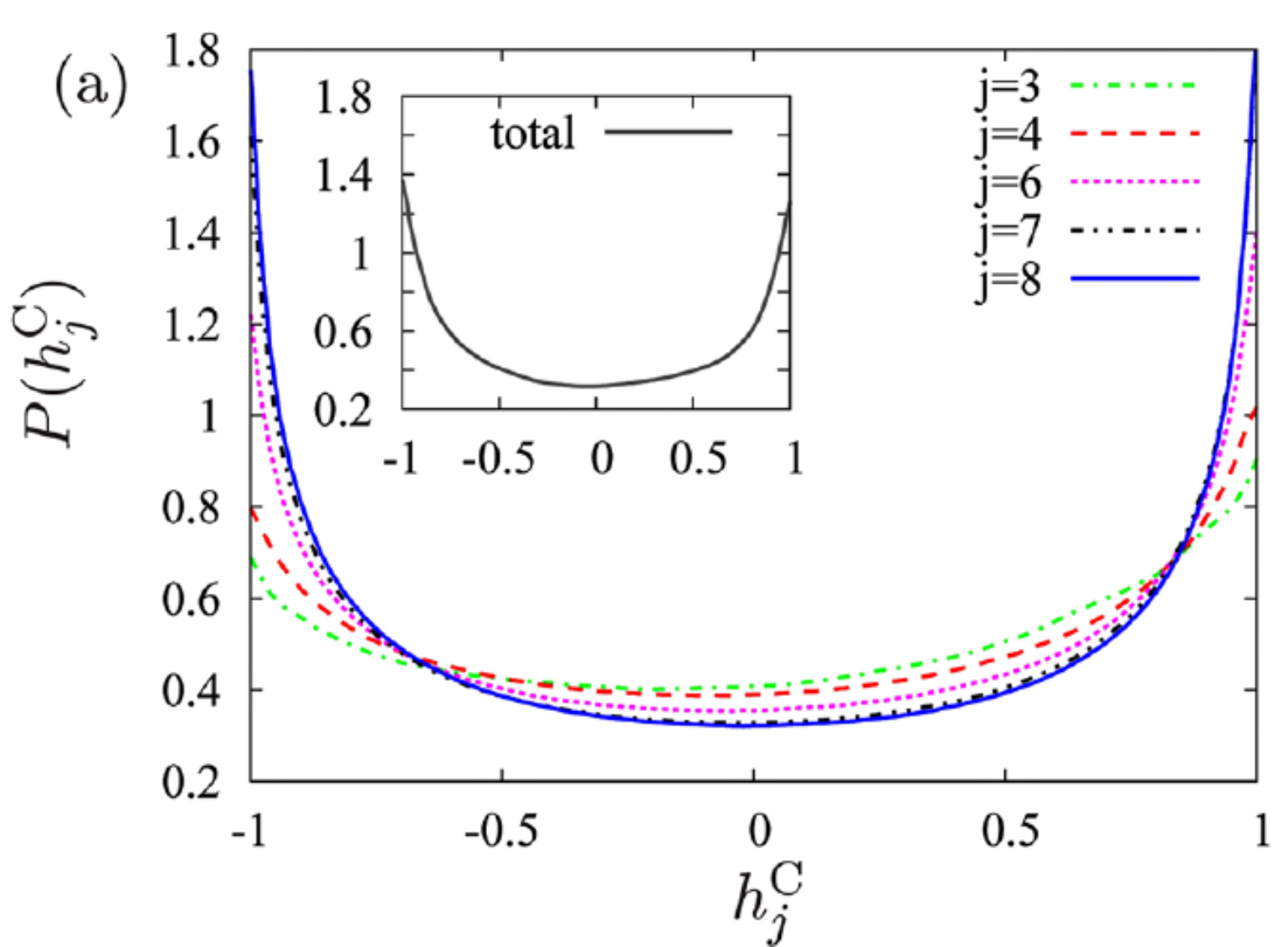}
\includegraphics[width=0.48\linewidth]{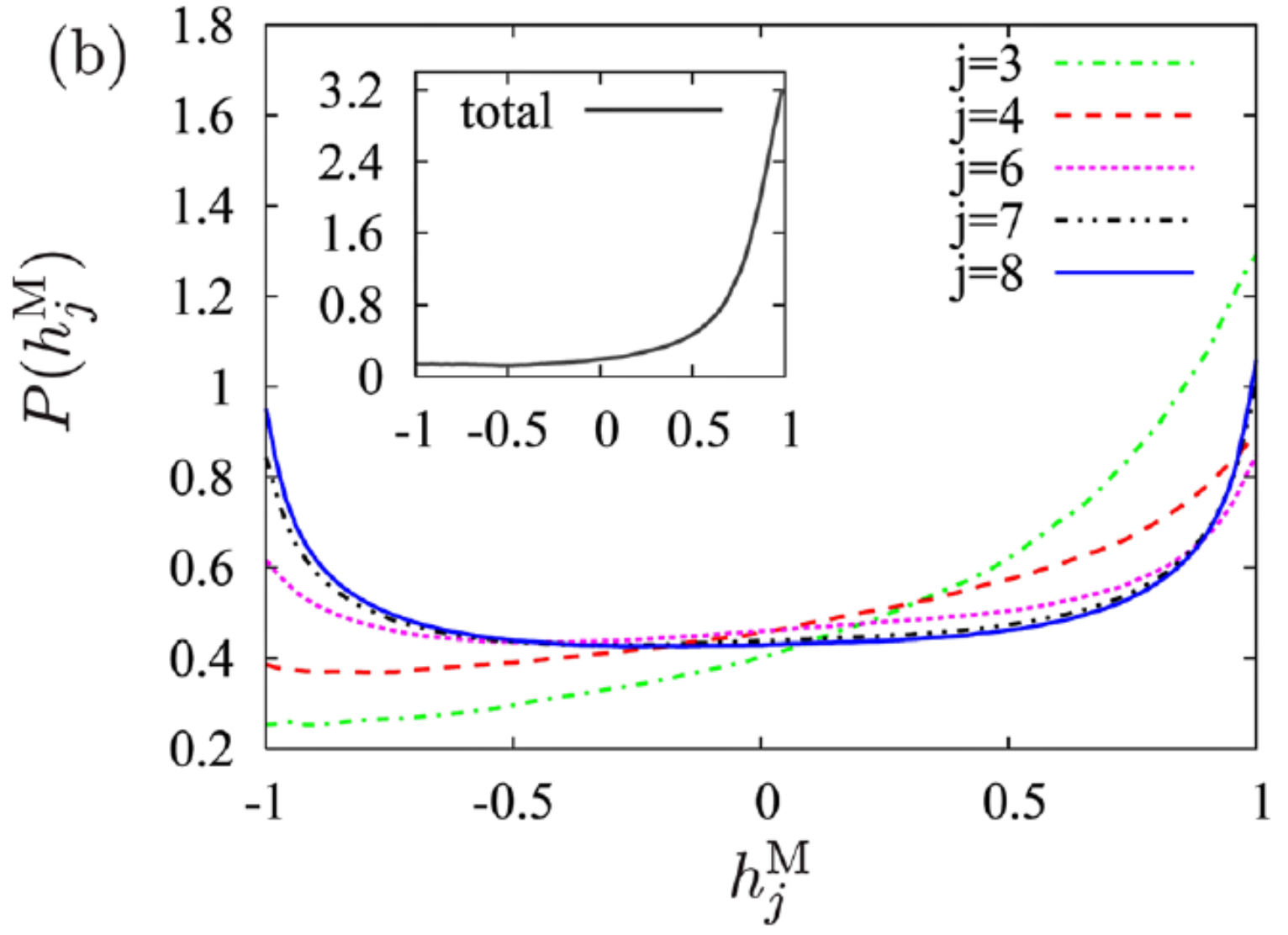}
\end{center}
\caption{3D-MHD: Scale-dependent PDFs of the relative helicities. Cross helicity $h^C_j$ (left) and magnetic helicity $h^M_j$ (right).  
The insets show the PDFs of the corresponding total relative helicities From~\cite{YSOKF11}.}
\label{fig:3dmhd_helicity}
\end{figure}

To illustrate the scale-dependent geometric statistics we consider homogeneous magnetohydrodynamic turbulence at unit Prandtl number without mean magnetic field. The flow has been computed by direct numerical simulation at resolution $512^3$ with random forcing and for further details we refer to \cite{YSOKF11}.
Figure~\ref{fig:3dmhd_helicity} shows the PDFs of the relative scale-dependent cross and magnetic helicity, $h^C_j$ and $h^M_j$.
Figure~\ref{fig:3dmhd_helicity} (left) exhibits two peaks at $h^C_j = \pm 1$ which corresponds to a pronounced scale-dependent dynamic alignment.
The peaks even become larger for smaller scales and thus the probability of alignement (or anti-alignement) of the velocity and the magnetic field increases.
Figure~\ref{fig:3dmhd_helicity} (right) illustrates that the distribution of the scale-dependent magnetic helicity becomes more symmetric at small scales.
The inset shows that the total relative magnetic helicity is strongly skewed with a peak at $+1$, which is due to the presence of substantical mean magnetic helicity.
%
%


\section{Extraction of coherent structures using wavelets}

In this section we illustrate the extraction of coherent structures using an algorithm which is based on wavelet denoising.
We first describe it for one-dimensional scalar-valued signals and illustrate its performance on an academic test signal.
We then generalize the algorithm to higher dimensions and to vector-valued fields.
Finally, different applications to experimental and numerical data are shown: 
\begin{itemize}
\item
a scalar-valued signal varying in time measured by a Langmuir probe in the scrape-off layer of the tokamak Tore Supra (Cadarache, France), 
\item
a two-dimensional academic example of the synthetic emissivity of a radiating toric shell with additive noise, 
\item
experimental movies obtained by a fast camera implemented in Tore Supra,
\item
two-dimensional vorticity fields computed for resistive drift-wave turbulence (Hasegawa-Wakatani model) using a pseudo-spectral method,
\item
three-dimensional vorticity and current density fields computed for resistive MHD turbulence (incompressible MHD equations) using a pseudo-spectral method.
\end{itemize} 

\subsection{Extraction algorithm}

\subsubsection{Principle}

We propose a wavelet-based method to extract coherent structures that emerge out of turbulent flows, both  in fluids ({\it e.g.}, vortices, shock waves in compressible fluids, ...) and in plasmas ({\it e.g.}, bursts, blobs, ...). The goal is to study their role regarding the transport and mixing properties of flows in the turbulent regime.

For this, we use the wavelet representation that keeps track of both time and scale, instead of the Fourier representation that keeps track of frequency only.
Since there is not yet an universal definition of the coherent structures encountered in turbulent flows, we use an apophatic method (introduced in Hinduist theology several thousands years ago) where one does not try to define what an entity ({\it e.g.}, a phenomenon, a noumenon, ...) is but rather what it is not.
We thus agree on the minimal and hopefully consensual statement : 
{\it 'coherent structures are not noise'}, and propose to define them as : {\it 'coherent structures are what remains after denoising'}. 

The mathematical definition of noise states that a signal is a noise if it cannot be compressed in any functional basis. 
As a result the shortest description of a noise is itself.
Note that in most of the cases the experimental noise generated by a measure device does not fit the definition of mathematical noise since it could be compressed in at least one functional basis ({\it e.g.}, parasite frequencies can be removed in the Fourier basis).

This new way of defining coherent structures allows to process signals and fields, but also their cuts or projections ({\it e.g.}, a probe located at one point provides a one dimensional cut of a four dimensional space-time field). 
Indeed, the algorithms commonly used to extract coherent structures cannot work for cuts or projections, because they require a template of the structures to extract (one would need to take into account how the probe sees all possible translations and distortions of the coherent structures).
The strength of our algorithm is that it treats fields and projections the same way.

Since we assume that coherent structures are what remains after denoising, we need a model, not for the structures themselves, but for the noise. 
Applying `Ockham's Razor principle' (or the `law of parsimony'), we choose as first guess the simplest possible model: we suppose the noise to be additive, Gaussian and white ({\it i.e.}, uncorrelated).
We then project the turbulent signal (in 1D), or turbulent field (in higher dimensions), into wavelet space and retain only the coefficients having their modulus larger than a given threshold.
As threshold value we follow Donoho and Johnstone's proposition of a threshold value that depends on the variance of the Gaussian noise we want to remove and on the chosen sampling rate \cite{DoJo94}.
Since the noise variance is not known {\it a priori} for turbulent signals (the noise being produced by their intrinsic nonlinear dynamics), we designed a recursive method \cite {AzFaSc04} to estimate it from the variance of the weakest wavelet coefficients, {\it i.e.}, those whose modulus is below the threshold value. 
After applying our algorithm we obtain two orthogonal fields: the coherent field retaining all coherent structures and the incoherent field corresponding to the noise.
We then check {\it a posteriori} that the latter is indeed noise-like ({\it i.e.},  spread all over physical space), Gaussian and uncorrelated ({\it i.e.}, also spread all over Fourier space), and thus confirm the hypotheses we have {\it a priori} chosen for the noise.
\subsubsection{Wavelet denoising}
We consider a signal $s(t)$ sampled on $N = 2^J$ points that we want to denoise, assuming the noise to be additive, Gaussian and white.
We first project $s(t)$ onto an orthogonal wavelet basis and then filter out some of the wavelet coefficients thus obtained, $\widetilde s_{ij}$. 
We retain only the wavelet coefficients whose modulus is larger than a threshold value.
The main difficulty is to estimate it {\it a priori} and we encounter two possible cases:
\begin{itemize}
\item
If we {\it a priori} know the noise's variance $\sigma^2$, the optimal threshold value is given by Donoho and Johnstone's formula \cite{DoJo94} 
\begin{equation}
\label{threshold}
\epsilon = (2  \sigma^2 \ln N)^{1/2} \; .
\end{equation}
In 1994 they proved \cite{DoJo94} that such a wavelet thresholding method is optimal to denoise signals in presence of additive Gaussian white noise, because it minimizes the maximal $L^2$-error (between the denoised signal and the noise-free signal) for functions whose regularity is inhomogeneous, such as bursty or intermittent turbulent signals.
\item
If we do not {\it a priori} know the variance of the noise, that is the most usual case, one should use the wavelet-based recursive algorithm we proposed in 1999 \cite{FSK99, AzFaSc04}.
This algorithm first estimates the variance of the noise by considering the variance of the noisy signal $\sigma_0^2$ and computes the corresponding threshold 
\begin{equation}
\label{threshold}
\epsilon_0 = (2  \sigma_0^2 \ln N)^{1/2} \; .
\end{equation}

The algorithm splits the wavelet coefficients into two classes: the weak coefficients whose modulus is smaller than the threshold,  and the remaining strong coefficients.
It then computes the variance of the weak coefficients $\sigma_n$ to obtain a better estimation of the variance of the noise  (estimated from the wavelet coefficients using Parseval's theorem) 

\begin{equation}
\label{sigman}
\sigma^2_n =\frac{1}{N} \, \sum_{(j,i) \in {\cal I}^J, |\widetilde{s}_{ji}| < \epsilon_n} |\widetilde{s}_{ji}|^2
\end{equation}

where ${\cal I}^J = \{ 0 \le j < J , i = 0, ..., 2^j -1  \}$ is the index set of the wavelet coefficients.
The algorithm then replaces $\epsilon_0$ by $\epsilon_n = (2  \sigma_n^2 \ln N)^{1/2}$, that yields a better estimate of the threshold.
This procedure is iterated until it reaches the optimal threshold value, when $\epsilon_{n+1} \approx \epsilon_n$.

In \cite{AzFaSc04} we proved that this algorithm converges for signals having a sufficiently sparse representation in wavelet space, such as the intermittent signals encountered in turbulence.
We also showed that the larger the signal to noise ratio is, the faster the convergence.
Hence, if the signal $s(t)$ is only a noise it converges in one iteration and retains $\epsilon_0$ as the optimal threshold. 
\end{itemize}

Using the optimal threshold we then separate the wavelet coefficients $\widetilde s_{ij}$ into two contributions: the coherent coefficients $\widetilde s^C_{ij}$ whose modulus is larger than $\epsilon$ and the remaining incoherent coefficients $\widetilde s^I_{ij}$.
Finally, the coherent component $s^C(t)$ is reconstructed in physical space using the inverse wavelet transform,
while the incoherent component is obtained as $s^I(t)=s(t)-s^C(t)$.

\subsubsection{Extraction algorithm for one-dimensional signals}
We detail the iterative extraction algorithm for the one-dimensional case and quote it from \cite{AzFaSc04}:

\smallskip

\noindent {\bf Initialization}

\begin{itemize}
\item[$\bullet$] given the signal $s(t)$ of duration $T$, 
			 sampled on an equidistant grid $t_i = i T/N$ for $i = 0, N-1$, with $N=2^J$,
\item[$\bullet$] set $n=0$ and perform a wavelet decomposition, {\it i.e.}, apply the Fast Wavelet
                 Transform \cite{Mallat98} to $s$ to obtain the wavelet coefficients
                 $\widetilde{s}_{ji}$ for $(j,i) \in {\cal I}^J$,
\item[$\bullet$] compute the variance $\sigma^2_0$ of ${s}$ as a rough estimate of the variance of 
			 the incoherent signal ${s^I}$ 
                 and compute the corresponding threshold $\epsilon_0=\left( 2 \ln N \sigma^2_0 \right)^{1/2}$, 
                 where $\sigma^2_0 = \frac{1}{N}\sum_{(j,i) \in {\cal I}^J} |\widetilde{s}_{ji}|^2\label{eqn:0}$,
\item[$\bullet$] set the number of coefficients considered as noise to $N_{I}=N $,
{\it i.e.}, to the total number of wavelet coefficients.
\end{itemize}

\smallskip

\noindent {\bf Main loop}

\noindent Repeat

\begin{itemize}
\item   set $N_I^{old}=N_{I}$ and count the number of wavelet coefficients smaller than $\epsilon_n$,
which yields a new value for $N_I$,
\item   	compute the new variance $\sigma^2_{n+1}$ from the wavelet coefficiens smaller than $\epsilon_n$, {\it i.e.},
		\mbox{$\sigma^2_{n+1}=\frac{1}{N} \sum_{(j,i) \in {\cal I}^J} |\widetilde{s}^I_{ji}|^2$}, 
		where
               \begin{eqnarray}
                \widetilde{s}^I_{ji} \, = \, \left \{
                \begin{array}{ll}
                  \widetilde{s}_{ji} &  \mbox{\rm for} \; |\widetilde{s}_{ji}| \le \epsilon_n \\
                  0  &  \mbox{\rm else}, \\
                \end{array}
                \right.
               \end{eqnarray}
and the new threshold 
$\epsilon_{n+1}= (2\ln N \sigma^2_{n+1})^{1/2}$,
\item  set $n=n+1$
\end{itemize}

\noindent
until ($N_I$==$N_I^{old}$).

\smallskip

\noindent {\bf Final step}
\begin{itemize}
\item reconstruct the coherent signal $s^C$ from the coefficients $\widetilde{s}^C_{ji}$
	using the inverse Fast Wavelet Transform,
	where
               \begin{eqnarray}
                \widetilde{s}^C_{ji} \, = \, \left \{
                \begin{array}{ll}
                  \widetilde{s}_{ji} &  \mbox{\rm for} \; |\widetilde{s}_{ji}| > \epsilon_n \\
                  0  &  \mbox{\rm else} \\
                \end{array}
                \right. 
               \end{eqnarray}

\item finally, compute pointwise the incoherent signal $s^I(t_i)={s}(t_i)-s^C(t_i)$ for $i=0, ..., N-1$.
\end{itemize}

\noindent {\bf End}

\medskip

Note that the signal is split into $s(t)  =  s^C(t) + s^I(t)$ and its energy into $\sigma^2 = \sigma_C^2 + \sigma_I^2$, since the coherent and incoherent components are orthogonal, {\it i.e.},
$\langle s^C , s^I \rangle = 0$.

We use the Fast Wavelet Transform (FWT) \cite{Mallat98} that is computed with $(2 M N)$ multiplications, $M$  being the length of the discrete filter defining the 
orthogonal wavelet used.
Remark: for all applications presented in this paper, we use Coiflet 12 wavelets \cite{Daub92}, unless otherwise stated.
As long as the filter length $M <  \frac{1}{2} \log_2 N$, the FWT is faster than the FFT (Fast Fourier Transform) computed with $N \log_2 N$ operations.
Consequently, the extraction algorithm requires $(2 n M N)$ operations, $n$ being the number of iterations, that is small, 
typically less than  $\log_2 N$.

This algorithm defines a sequence of estimated thresholds 
$\left(\epsilon_n\right)_{n \in \N}$ and the corresponding sequence of estimated variances $\left(\sigma^2_n \right)_{n \in \N}$.
In \cite{AzFaSc04} we proved that this sequence converges after a finite number of iterations by applying a fixed point type argument to the iteration function

\begin{equation}\label{for:iteration function}
{\cal F}_{{s},N}(\epsilon_{n+1})=\left(\frac{2\ln N}{N} \sum_{(j,i) \in {\cal I}^J} |\widetilde{s}^I_{ji}(\epsilon_n)|^2  \right)^{1/2} \; .
\end{equation}

The algorithm stops after $n$ iterations, when ${\cal F}_{{s},N}(\epsilon_n)=\epsilon_{n+1}$, since the number of samples $N$ is finite.
In \cite{AzFaSc04} we have also proved that the convergence rate depends on the signal to noise ratio 
($SNR=10 \log_{10}  (\sigma^2 /  \sigma_I^2$)), 
since the smaller the SNR, {\it i.e.}, the stronger the noise, the faster the convergence is.
Moreover, if the algorithm is applied to a Gaussian white noise, it converges in one iteration only.
If it is applied to a signal without noise, the signal is fully preserved.
In \cite{AzFaSc04}  we have also proven the algorithm's idempotence, {\it i.e.}, if it is applied several times the noise is
eliminated the first time and the coherent signal will remain the same if the algorithm is reapplied several times.
This would be the case for a Gaussian filter which, in contrast, is not idempotent.

\subsubsection{Application to an academic test signal}

\begin{figure}[htbp!]
\begin{center}
\includegraphics[height=4.4cm, width=10cm]{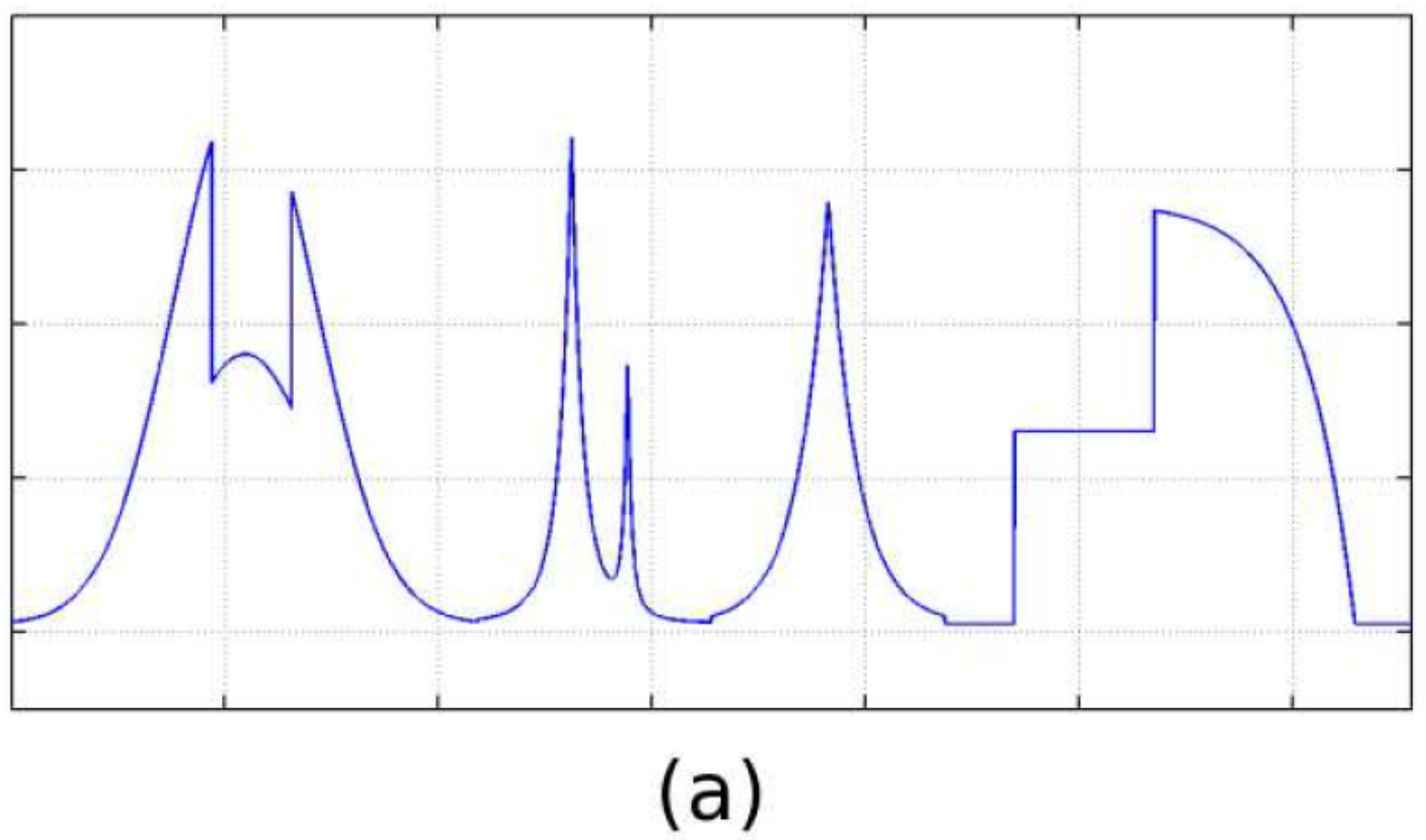} 
\includegraphics[height=4.4cm, width=10cm]{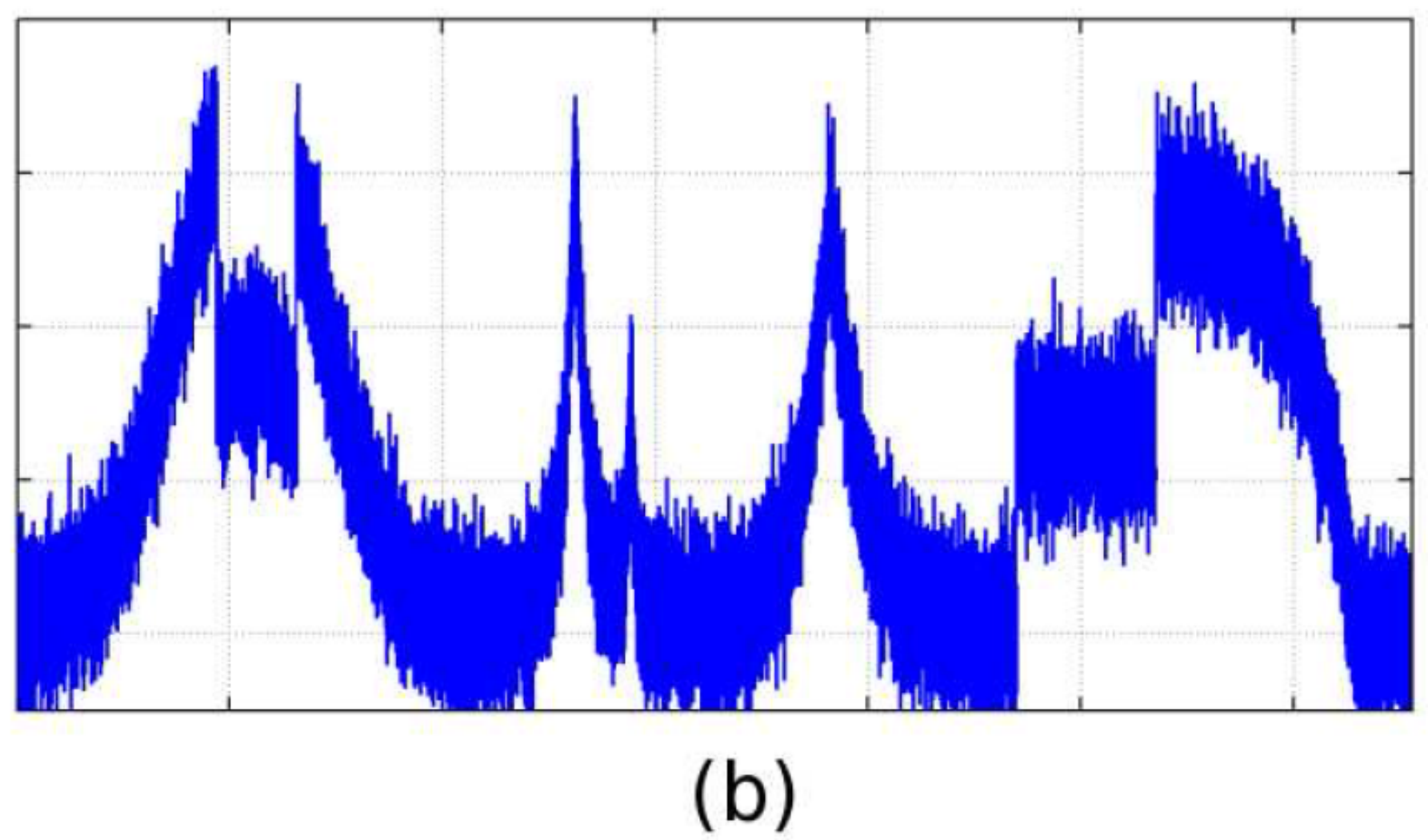} 
\includegraphics[height=4.4cm, width=10cm]{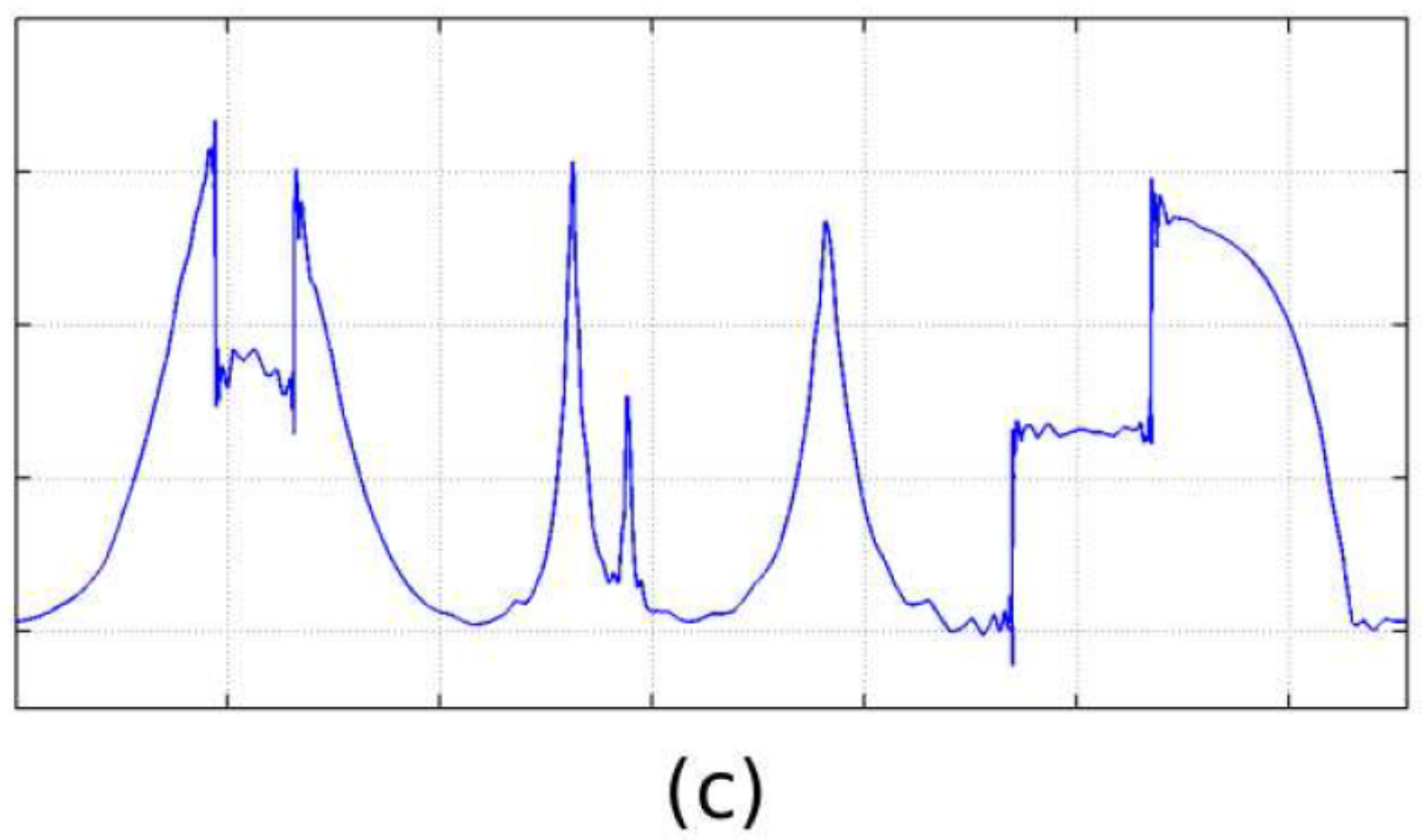} 
\end{center}
\caption{Denoising of a piecewise regular signal using iterative wavelet thresholding. 
Top: original academic signal $a(t)$. Middle: Noisy signal $s(t)$ with a SNR = 11 dB. Bottom: Denoised signal $s^C(t)$ with a SNR = 28 dB. 
}
\label{fig:ursi}
\end{figure}

To illustrate the performance of the iterative algorithm we consider a one-dimensional noisy
test signal ${s}(t)$ sampled on $N=2^{13}=8192$ points (Fig.~\ref{fig:ursi}, middle).
It is made by adding a Gaussian white noise $w(t)$, of mean zero and variance $\sigma^2_w=25$, to a piecewise regular academic signal ${a}(t)$ presenting several discontinuities,  in the function or in its derivatives
(Fig.~\ref{fig:ursi}, top).
The signal to noise ratio is $SNR = 10 \log_{10} (\sigma^2_a / \sigma^2_w)=11 \, dB$.
After applying the extraction algorithm we estimate the noise variance to be $25.6$ and we obtain a coherent signal $s^C(t)$ very close to the original academic signal $a(t)$ (Fig.~\ref{fig:ursi}, bottom).
The incoherent part $s^I(t)$ is homogeneous and noise like with flatness $ 3.03$, which corresponds to quasi--Gaussianity.
%
%
In Fig.~\ref{fig:ursi} (bottom) we observe that the coherent signal retains all discontinuities and peaks present in the academic signal $a(t)$,
which is an advantage with respect to standard denoising techniques, {\it e.g.}, low pass Fourier filtering, which have smoothed them.
In the vicinity of the discontinuities we observe slight overshoots, which are more local than the classical Gibbs phenomena and could for example be removed using the translation invariant wavelet transform \cite{Mallat98}.

\subsubsection{Extension of the algorithm to higher dimensional scalar and vector-valued fields}

The extraction algorithm was described in section~3.1.3 for one-dimensional scalar-valued signals $s(t)$ varying in time.
First, it can be extended to higher-dimensional scalar fields $s({\bm x})$ varying in space ${\bm x} \in \R^d$ where $d$ is the space dimension.
To this end the extraction algorithm only requires that the one-dimensional wavelets are replaced by their equivalent $d$-dimensional wavelets
using tensor product constructions, see, {\it e.g.},  \cite{Daub92, Mallat98, ScFa06}.

Second, the extraction algorithm can also be extended to vector-valued fields ${\bm v} = (v^{(1)}, ..., v^{(d)} )$ where each component $v^{\ell}, \ell= 1, ...,d$
is a scalar valued field. The extraction algorithm is then applied to each component of the vector field.
For thresholding the wavelet coefficients we consider the vector ${\widetilde {\bm v}}_{j,\mu,{\bm i}}$ in eq.~(\ref{OWS}).
Assuming statistical isotropy of the noise, the modulus of the wavelet coefficient vector is computed.
The coherent contribution is then reconstructed from those coefficients whose modulus is larger than the threshold defined as
$\epsilon= (2/d  \;  \sigma^2 \ln N)^{1/2}$
where $d$ is the dimensionality of the vector field, $\sigma$ the variance of the noise and $N$ the total number of grid points.
The iterative algorithm in section~3.1.3 can then be applied in a straightforward way.

To extract coherent structures out of turbulent flows we consider the vorticity field, which is decomposed in wavelet space.
Applying the extraction algorithm then yields two orthogonal components, the coherent and incoherent vorticity fields.
Subsequently the corresponding induced velocity fields can be reconstructed by applying
the Biot--Savart kernel, which is the inverse curl operator. 
For MHD turbulence, we consider in addition the current density and we likewise split it into two components,
the coherent and incoherent current density fields.
Using Biot--Savart's kernel we reconstruct the coherent and incoherent magnetic fields.

Note that the employed wavelet bases do not a priori constitute divergence-free bases.
Thus the resulting coherent and incoherent vector fields are not necessarily divergence free.
However, we checked that the departure from incompressibility only occurs in the dissipative range and remains negligible \cite{YKSOHF09}.  
Another solution would be to use directly div-free wavelets, but they are much more cumbersome to implement \cite{DFS10}.

\subsection{Application to 1D experimental signals from tokamaks}

\begin{figure}[ht!]
\begin{center}
\includegraphics[width= 0.55 \linewidth]{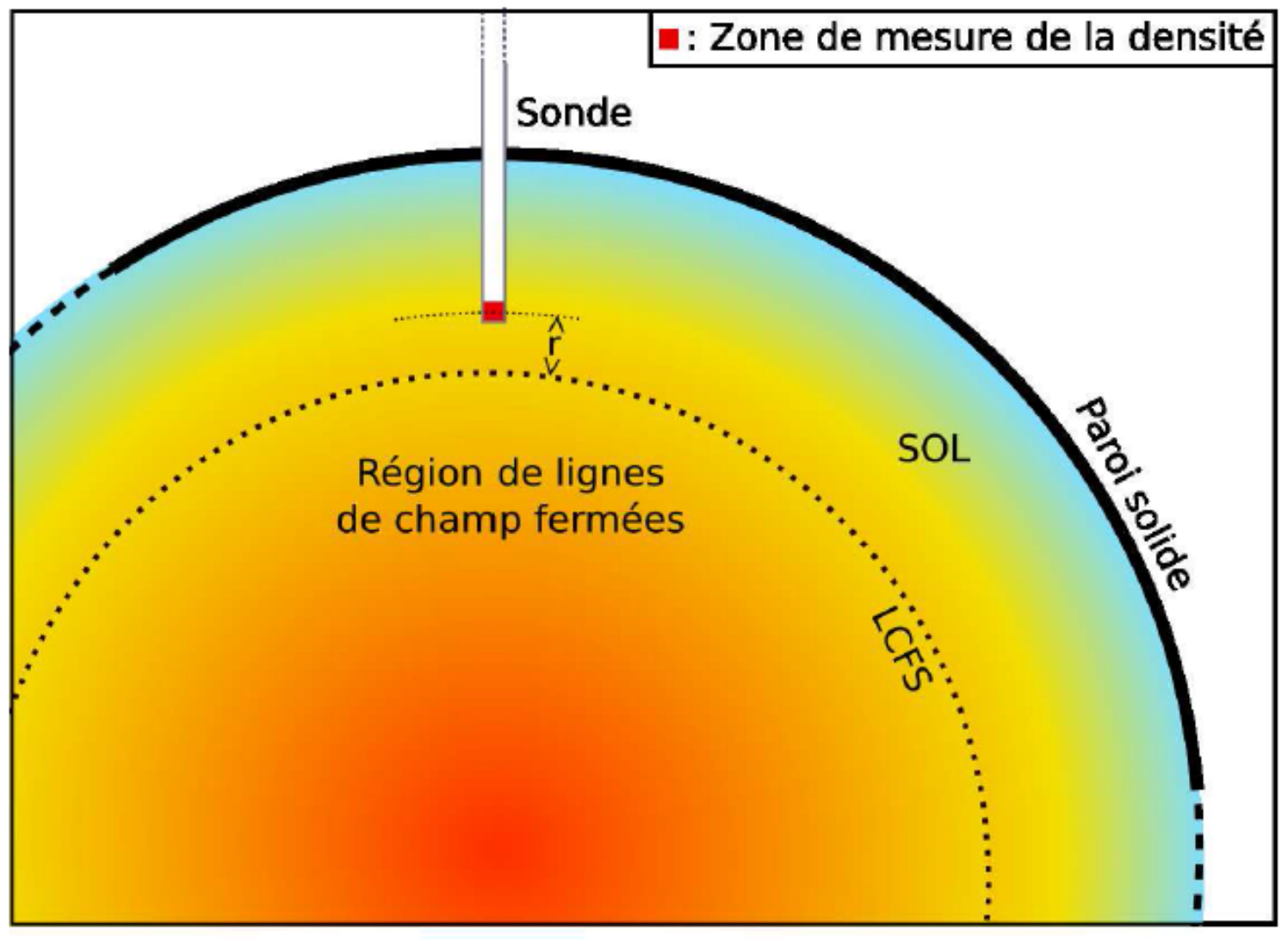} 
\includegraphics[width= 0.55 \linewidth]{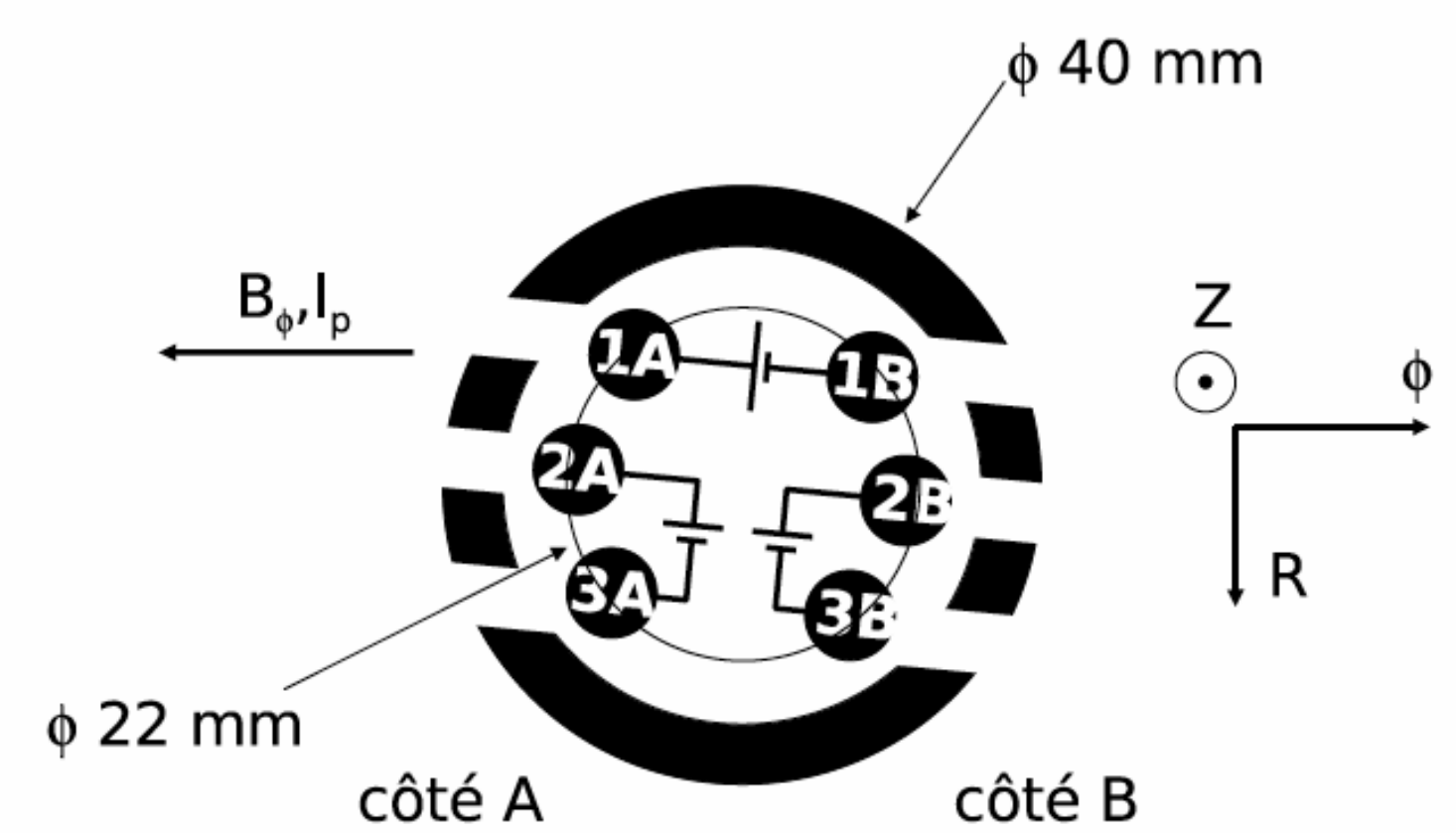} 
\end{center}
\caption{Left: Position of the reciprocating Langmuir probe in the scrape-off layer of the tokamak Tore Supra in Cadarache.
Right: Schematic top view of the probe. }
\label{fig:ursi_2+3}
\end{figure}

\begin{figure}[htbp!]
\begin{center}
\includegraphics[height=4.6cm]{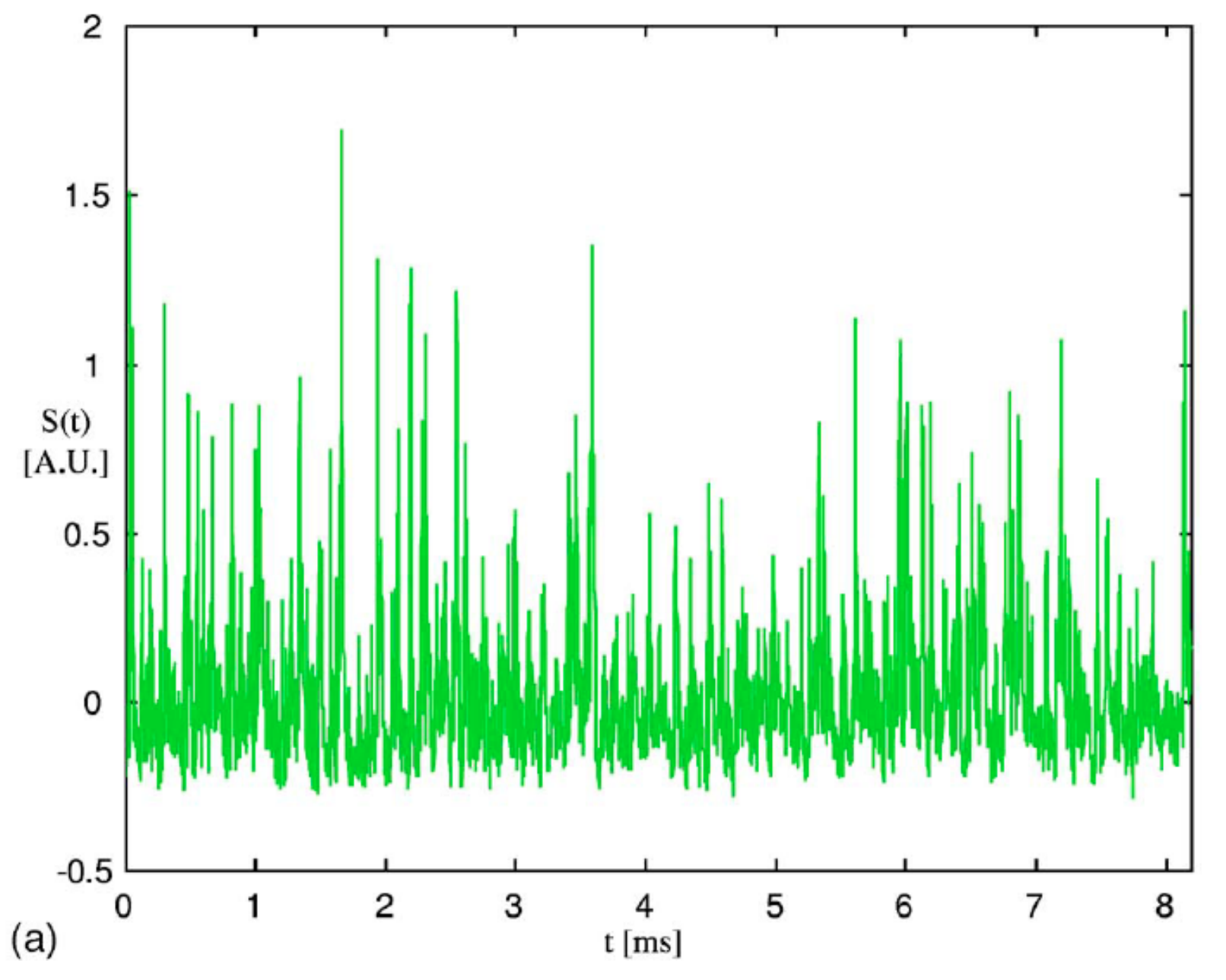} \\
\includegraphics[height=4.6cm]{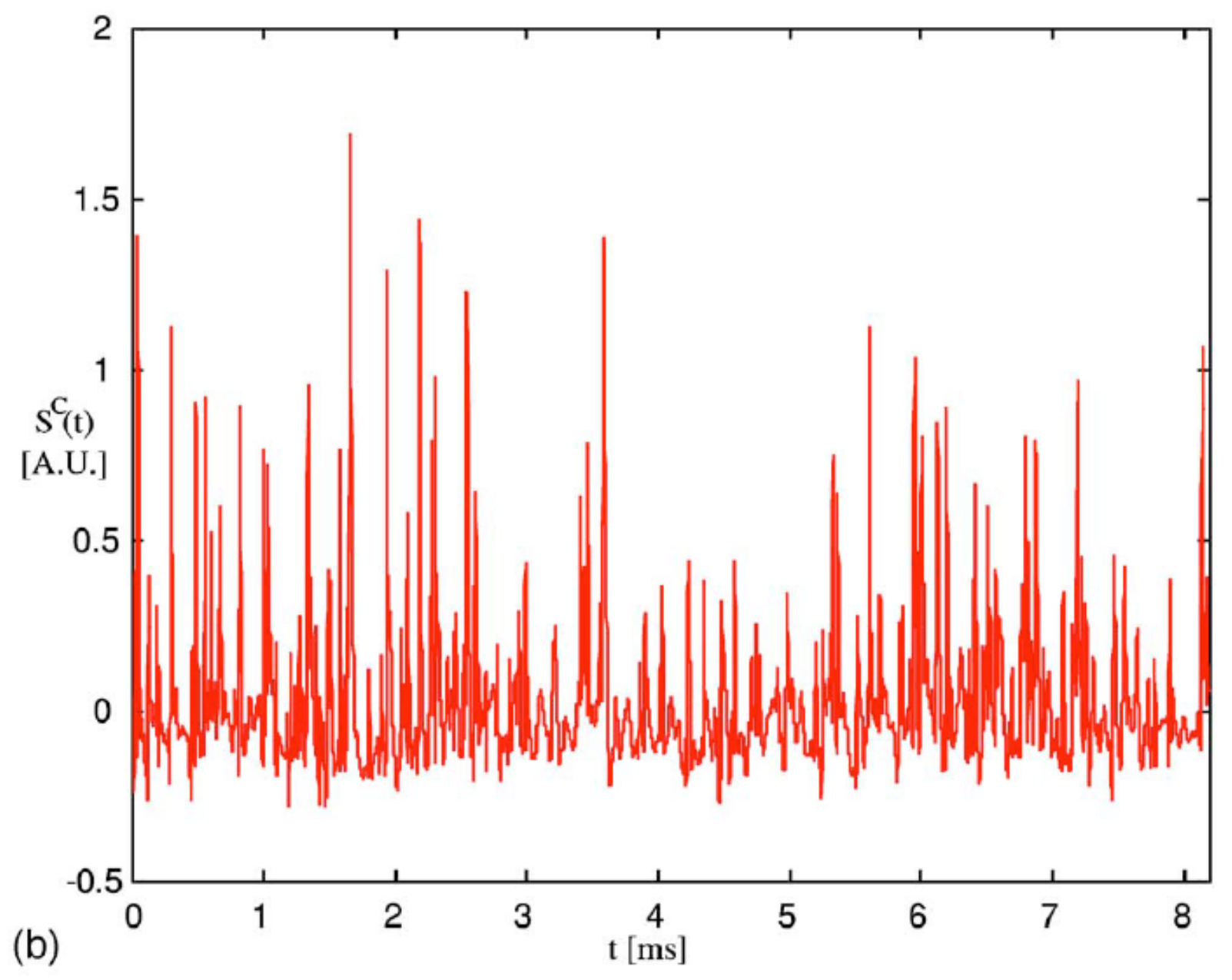}
\includegraphics[height=4.6cm]{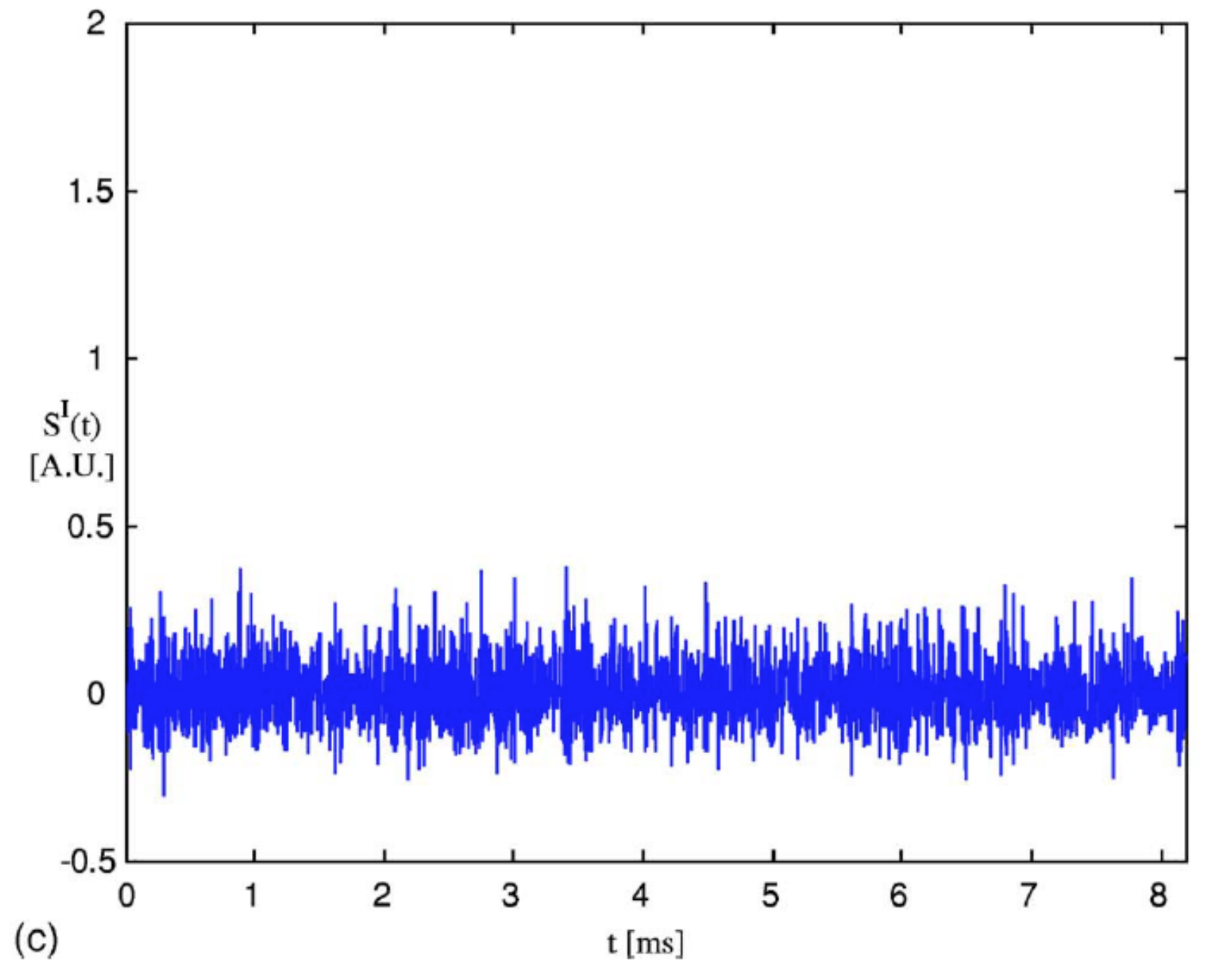}
\end{center}
\caption{Signal $s(t)$ of duration $8.192$ ms, corresponding to the saturation
current fluctuations measured at $1$ MHz in the scrape-off layer of the tokamak Tore Supra (Cadarache, France).
Top: total signal $s$, bottom left coherent part $s_C$, and bottom right incoherent part $s_I$.
From \cite{FaSD06}.}
\label{fig:pop06_1}
\end{figure}

In \cite{FaSD06} we presented a new method to extract coherent bursts from turbulent signals.
Ion density plasma fluctuations were measured by a fast reciprocating
Langmuir probe in the scrape-off layer of the tokamak Tore Supra (Cadarache, France), for a schematic view we refer to Fig.~\ref{fig:ursi_2+3}. 
%
%
The resulting turbulent signal is shown in Fig.~\ref{fig:pop06_1} (top).
To extract the coherent burst the wavelet representation is used which keeps track of both time and scale and 
thus preserves the temporal structure of the analyzed signal, in contrast to the Fourier representation which scrambles it among the phases
of all Fourier coefficients. 
Applying the  extraction algorithm described in section~3.1.3 the turbulent signal in Fig.~\ref{fig:pop06_1} (top) is
decomposed into coherent and incoherent components (Fig.~\ref{fig:pop06_1}, bottom).
Both signals are orthogonal to each other and their properties can thus be studied
independently. 
This procedure disentangles the coherent bursts, which contain most of the density variance, are
intermittent and correlated with non-Gaussian statistics, from the incoherent background
fluctuations, which are much weaker, non-intermittent, noise-like and almost decorrelated with
quasi-Gaussian statistics. 
%

\begin{figure}[htbp!]
\begin{center}
\includegraphics[height=4.6cm]{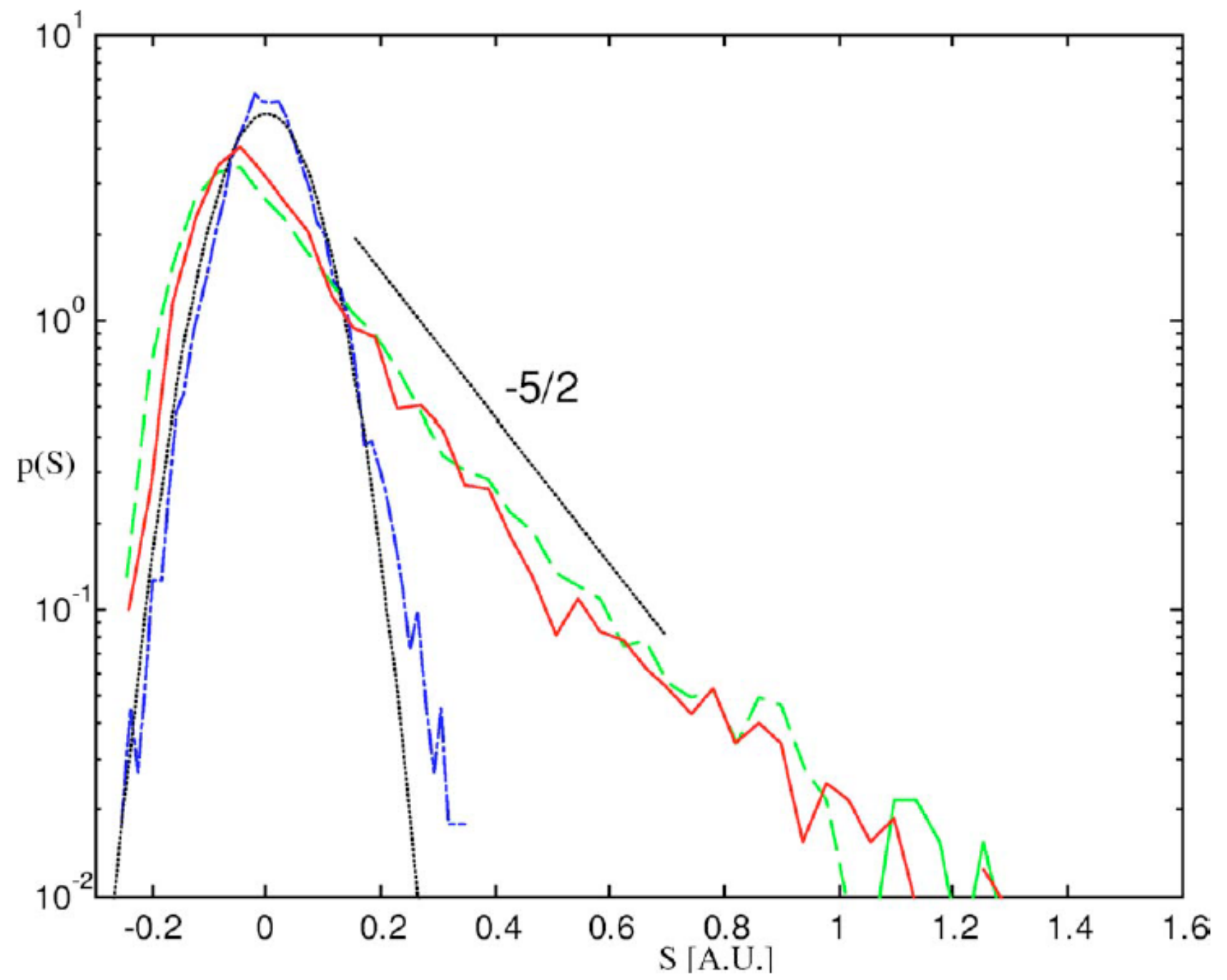} 
\end{center}
\caption{Probability density function $p(s)$ estimated using histograms with $50$ bins.
PDF of the total signal $s$ (green dashed line), of the coherent component $s_C$ (red solid line) and of the incoherent component $s_I$ (blue dotted line, together with a Gaussian fit with variance $\sigma^2_I$ (black dotted line). From \cite{FaSD06}.}
\label{fig:pop06_4}
\end{figure}
\begin{figure}[htbp]
\begin{center}
\includegraphics[height=4.5cm]{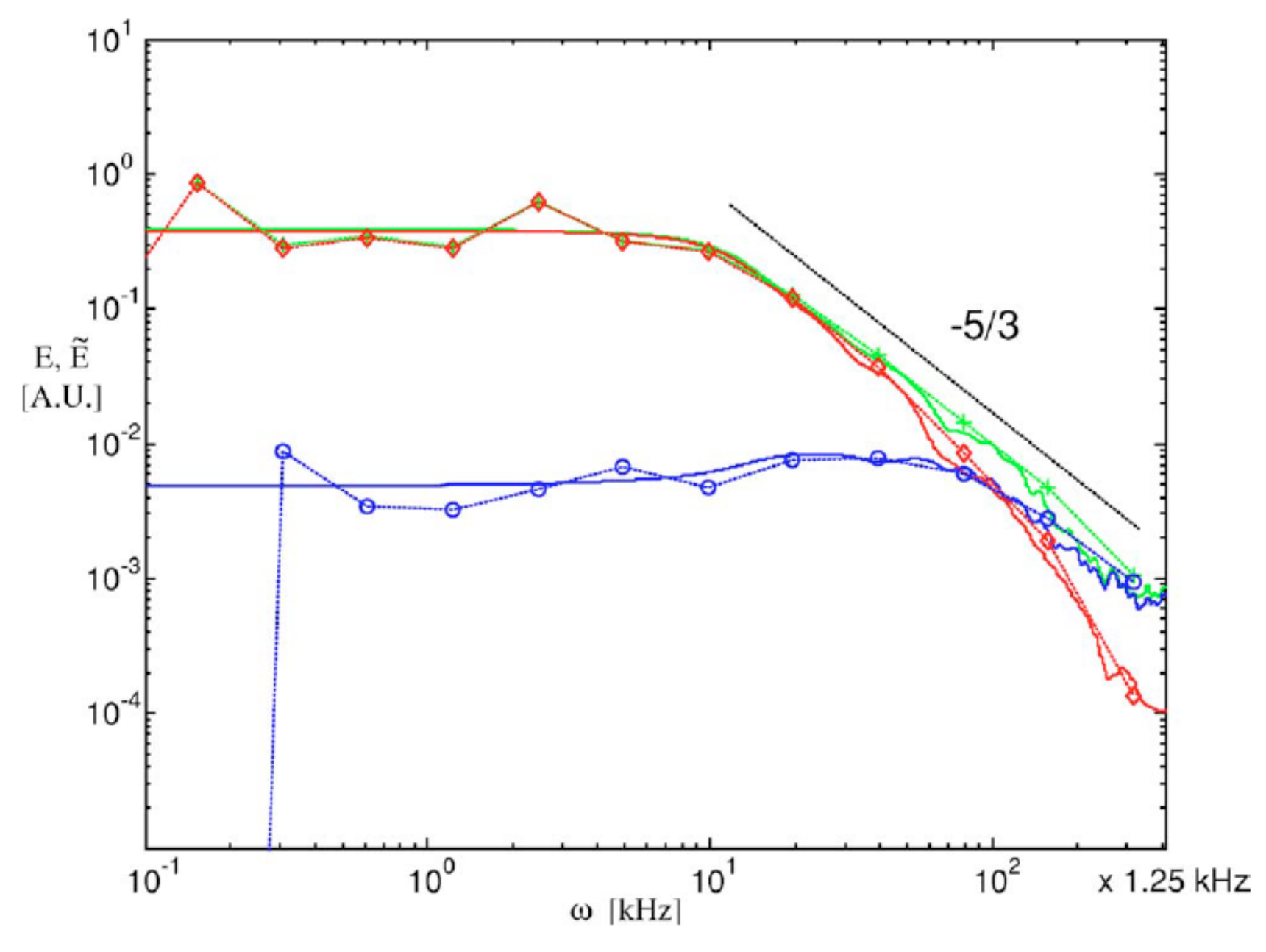} 
\includegraphics[height=4.5cm]{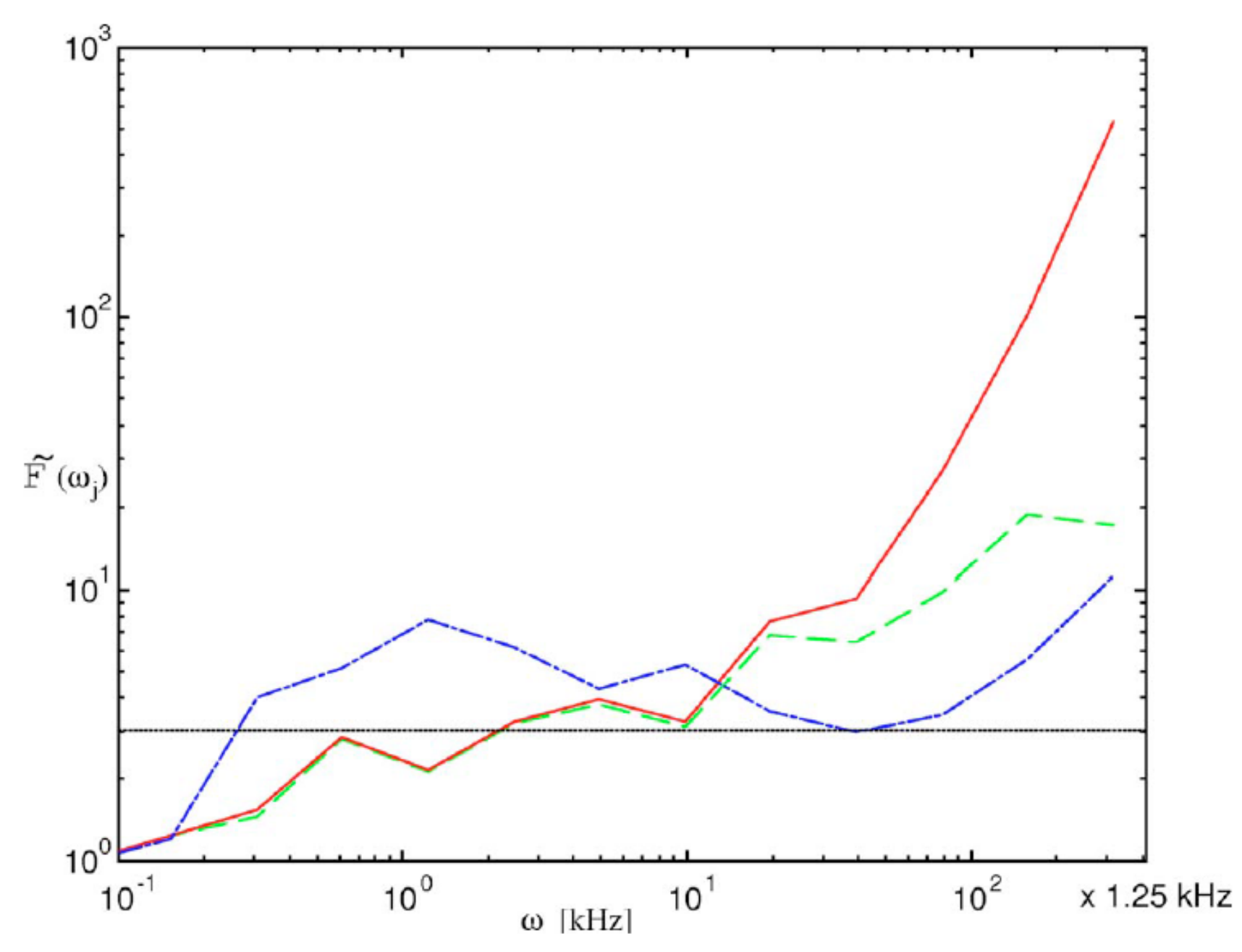} 
\end{center}
\caption{Left: wavelet spectra $\widetilde E(\omega_j)$ (lines with symbols) and modified periodograms $E(\omega)$ (lines) of the total signal $s$ (green and $+$) , coherent signal $s_C$ (red and $\diamond$) and incoherent signal $s_I$ (blue and $\circ$).
Right: corresponding scale dependent flatness $\widetilde F$ vs frequency $\omega_j$. 
The horizontal dotted line $\widetilde F (\omega_j) = 3$ corresponds to the flatness of a Gaussian process.
From \cite{FaSD06}.}
\label{fig:pop06_5+6}
\end{figure}
\begin{figure}[htbp]
\begin{center}
\includegraphics[width= \linewidth]{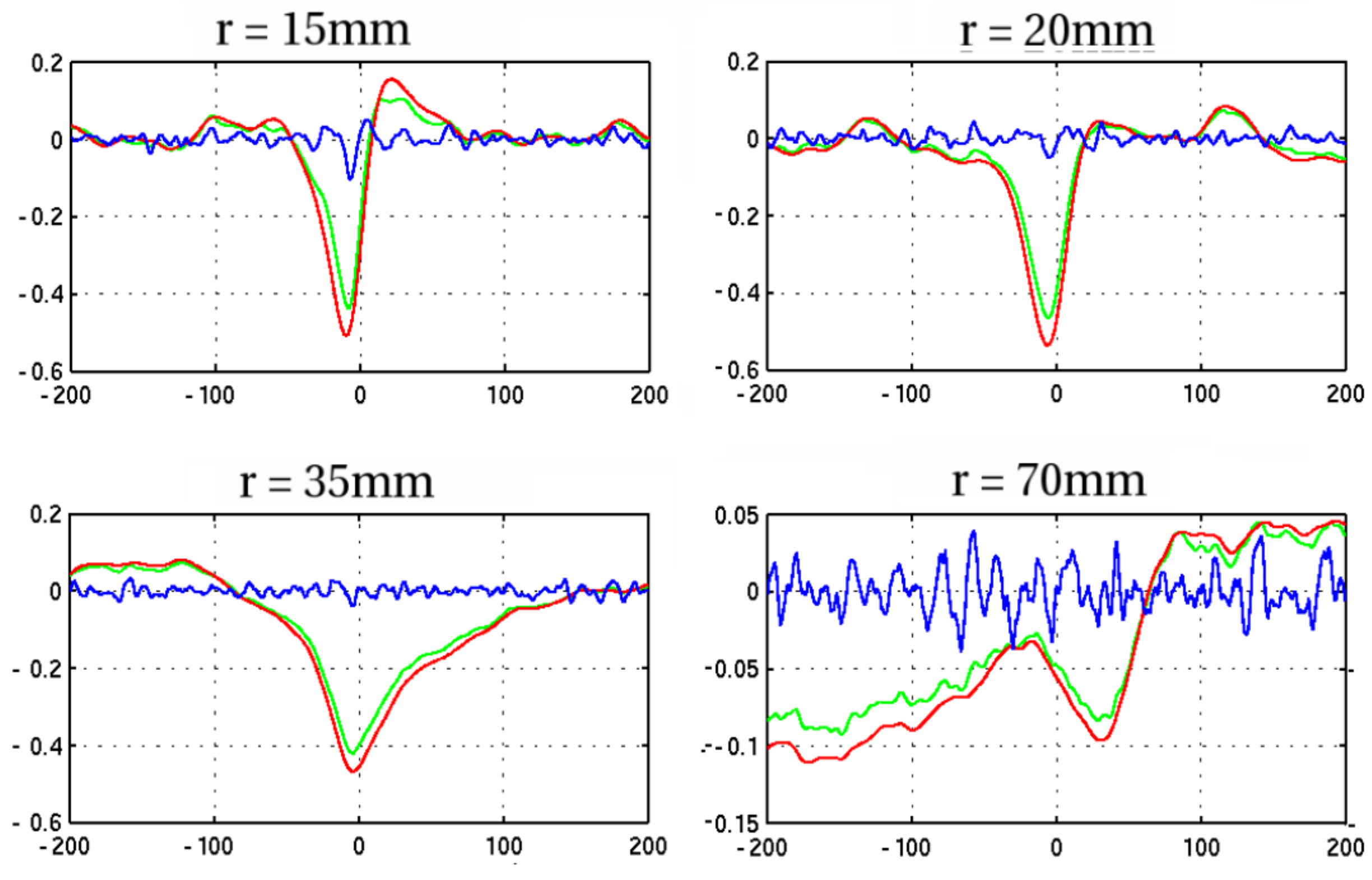} 
\end{center}
\caption{Energy flux: total (green), coherent (red) and incoherent (blue). 
The split is made using complex valued wavelets.}
\label{fig:ursi_4}
\end{figure}

%
The corresponding PDFs are shown in Fig.~\ref{fig:pop06_4} which confirm that the incoherent part
is indeed Gaussian like, while the total and coherent signal have similar skewed PDFs with algebraic heavy tails for positive signal values.
Diagnostics based on the wavelet representation were also introduced in ~\cite{FaSD06} which allow 
to compare the statistical properties of the original signals with their coherent and incoherent components. 
The wavelet spectra in comparison with classical Fourier spectra (obtained via modified periodograms)
in Fig.~\ref{fig:pop06_5+6} (left) confirm that the total and coherent signals have almost the same scale energy distribution
with a power law behavior close to $-5/3$.
Furthermore the wavelet spectra agree well with the Fourier spectra. 
The incoherent signal yields an energy equipartition for more than two magnitudes, which corresponds to decorelation
in physical space.
To quantify the intermittency we plot in Fig.~\ref{fig:pop06_5+6} (right) the scale dependent flatness of the different signals
which shows that the coherent contribution extracted from the total signal has the largest values at small scale ({\it i.e.}, high frequency) and is thus the most intermittent.
In \cite{FaSD06} we conjectured that the coherent bursts are responsible for turbulent
transport, whereas the remaining incoherent fluctuations only contribute to turbulent diffusion.
This is confirmed by the resulting energy flux of the total, coherent and incoherent parts given in Fig.~\ref{fig:ursi_4}.
%
%
Note that cross correlation between coherent and incoherent contributions of the electric potential and the saturation current are not shown.

\subsection{Application to 2D experimental movies from tokamaks}	


\subsubsection{Tomographic reconstruction using wavelet-vaguelette decomposition}

Cameras installed in tokamaks aquire images which are difficult to interpret, since the three-dimensional structure
of the plasma is mapped onto two space dimensions and thus flattened in a non-trivial way. 
This implies that the received flux cannot be directly related to the volumic emissivity of the plasma, which is a major limitation of such optical diagnostics.
The reason is that the photons collected by each pixel on the camera sensor have been emitted along a corresponding ray, rather than out of a single point in space.
Nevertheless the three-dimensional radiation can be related to the two-dimensional image using tomographic reconstruction, because the dominant structures in tokamak edge turbulence happen to be field-aligned filaments, commonly known as ‘blobs’.
They have a higher density than their surroundings, and their structure varies more slowly along magnetic field lines than in their orthogonal directions.

Mathematically the tomographic reconstruction corresponds to an inverse problem which has a formal solution under the assumed symmetry, but is ill-posed in the presence of noise.
Taking advantage of the slow variation of the fluctuations along magnetic field lines in tokamaks, this inverse problem can be modelled by a helical Abel transform, which is a Volterra integral operator of the first kind.
In \cite{NFBBSFM12} we proposed a tomographic inversion technique, based on a wavelet-vaguelette decomposition and
coupled with wavelet denoising to extract coherent structures, that allows to detect individual blobs on the projected movie and to analyse their behaviour.
The wavelet-vaguelette decomposition (WVD) was introduced by Tchamitchian~\cite{Tcha87} and used by Donoho~\cite{Dono95} to solve inverse problems in the presence of localized structures.
Tomographic inversion using the wavelet-vaguelette decomposition is as an alternative to SVD
(Singular Value Decomposition).
SVD and WVD regularize the problem by damping the modes of the inverse transform
to prevent amplification of the noise, {\it i.e.},  modes below a given threshold are eliminated.
For WVD the nonlinear iterative thresholding procedure (see section~3.1.3) is applied to the vaguelette coefficients.
Here Coiflets with two vanishing moments are used \cite{Daub92}.
%
However, in contrast to SVD, WVD takes in addition advantage of the spatial localization of coherent
structures present in the plasma.

The technicalities of WVD are described in detail in ~\cite{NFBBSFM12}, in the following we only
explain the principle.
The helical Abel transform related the plasma light  emissivity $S$ (a scalar-valued field) to the integral of the volume emissivity 
received by the camera  $I=KS$, where $K$ is a compact continuous operator.
The reconstruction of the plasma light emissivity $S$ from $I$ is an inverse problem
which becomes very difficult when $S$ is corrupted by noise,
since computing $K^{-1}$ is an ill-posed problem which amplifies the noise.
The vaguelettes are operator adapted wavelets and a biorthogonal set of basis functions is obtained
from the wavelet bases $\psi_\lambda$ by computing $K \psi_\lambda$ and ${K^{\star}}^{-1} \psi_\lambda$, where ${K^{\star}}^{-1}$ denotes the adjoint inverse operator \cite{Tcha87}.
Note that vaguelettes inherit the localization features of wavelets but may loose the translation and scale invariance, and thus the fast wavelet transform cannot be applied anymore.

\subsubsection{Application to an academic example}

\begin{figure}[htbp]
\begin{center}
\includegraphics[width=\linewidth]{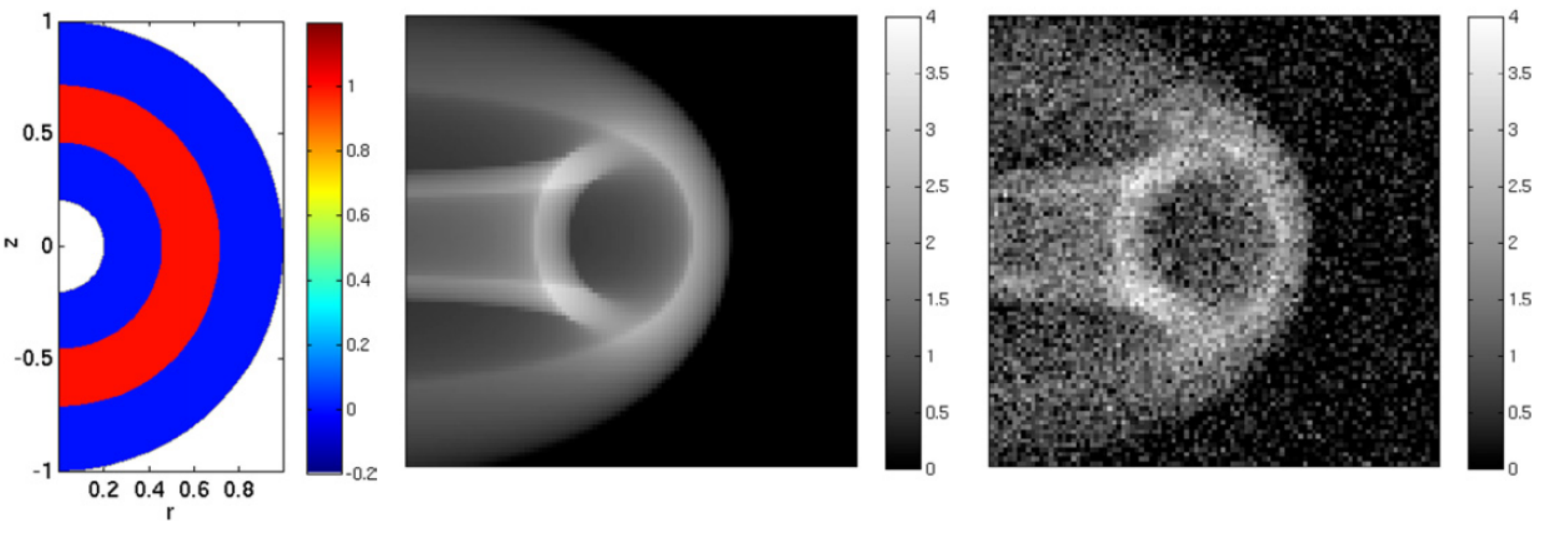}
\end{center}
\caption{Denoising WVD academic test case with a uniform radiating shell.
Left: source emission intensity $S$ in the poloidal plane.
Middle: corresponding noiseless image $I = KS$ in the image plane.
Right: noisy image obtained by adding Gaussian white noise with variance 0.5.
From \cite{NFBBSFM12}.}
\label{fig:nf1}
\end{figure}

To illustrate the method we first consider an academic test case with an given emissivity map $S$,
having a uniform radiating shell at constant value one and zero elsewhere. 
A two-dimensional cut in the poloidal plane is shown in Fig.~\ref{fig:nf1}, left.
Applying the helical Abel transform we generate the corresponding synthetic image $I = K S$ (Fig.~\ref{fig:nf1}, middle).
Then we add a Gaussian white noise with standard deviation $0.5$, which yields the synthetic noisy image (Fig.~\ref{fig:nf1}, right).
%

\begin{figure}[htbp]
\begin{center}
\includegraphics[height=5.5cm]{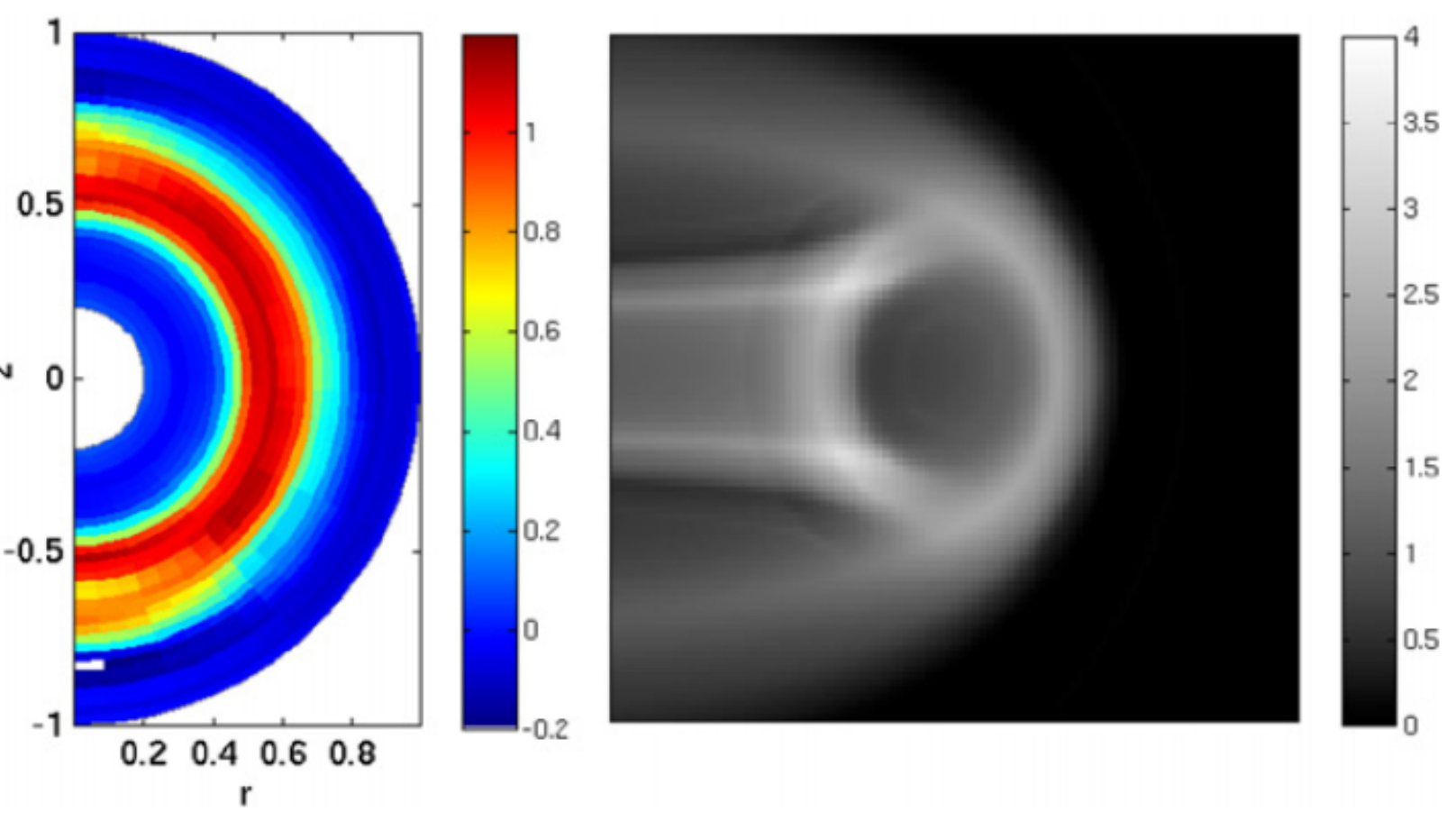}
\end{center}
\caption{Denoising WVD academic test case. WVD inversion results.
Left: reconstructed poloidal emissivity map $S_d$.
Right: denoised image $I_d = K S_d$.
From \cite{NFBBSFM12}.}
\label{fig:nf2}
\end{figure}

Applying the WVD reconstrution to the synthetic noisy image (Fig.~\ref{fig:nf1}, right) gives a denoised emissivity map,
a poloidal cut is shown in Fig.~\ref{fig:nf2}, left.
We observe that the main features are preserved, {\it i.e.}, the constant emissivity shell is well recovered,
besides some spurious oscillations close to discontinuities.
The corresponding denoised image $I_d = K S_d$ (Fig.~\ref{fig:nf2}, right) illustrates that the noise has been successfully removed. 
A comparison with the standard SVD technique in \cite{NFBBSFM12} (not shown here) illustrates 
the superiority of the wavelet-vaguelette technique.

\subsubsection{Application to fast camera data from tokamaks}

\begin{figure}[htbp]
\begin{center}
\includegraphics[height=4.6cm]{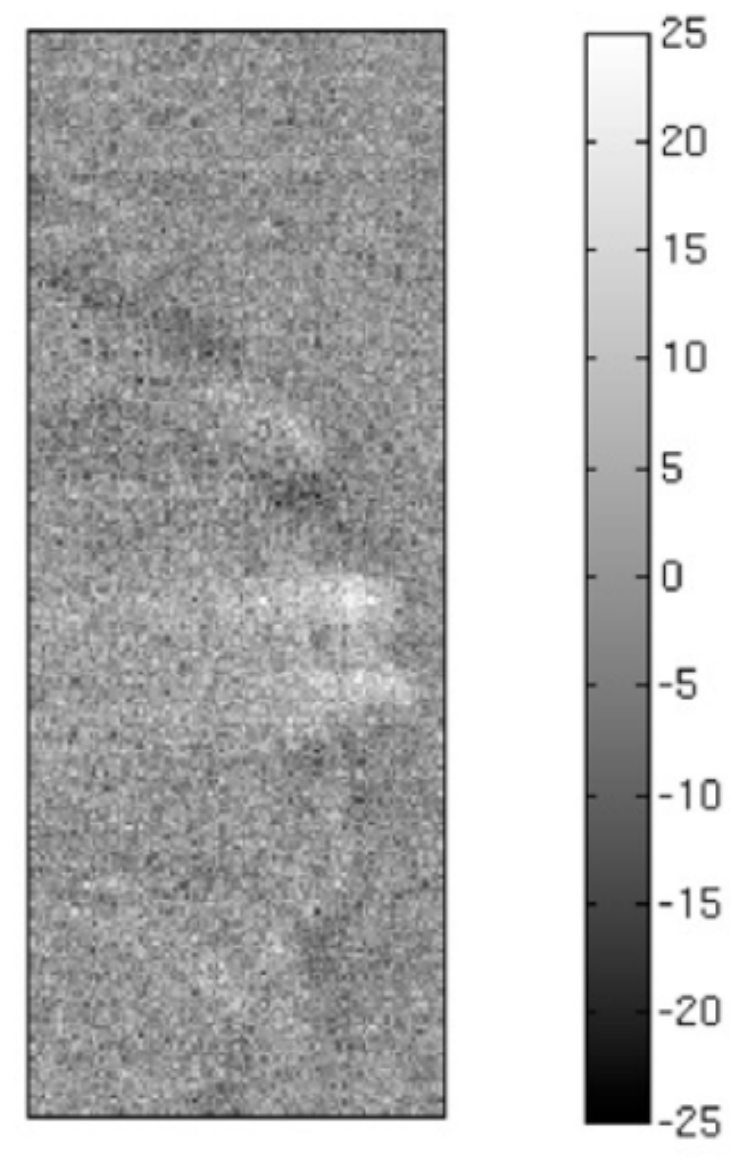} \hspace{0.4cm}
\includegraphics[height=4.9cm]{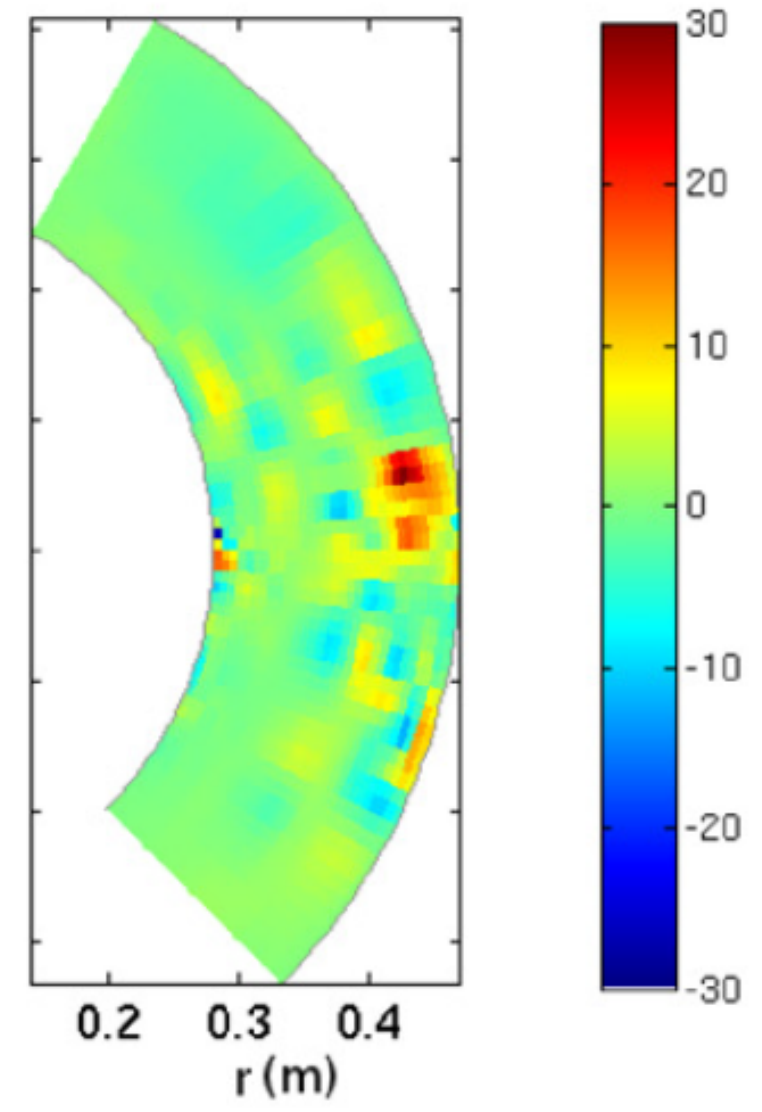} \hspace{0.4cm}
\includegraphics[height=4.7cm]{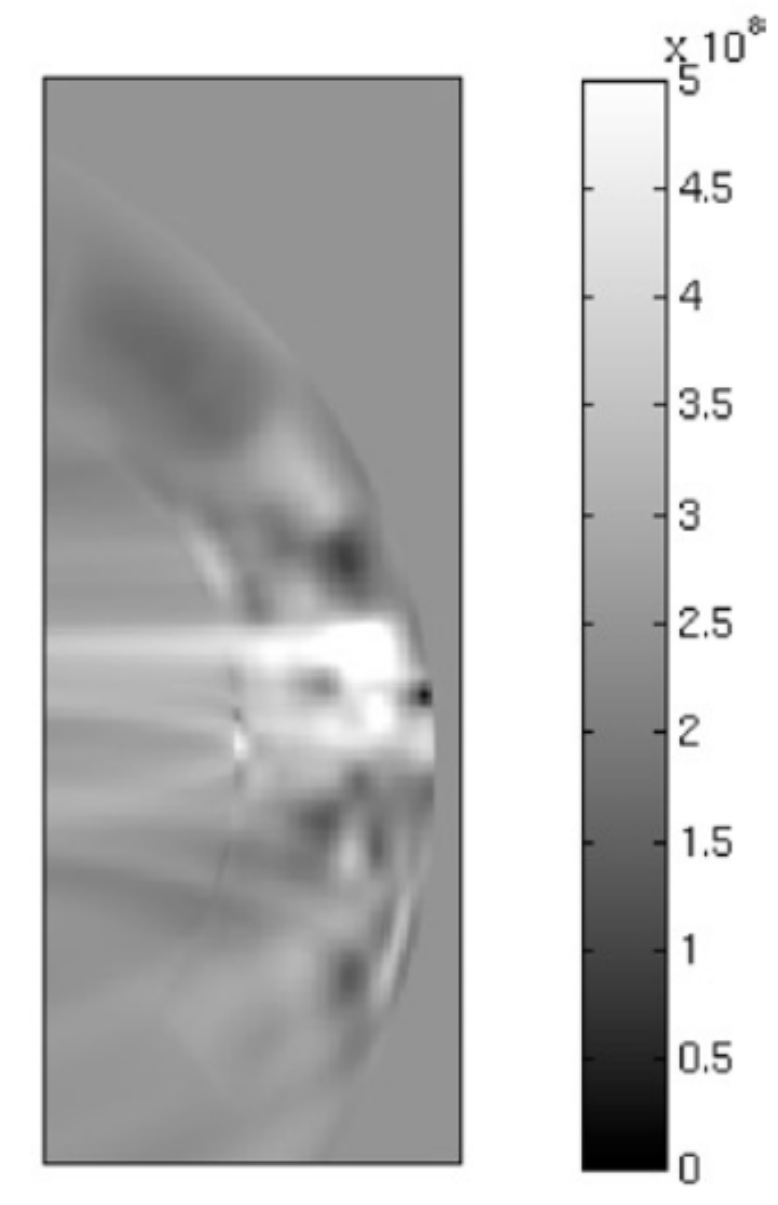}
\end{center}
\caption{WVD-inversion of a snapshot from a movie obtained from Tore Supra, discharge TS42967. 
Left: noisy frame used as input for the WVD algorithm. 
Middle: reconstructed emissivity map obtained as a result of WVD.
Right: denoised frame obtained by applying the operator $K$ to the reconstructed emissivity map.
From \cite{NFBBSFM12}.}
\label{fig:nf3}
\end{figure}

Now we present an application to an experimental movie acquired during the Tore Supra discharge TS42967,
where the plasma was fully detached and stabilized over several seconds using a feedback control.
The movie has been obtained using a fast camera recording at 40 kHz.
Moreover, the time average of the whole movie was subtracted
from each frame, which helps us to decrease the effect of
reflection on the chamber wall. 
The algorithm is then applied directly to the fluctuations in the signal instead of the full signal.
The experimental conditions can be found in  \cite{NFBBSFM12}.
One frame of the movie is shown in Fig.~\ref{fig:nf3}, left and used as input for the WVD reconstruction algorithm.
The resulting emissivity map in the poloidal plane, in Fig.~\ref{fig:nf3}, middle, shows the presence of localized blobs,
which propagate counterclockwise as observed in the movies, not shown here.
Thus their propagation velocity can be determined.
The corresponding denoised movie frame $I_d$ (Fig.~\ref{fig:nf3}, right) is obtained by applying the operator $K$ to the inverted emissivity map $S_d$.
We observe that the noise has been removed and the local features such as blobs and fronts have been extracted.

\subsection{Application to 2D simulations of resistive drift-wave turbulence}		

%
At the edge of the plasma in tokamaks drift-waves play an important role in the dynamics and transport.
In \cite{BFBFS08} we considered a two-dimensional slab geometry and performed direct numerical
simulations using a two-field model, the Hasegawa--Wakatani system which describes  the main features of
resistive drift-waves.
The evolution equations for the plasma density fluctuations and the electrostatic potential fluctuations are coupled via the adiabaticity parameter which models the intensity of the parallel electron resistivity.
A Poisson equation relates the vorticity with the electrostatic potential.
The wavelet-based coherent vortex extraction method (see section~3.1.3) 
is then applied in~\cite{BFBFS08} to assess the role of coherent vorticity for radial transport
and to identify only the active degrees of freedom which are responsible for the transport. 

\begin{figure}[htbp]
\begin{center}
\includegraphics[width= 0.8\linewidth]{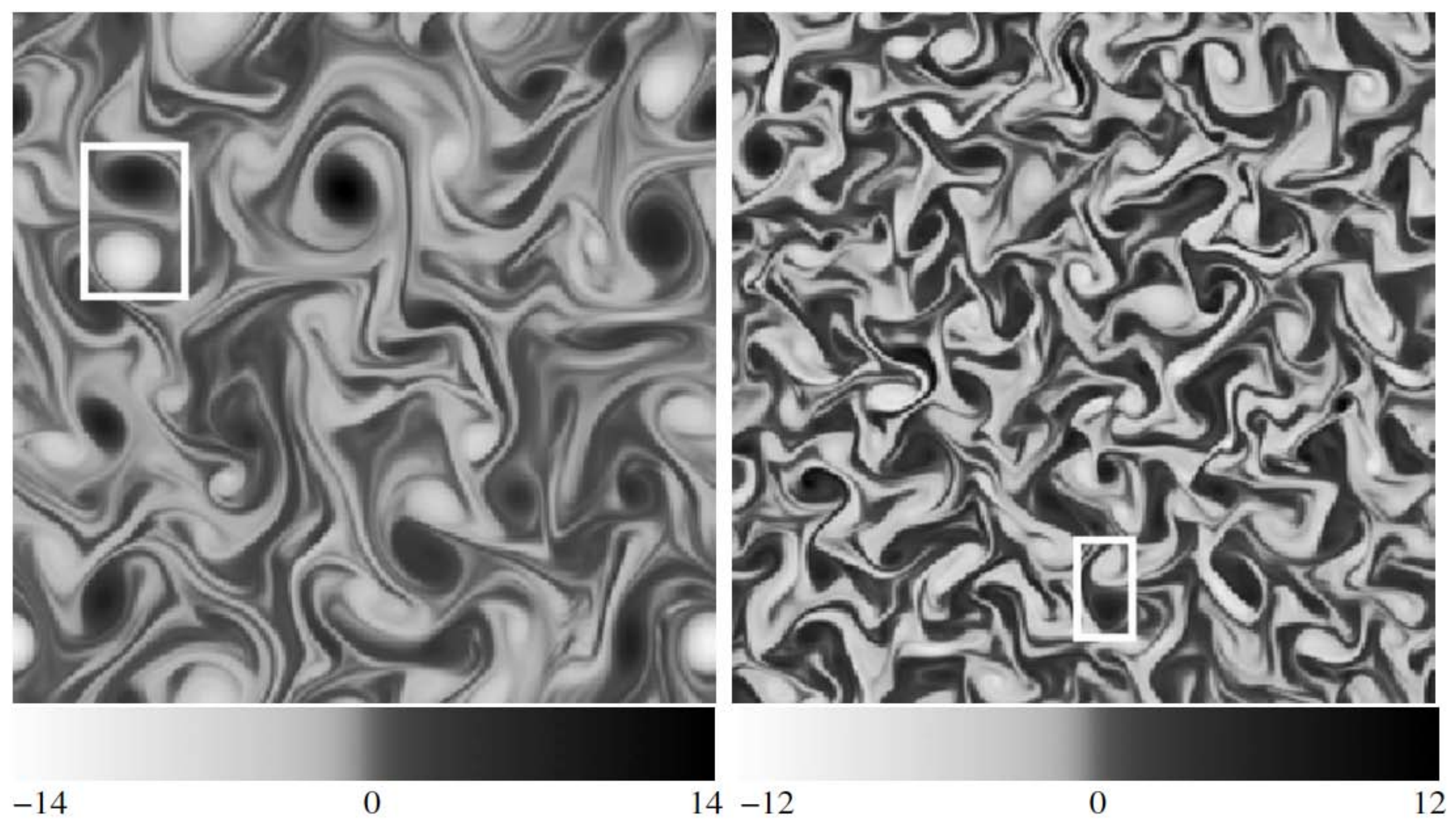}
\end{center}
\caption{Snapshots of the vorticity field for the quasi-hydrodynamic case (left) and for the quasi-adiabatic case (right).
Abscissa and ordinate correspond to the radial and poloidal position, respectively.
The white rectangles indicate the selected dipoles.
From \cite{BFBFS08}.}
\label{fig:pop08_0}
\end{figure}

\begin{figure}[htbp!]
\begin{center}
\includegraphics[width= \linewidth]{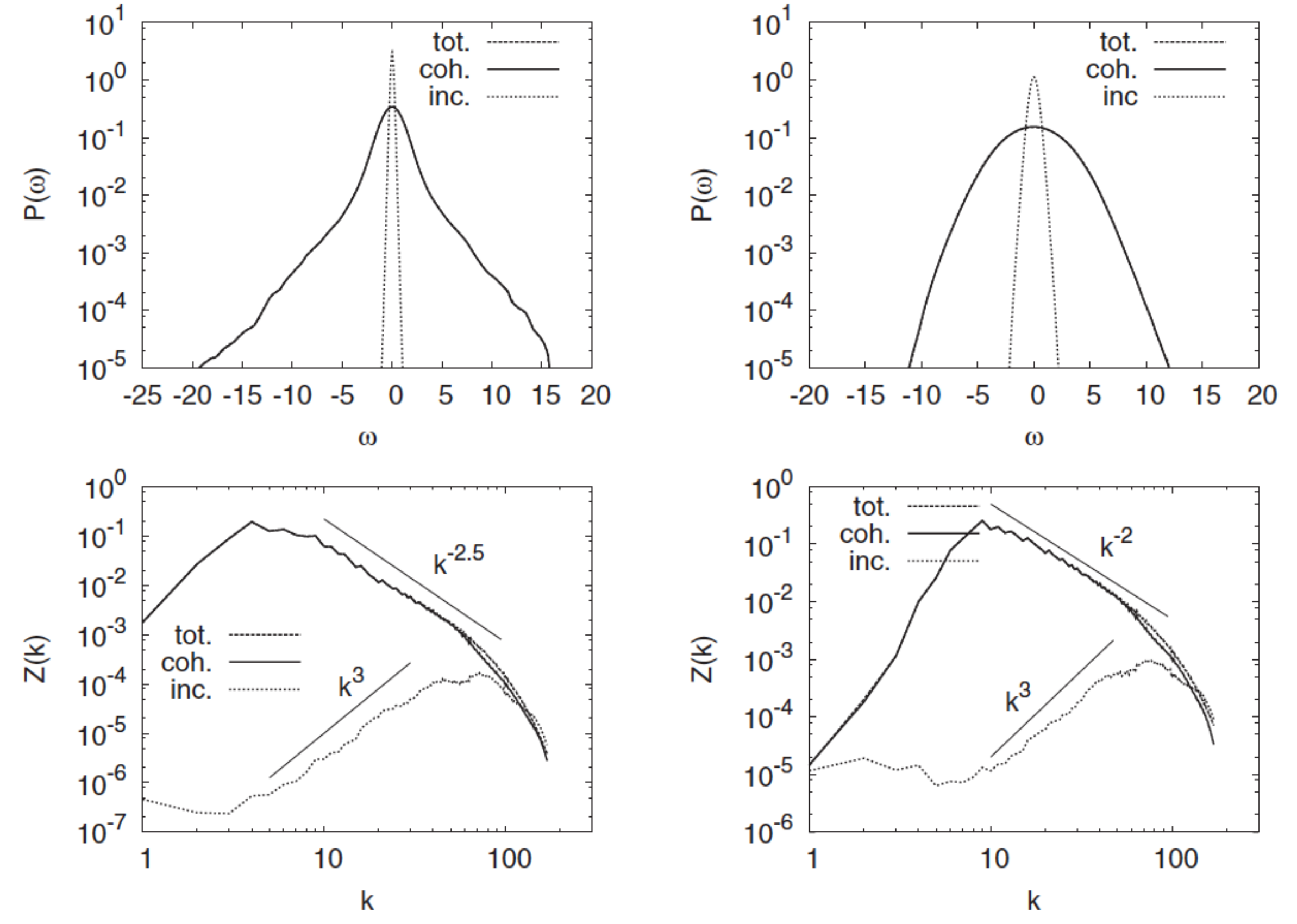}
\end{center}
\caption{Top: PDFs of the vorticity. Bottom:
Fourier spectrum of the enstrophy versus wavenumber. 
Left: quasi-hydrodynamic case. 
Right: quasi-adiabatic case. 
Dashed line: total field, solid line: coherent part, dotted line: incoherent part. 
Note that the coherent contribution (solid) superposes the total field (dashed), which is thus hidden under the solid line in all four figures.
The straight lines indicating power laws are plotted for reference.
From \cite{BFBFS08}.}
\label{fig:pop08_1}
\end{figure}

\begin{figure}[htbp]
\begin{center}
\includegraphics[width= \linewidth]{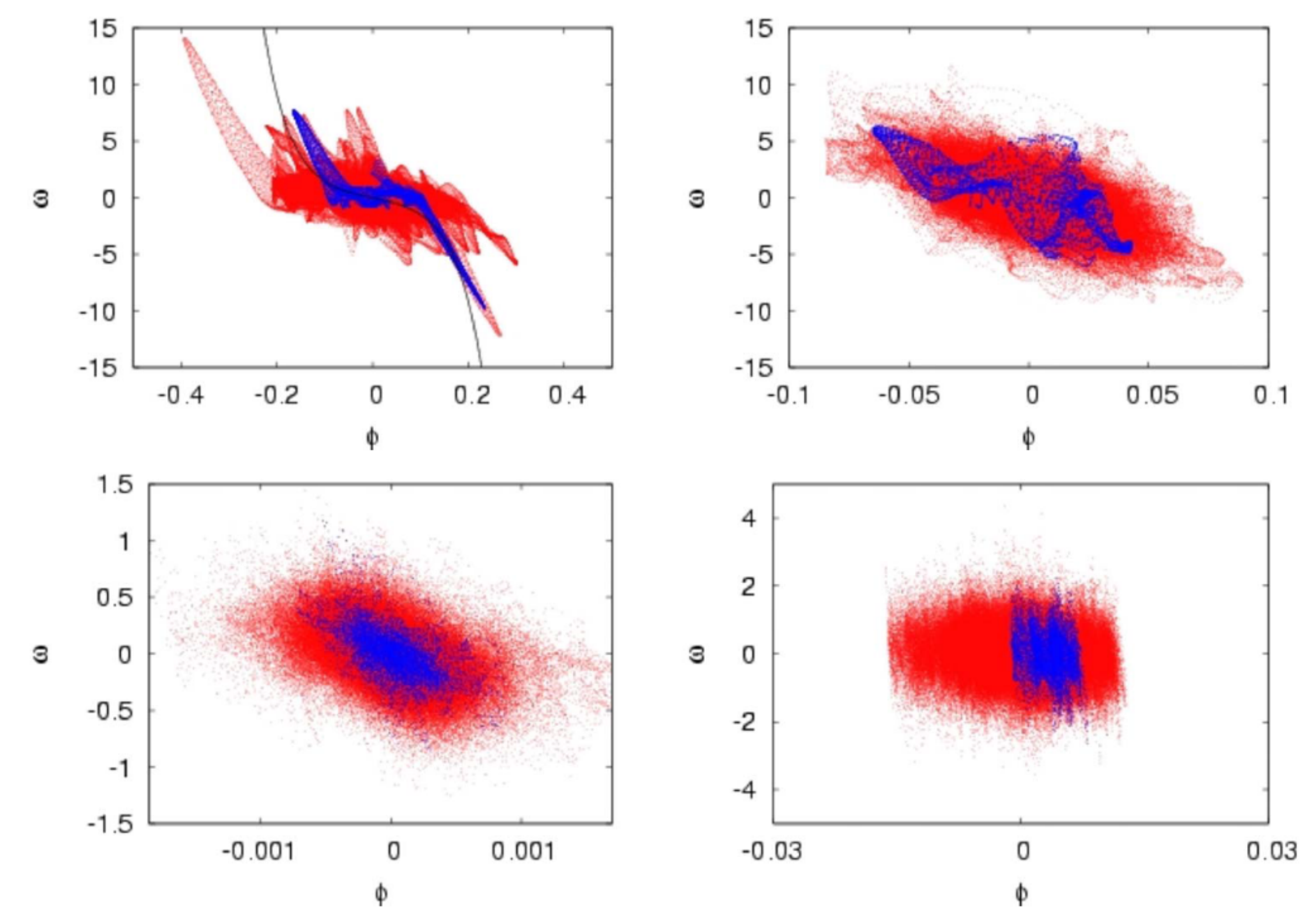}
\end{center}
\caption{Scatter plot of vorticity against electrostatic potential for the coherent part (top) and incoherent part (bottom).
Left: quasi-hydrodynamic case; right: quasi-adiabatic case.
The red dots correspond to the total field, the blue dots correspond to a selected vortex dipole in Fig.~\ref{fig:pop08_0}.
From \cite{BFBFS08}.}
\label{fig:pop08_2}
\end{figure}

Visualizations of the vorticity field for two regimes, the quasi-hydrodynamic case and the quasi-adiabatic case,
corresponding respectively to low and high collisionality of the plasma, are given in Fig.~\ref{fig:pop08_0}.
In both cases coherent vortices can be observed and a dipolar structure is framed by the white rectangles.
Applying the CVE algorithm we split the vorticity fields into coherent and incoherent contributions.
In the quasi-hydrodynamics case we find that $1.3 \%$ of the wavelet coefficients are sufficient to retain $99.9 \%$ of the energy, while in the quasi-adiabatic case $1.8 \%$ of the modes retain $99.0 \%$ of the energy.
%
%
The statistical properties of the total, coherent and incoherent vorticity fields are assessed in Fig.~\ref{fig:pop08_1}
by plotting the vorticity PDFs and the Fourier enstrophy spectra for the two cases.
For the quasi-hydrodynamic vorticity the PDFs of the total and the coherent field are slightly skewed and exhibit a non--Gaussian distribution, while for the quasi-adiabatic case a symmetric almost Gaussian like distribution can be observed.
The variances of the incoherent parts are strongly reduced in both cases with respect to the total fields and the PDFs have a Gaussian-like shape.
The enstrophy spectra illustrate that coherent and incoherent contributions exhibit a multiscale behavior.
The spectra of total and coherent vorticity agree well all over the inertial range. The spectra of the incoherent contributions  have a powerlaw behavior close to $k^3$ which corresponds to an equipartition of kinetic energy.
%
%
In \cite{BFBFS08} it is furthermore shown that the radial density flux, {\it i.e.}, more than $98 \%$, is indeed carried by these coherent modes. 
In the quasi-hydrodynamic regime, coherent vortices exhibit depletion of the
polarization-drift nonlinearity as shown in the scatter plot of vorticity against the electrostatic potential in Fig.~\ref{fig:pop08_2}. 
Moreover vorticity strongly dominates over strain, in contrast to the quasiadiabatic regime.
Details can be found in \cite{BFBFS08}.
%

\subsection{Application to 3D simulations of resistive MHD turbulence}		

\begin{figure}[htbp]
\begin{center}
\includegraphics[width= \linewidth]{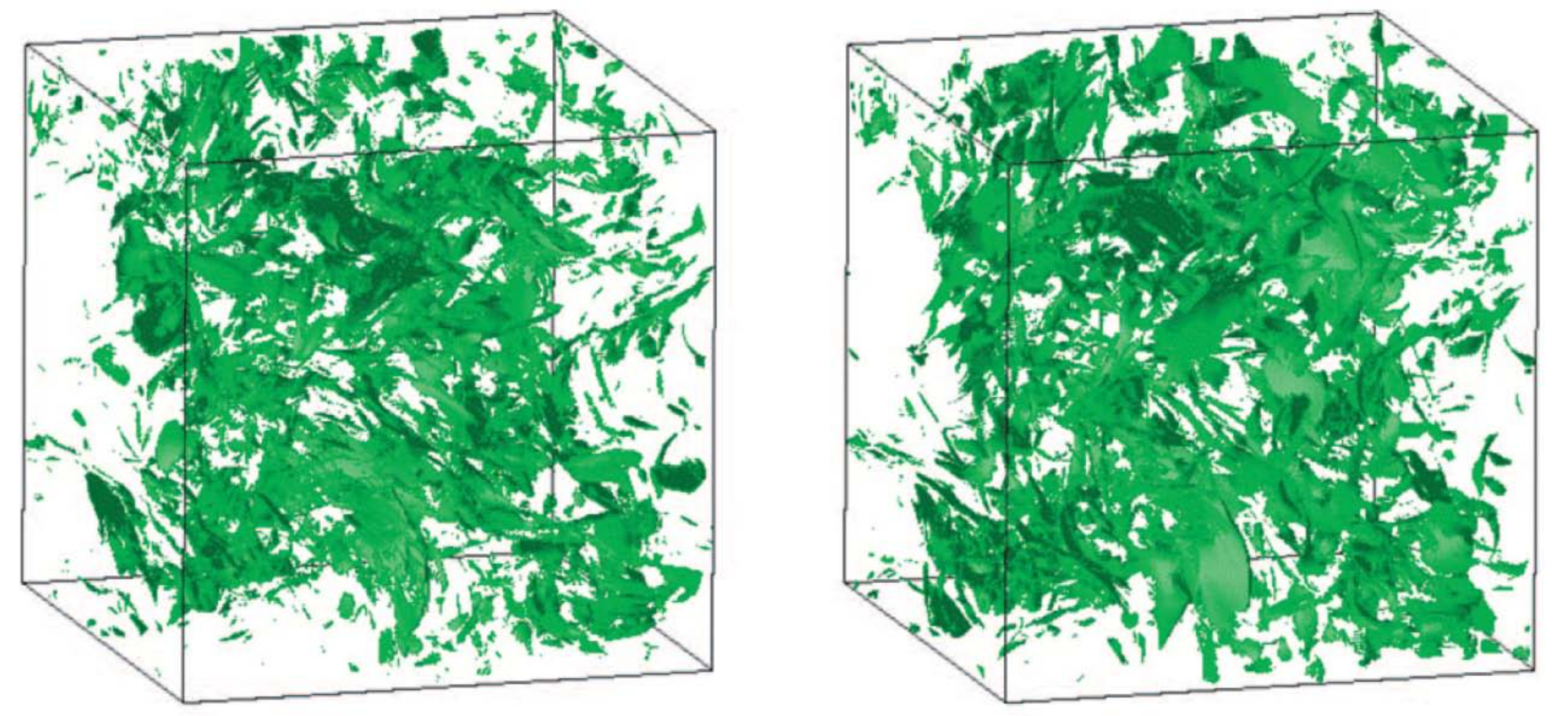} \\
\includegraphics[width= \linewidth]{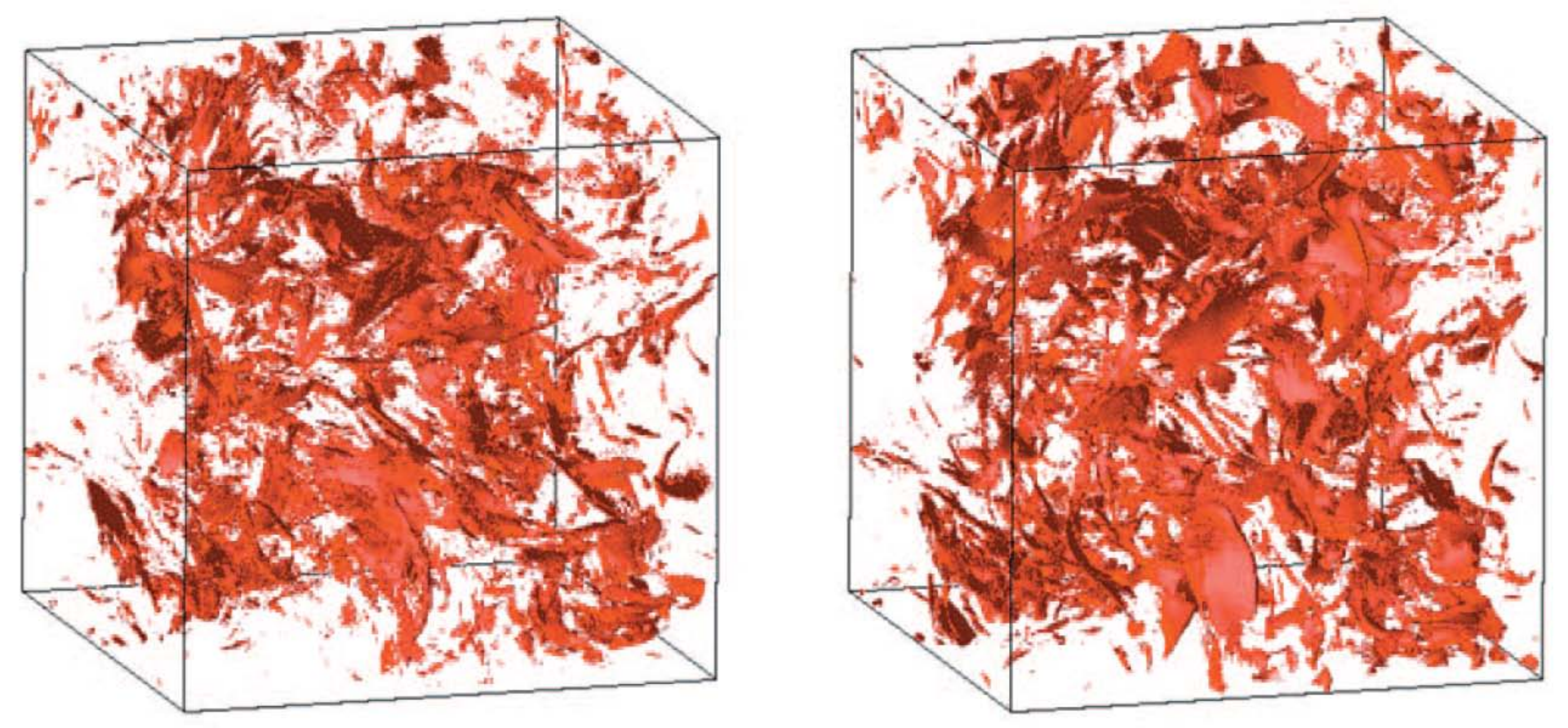} \\
\includegraphics[width= \linewidth]{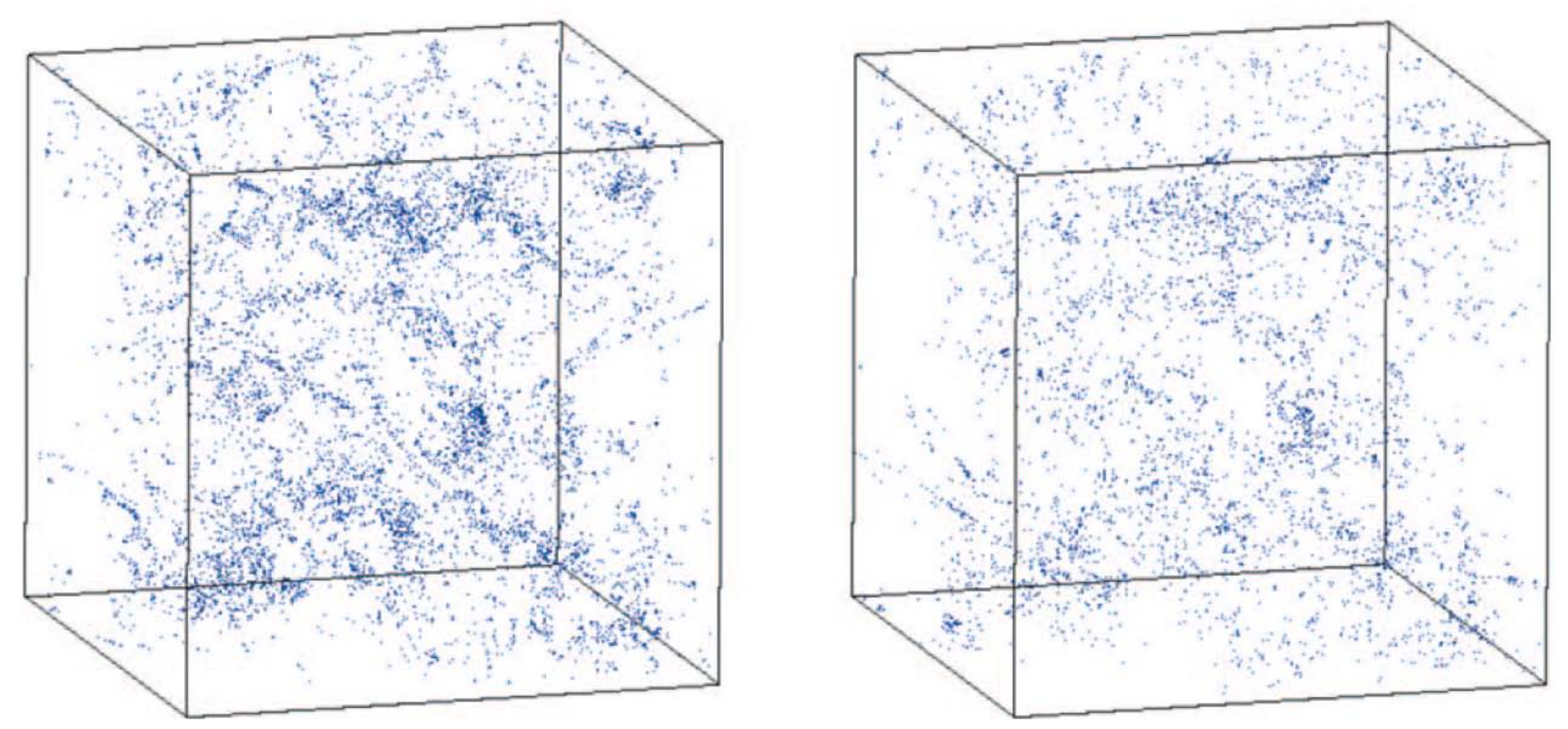}
\end{center}
\caption{Isosurfaces of vorticity (left) and current density (right)
of the total (top), coherent (middle) and incoherent contributions (bottom).
From \cite{YKSOHF09}.}
\label{fig:pop09_1-3}
\end{figure}
\begin{figure}[htbp]
\begin{center}
\includegraphics[width= 0.48\linewidth]{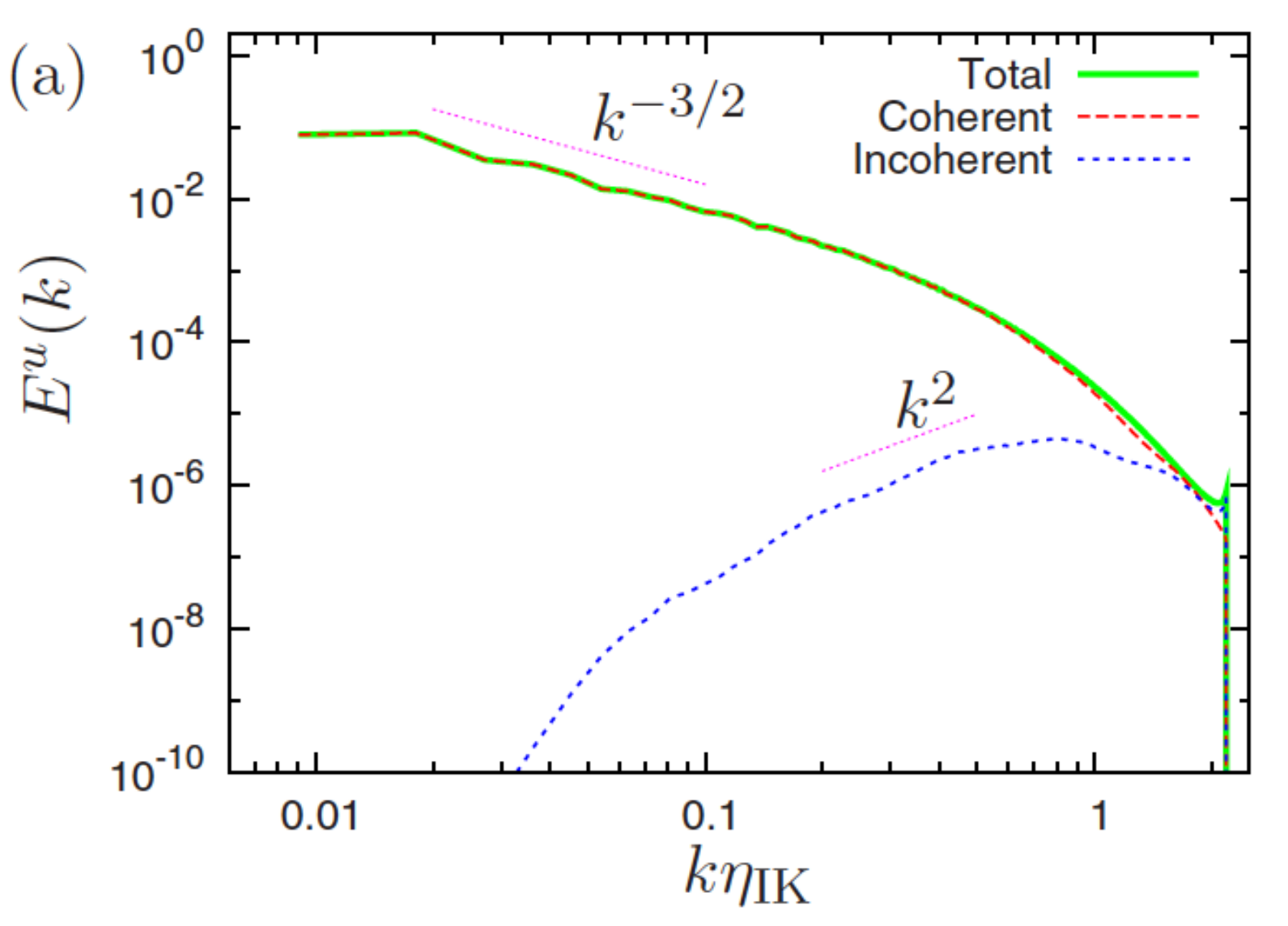}
\includegraphics[width= 0.48\linewidth]{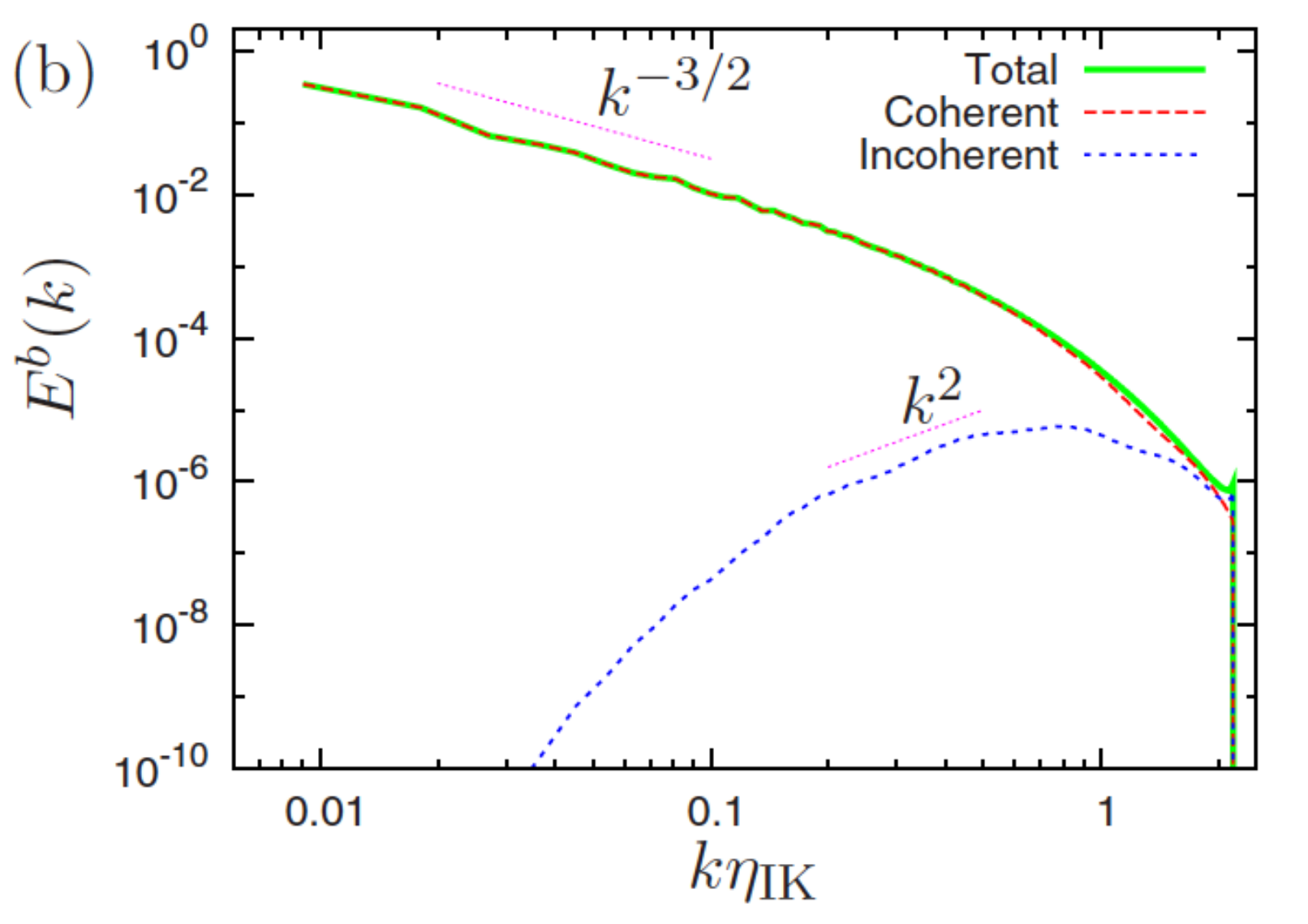}
\end{center}
\caption{Kinetic (a) and magnetic (b) energy spectra of the total, coherent and incoherent fields. The wavenumber is normalized with the Iroshnikov-Kraichnan scale.
From \cite{YKSOHF09}.}
\label{fig:pop09_6+7}
\end{figure}
\begin{figure}[htbp!]
\begin{center}
\includegraphics[width= \linewidth]{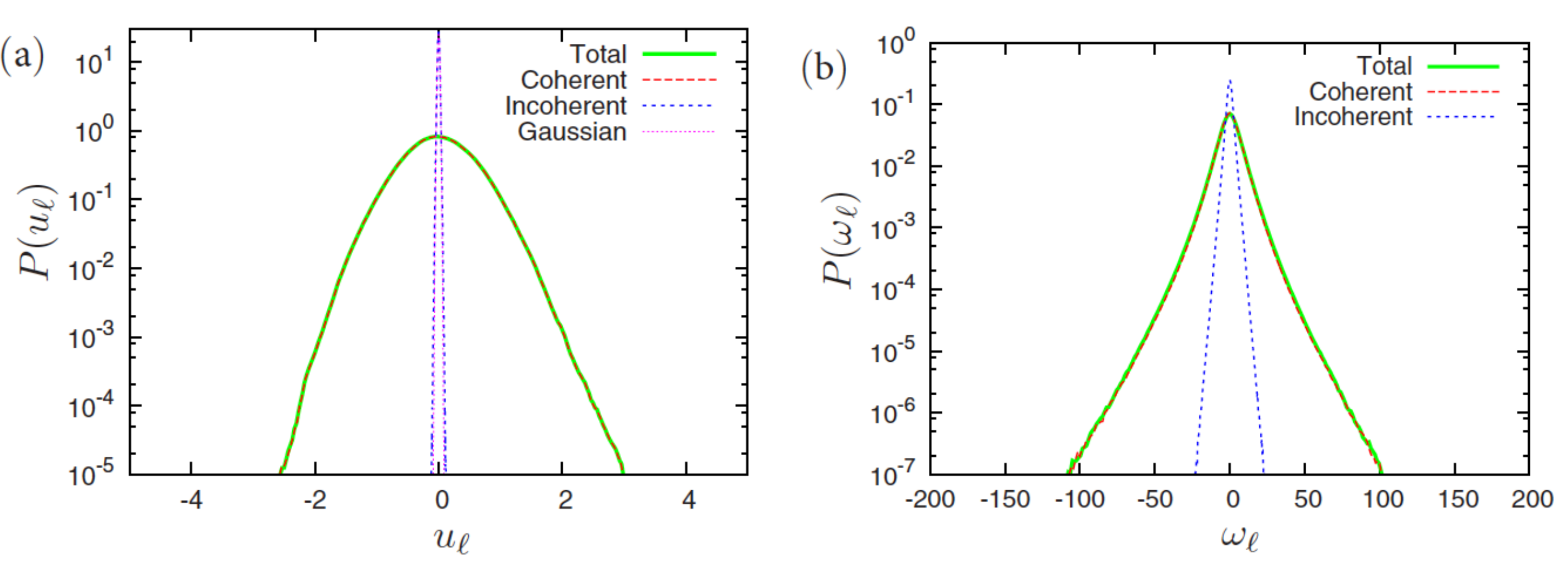}\\ 
\includegraphics[width= \linewidth]{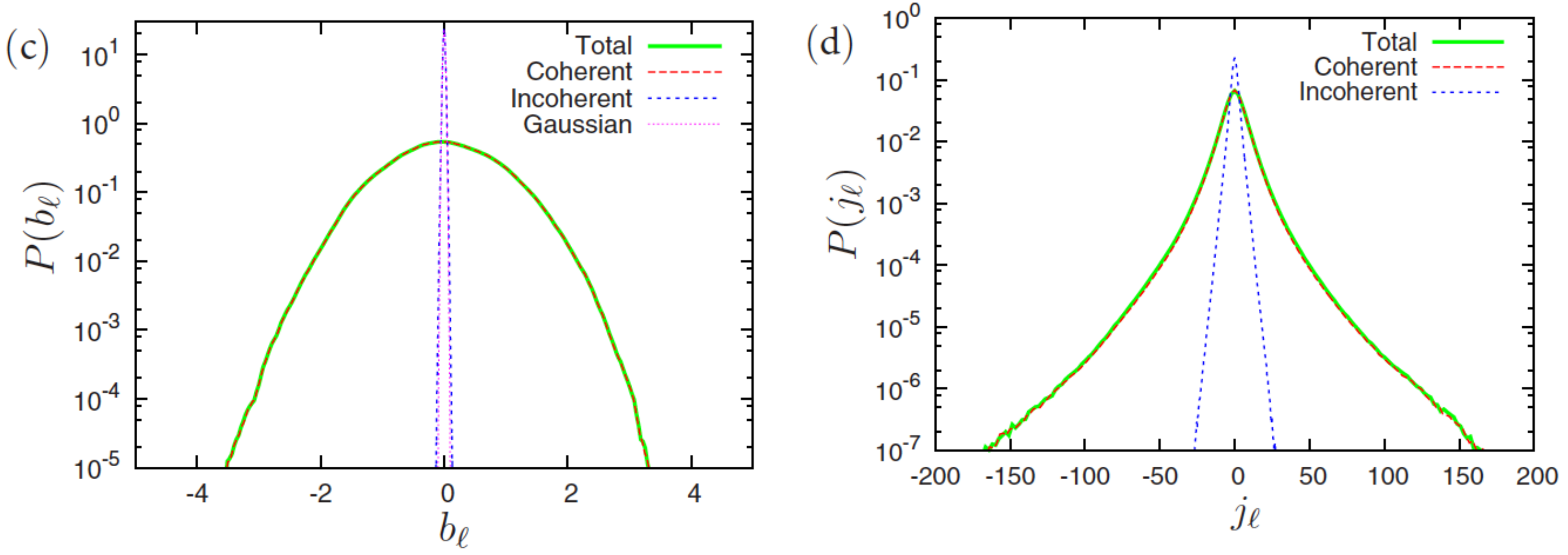}
\end{center}
\caption{PDFs of the $\ell$-th component of the velocity (a), vorticity (b), magnetic field (c) and current density (d)
for the total, coherent and incoherent contributions.
From \cite{YKSOHF09}.}
\label{fig:pop09_4+5}
\end{figure}
\begin{figure}[htbp]
\begin{center}
\includegraphics[width= 0.75\linewidth, height=5.0cm]{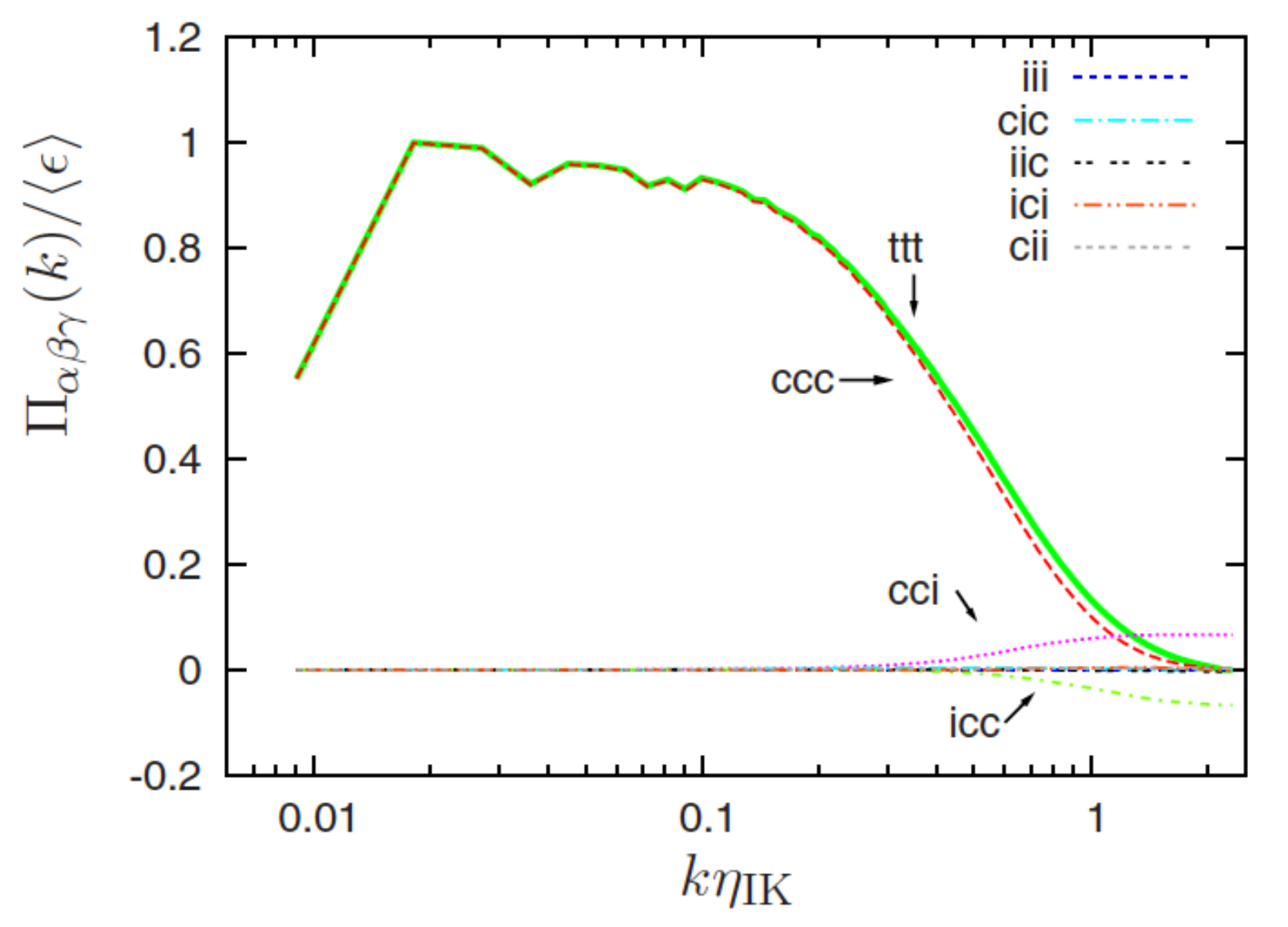}
\end{center}
\caption{Contributions to the energy flux normalized by the energy dissipation rate versus the
wavenumber, which is normalized with the Iroshnikov-Kraichnan scale.
From \cite{YKSOHF09}.}
\label{fig:pop09_8}
\end{figure}
\begin{figure}[htbp]
\begin{center}
\includegraphics[width= \linewidth]{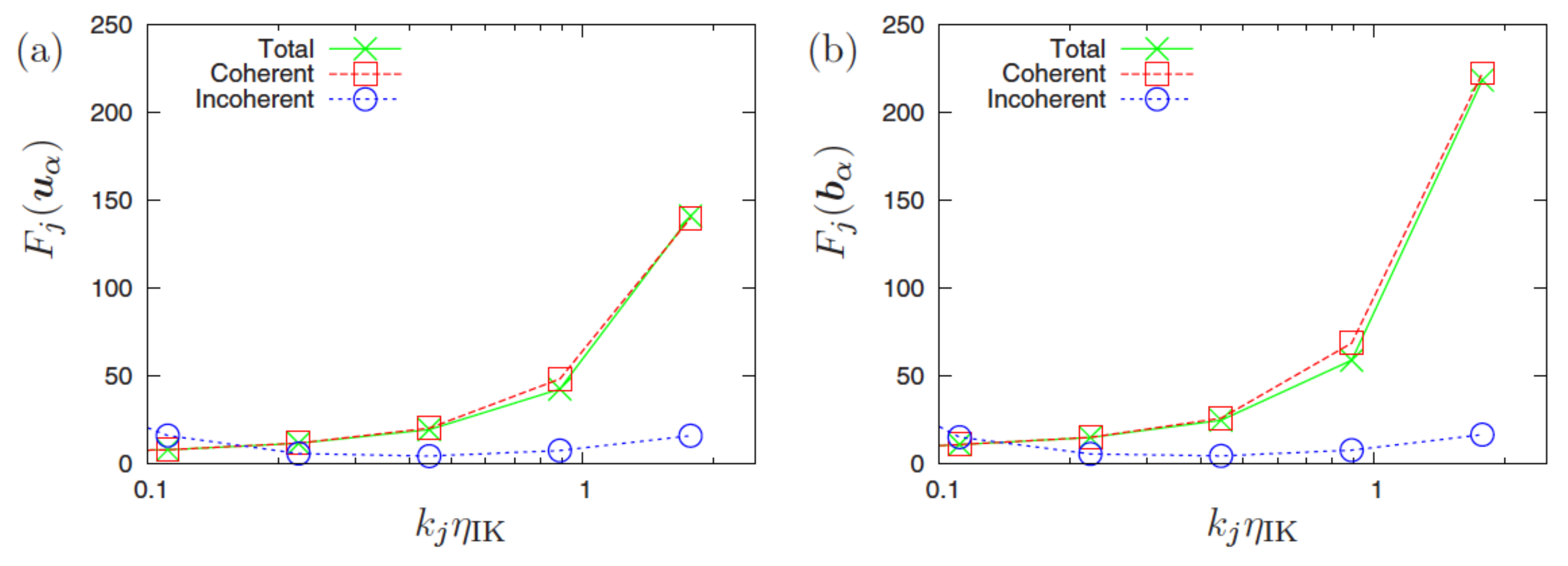}
\end{center}
\caption{Scale-dependent flatness of velocity (a) and magnetic field (b) versus the
wavenumber, which is normalized with the Iroshnikov-Kraichnan scale.
From \cite{YKSOHF09}.}
\label{fig:pop09_9}
\end{figure}

In \cite{YKSOHF09} we proposed a method for extracting coherent vorticity sheets and current sheets out of three-dimensional homogeneous magnetohydrodynamic (MHD) turbulence.
To this end the wavelet-based coherent vortex extraction method (see section~3.1.3) has been applied to vorticity and current density fields computed by direct numerical simulation (DNS) of forced incompressible MHD
turbulence without mean magnetic field at resolution of $512^3$.
Coherent vorticity sheets and current sheets are extracted from the DNS data at a given time instant.
A visualization of isosurfaces of vorticity and current density of the total, coherent and incoherent fields is shown in 
Fig.~\ref{fig:pop09_1-3}.
%
%
The coherent vorticity and current density are found to preserve both the vorticity sheets
and the current sheets present in the total fields while retaining only a few percent of the degrees of
freedom. 
The incoherent vorticity and current density are shown to be structureless and of mainly
dissipative nature. 
The spectral distributions in Fig.~\ref{fig:pop09_6+7} of kinetic and magnetic energies of the coherent fields
only differ in the dissipative range, while the corresponding incoherent fields exhibit
quasi-equipartition of energy, corresponding to a $k^2$ slope. 
%
%
The probability distribution functions (PDFs) of total and coherent fields, for
both vorticity and current density, in Fig.~\ref{fig:pop09_4+5} coincide almost perfectly, while the incoherent vorticity and current density fields have strongly reduced variances. 
%
%
The energy flux shown in Fig.~\ref{fig:pop09_8} confirms that the nonlinear dynamics is indeed fully
captured by the coherent fields only.
%
%
The scale-dependent flatness of the velocity and the magnetic field in Fig.~\ref{fig:pop09_9} illustrate that the total and coherent fields have similar scale dependent high order moments and reflect strong intermittency characterized by the strong increase of the flatness for decreasing scale. 
The flatness values of the incoherent contributions, of both the velocity and the magnetic field
are are much smaller and do not increase significanlty for decreasing scale, {\it i.e.}, they are not intermittent.
%

\section{Wavelet-based simulation schemes}

In the following two wavelet-based methods for solving kinetic plasma equations are presented:
an application of nonlinear wavelet denoising to improve the convergence of particle-in-cell schemes (PIC)
and a particle-in-wavelet scheme for solving the Vlasov-Poisson equation directly in wavelet space.
We also present the Coherent Vorticity and Current sheet Simulation (CVCS) method which extends the Coherent Vorticity Simulation (CVS) \cite{FSK99,FS01} developed for the Navier--Stokes equations to the resistive non-ideal MHD equations.
Numerical examples illustrate the properties and the efficiency of the different methods.

\subsection{Improving particle-in-cell (PIC) schemes by wavelet denoising}


Plasma simulations using particles are characterized by the presence of noise, a typical feature of Monte--Carlo type simulations.
The number of particles, which is restricted by the computational resources, limits the statistical sampling and thus
the accuracy of the reconstructed particle distribution function.

The discretization error generically known as ‘‘particle noise” due to its random-like character of the method
quantifies the difference between the distribution function reconstructed from a simulation using $N_p$ particles and the exact distribution function.
The weak scaling of the error with the number of particles, $ \propto 1/ \sqrt{N_p} $,  however limits the reduction of particle noise by increasing the number of computational particles in practical applications.
This has motivated the development of various noise reduction techniques, see, {\it e.g.}, \cite{NCSFC10}, which is of importance in the validation and verification of particle codes.

In  \cite{NCSFC10} we proposed a wavelet-based method for noise reduction in the reconstruction of particle distribution
functions from particle simulation data, called wavelet-based density estimation (WBDE). 
The method was originally introduced in \cite{DJKP96} in the context of statistics to estimate probability densities given a finite number of independent measurements.
WBDE, as used in \cite{NCSFC10}, is based on a truncation of the wavelet representation of the Dirac delta function associated with each particle.
The method yields almost optimal results for functions with unknown local smoothness without compromising computational
efficiency, assuming that the particles’ coordinates are statistically independent.
It can be viewed as a natural extension of the finite size particles (FSP) approach,
with the advantage of estimating more accurately distribution functions that have localized
sharp features.
The proposed method preserves the moments of the particle distribution
function to a good level of accuracy, has no constraints on the dimensionality of the
system, does not require an a priori selection of a global smoothing scale, and is able to
adapt locally to the smoothness of the density based on the given discrete particle data.
Indeed, the projection space is determined from the data itself, which allows for
a refined representation around sharp features, and could make the method more precise than PIC for a given
computational cost.
Moreover, the computational cost of the denoising stage is of the same order as one time
step of a FSP simulation. 

The underlying idea of WBDE is to expand the sampled particle distribution function, represented by a histogram, 
into an orthogonal wavelet basis using the fast wavelet transform.
We define the empirical density associated to the particles positions $x_n$ for $n= 1, ..., N_p$
where $N_p$ is the number of particles,
\begin{equation} \label{eqn:particledensity}
\rho^{\delta} (x) \, = \, \frac{1}{N_p} \, \sum_{n=1}^{N_p} \, \delta(x -x_n)
\end{equation}
and where $\delta$ is the Dirac measure.
We then project $\rho^{\delta} (x)$ onto an orthogonal wavelet basis
retaining only scales $j$ such that $L \le j \le J$ where the scales $L$ and $J$ 
denote the largest and smallest retained scales, respectively \cite{DJKP96}. 
The remaining wavelet coefficients are then  thresholded retaining only those whose modulus is larger than the scale-dependent threshold $K \sqrt{j/N_p}$, where $K$ is a constant which depends on the regularity of the solution \cite{DJKP96}.
Finally the denoised particle density is obtained by applying an inverse fast wavelet transform.
In \cite{NCSFC10} Daubechies wavelets with 6 vanishing moments were used.

\begin{figure}[htbp]
\begin{center}
\includegraphics[width= \linewidth]{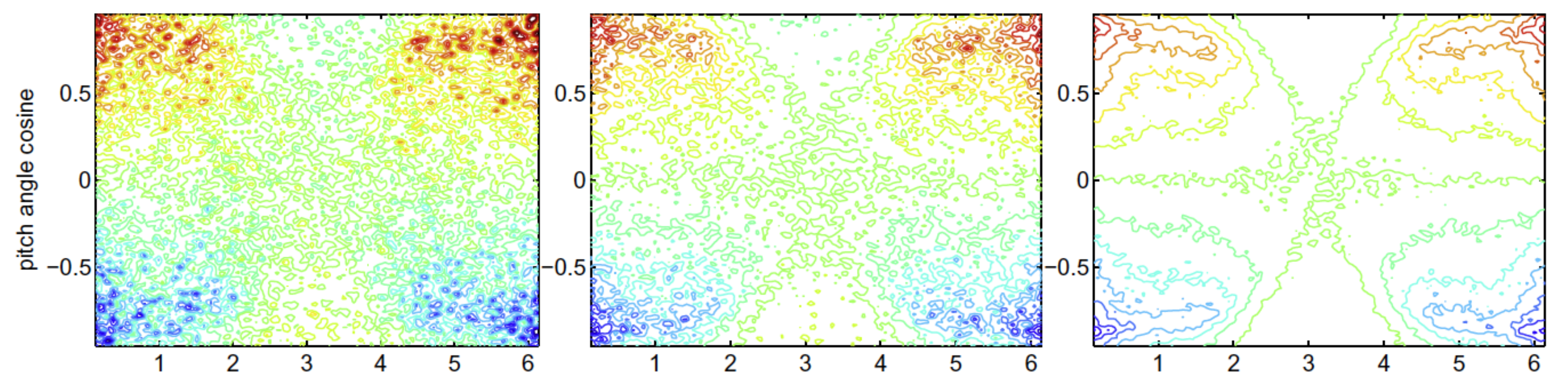}\\
\includegraphics[width= \linewidth]{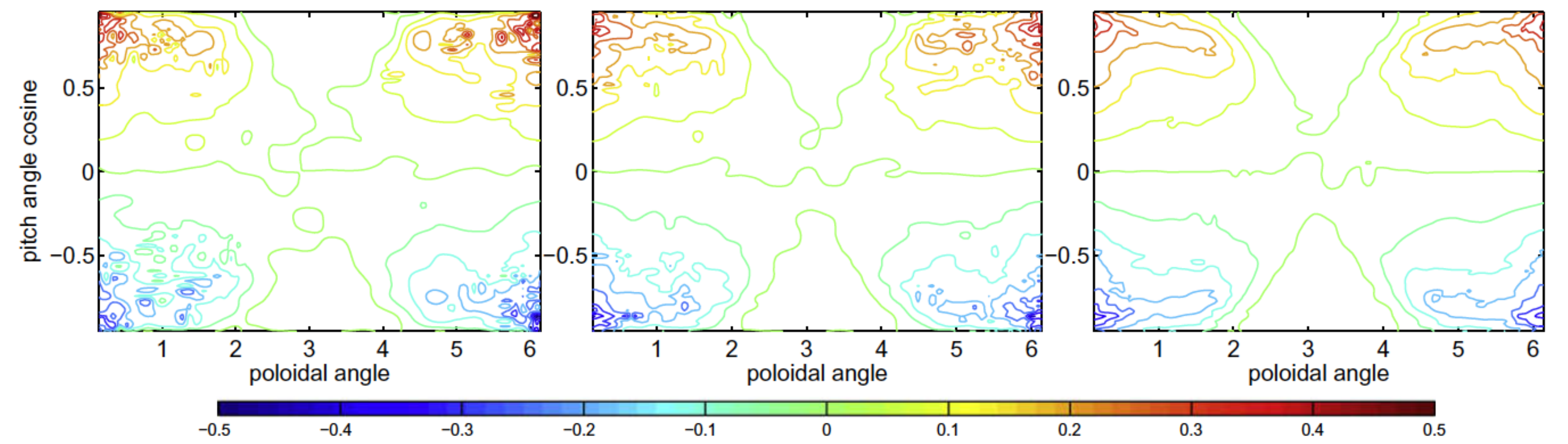}
\end{center}
\caption{Contour plots of estimates of $\delta f$ for the collisional guiding center transport particle data: histogram method (top row) and WBDE method (bottom row). The left, center and right columns correspond to $N_p = 32 \cdot 10^3$ (left), $N_p = 128 \cdot 10^3$ (middle) and $N_p = 1024 \cdot 10^3$ (right), respectively. The plots show $17$ isolines equally spaced within the interval $[0.5, 0.5 ]$.
From \cite{NCSFC10}.}
\label{fig:jcp1+2}
\end{figure}

\begin{figure}[htbp]
\begin{center}
\includegraphics[height=5.6cm]{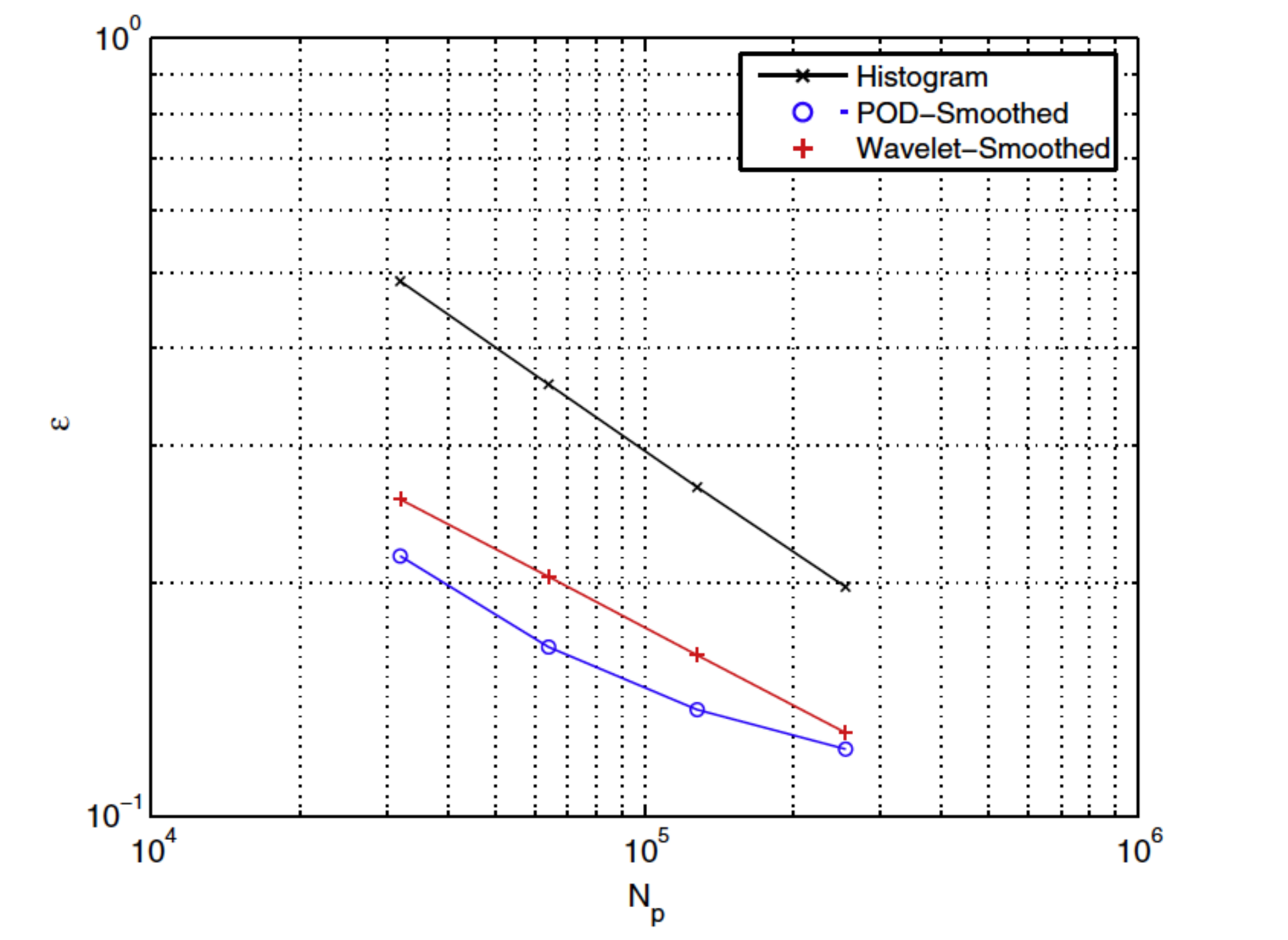}
\end{center}
\caption{RMS error estimate for collisional guiding center transport particle data according to the histogram, 
the POD, and the wavelet methods. The reference density is computed with $N_p = 1024 \cdot 10^3$.
From \cite{NCSFC10}.}
\label{fig:jcp3}
\end{figure}

In \cite{NCSFC10} we treated three cases in order to test how the efficiency of the denoising algorithm depends 
on the level of collisionality of the plasma. 
A strongly collisional, weakly collisional and collisionless regimes were considered. 
For the strongly collisional regime we computed particle data of force-free collisional relaxation involving energy and pinch-angle scattering. 
The collisionless regime is studied using PIC-data corresponding to bump-on-tail and two-stream instabilities in the Vlasov--Poisson system.
The third case of a weakly collisional regime is illustrated here using guiding-center particle data of a magnetically confined plasma in toroidal geometry. The data was generated with the code DELTA5D.
Figure~\ref{fig:jcp1+2} shows contour plots of the histogram (top row) and the reconstructed densities using WBDE for increasing number of particles.
It can be seen that the WBDE results in efficiently denoised densities and that the error has been reduced by a factor two
with respect to the raw histograms as shown in Fig.~\ref{fig:jcp3}.

\subsection{Particle-in-wavelets scheme (PIW)}


In \cite{NSSF11} we proposed a new numerical scheme, called particle-in-wavelets, for the Vlasov--Poisson
equations describing the evolution of the particle distribution function $f$ in collisionless plasma, and assessed its efficiency in the simplest case of one spatial dimension. 
In non-dimensional form the equations read
\begin{eqnarray} \label{eqn:vlasovpoisson}
\partial_t f + v \partial_x f + \partial_x \phi \partial_v f  \, &=& \, 0 \\
\partial_{xx} \phi + 1 - 2 \pi \int_{\R} \, f(x,v,t) \, dv    \, &=& \, 0
\end{eqnarray}
where $\phi$ is electric potential.
The particle distribution function $f$ is discretized using tracer particles, and the charge distribution is reconstructed 
using wavelet-based density estimation (WBDE), discussed in the previous section. 
The latter consists in projecting the Delta distributions corresponding to the particles
onto a finite dimensional linear space spanned by a family of wavelets, which is chosen adaptively.
A wavelet-Galerkin Poisson solver is used to compute the electric potential
once the wavelet coefficients of the electron density $\rho(x,t) = \int_{\R} f(x,v,t) dv$ have been obtained by WBDE.
The properties of wavelets are exploited for diagonal preconditioning of the linear system in wavelet space, which is solved by an iterative method, here conjugated gradients.
Similar to classical PIC codes the interpolation method is compatible with the charge assignment scheme.
Once the electric field $E(x,t) = -\partial_x \phi(x,t)$ has been interpolated at the particle positions the characteristic trajectories,
defined by $x'(t) = v(t)$ and $v'(t) = -E(x(t),v(t),t)$ can be advanced in time using the Verlet integrator.
%
\begin{figure}[htbp!]
\begin{center}
\includegraphics[width= \linewidth]{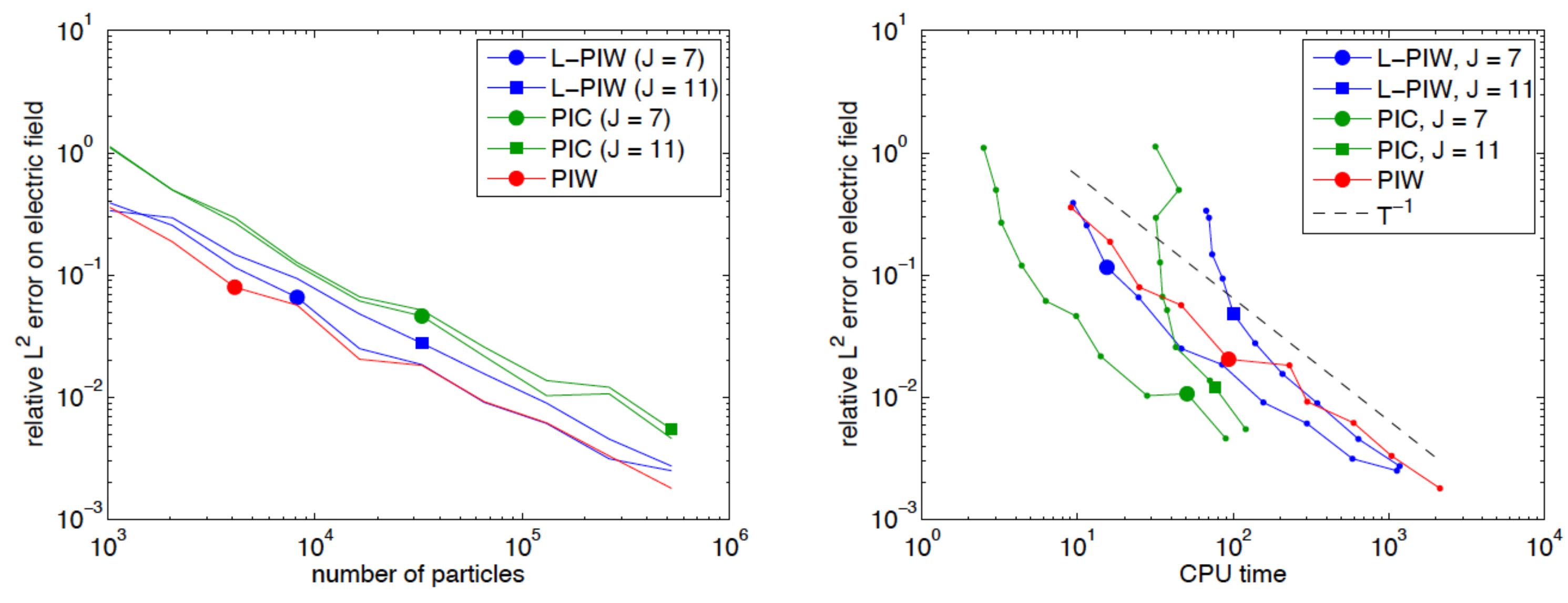}
\end{center}
\caption{Comparisons between PIW and PIC for the two-stream instability test case. 
Relative $L^2$ error of the electric field at $t = 30$, as 
a function of the number of particles (left) and the corresponding computing time (right).
Note that L-PIW is a variant of PIW where only linear filtering has been applied. 
From \cite{NSSF11}.}
\label{fig:esaim1}
\end{figure}

To demonstrate the validity of the PIW scheme, numerical computations of Landau damping and of the two-stream instability 
have been performed in \cite{NSSF11}.
The stability and accuracy have been assessed with respect to reference computations obtained with a precise semi-Lagrangian scheme \cite{SRBG99}.
We showed that the precision is improved roughly by a factor three compared to a classical PIC scheme, for a given number of particles \cite{NSSF11}, as illustrated in Fig.~\ref{fig:esaim1} for the two-stream instability. 
We observe that PIW remains uniformely more precise for any number of particles thanks to its adaptive properties (Fig.~\ref{fig:esaim1}, left). The total CPU time measured in seconds scaled for the PIW code inversely proportional to the number of particles,
while for PIC and L-PIW the scaling changes when the number of particles is too low for a given spatial resolution.
However, note that the actual CPU time may depend on the implementation, since the PIC code is written in Fortran,
while the PIW code is written in C++, although the same computer was used for both codes.

\subsection{Coherent Vorticity and Current sheet Simulation (CVCS)}

\begin{figure}[htbp!]
\begin{center}
\includegraphics[width= \linewidth]{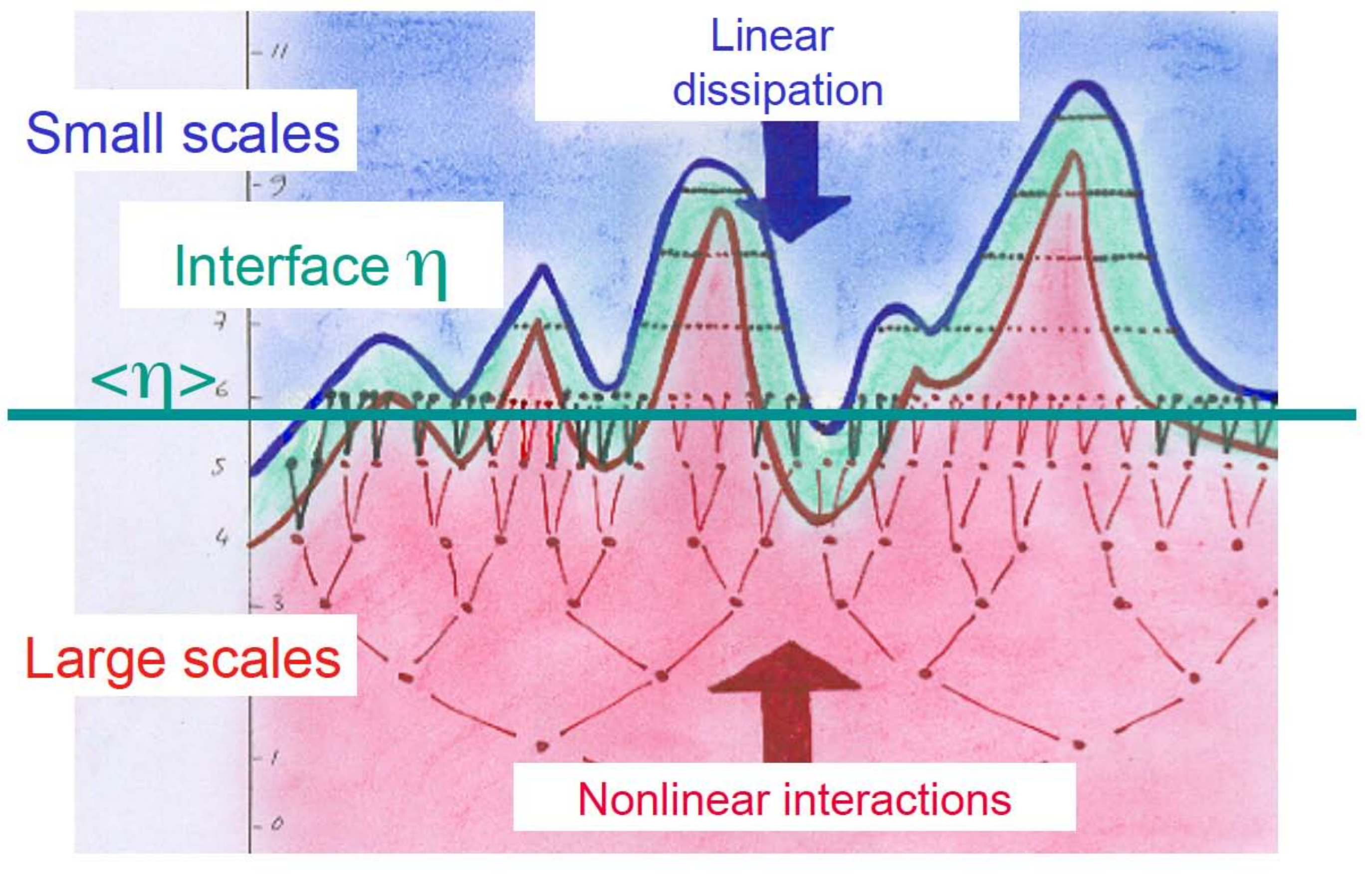}
\end{center}
\caption{Illustration of the safety zone in wavelet coefficient space used in CVS. The degrees of freedom retained by CVE are drawn in red, the adjacent coefficients of the safety zone are drawn in green, while the coefficients in blue correspond to the inactive degrees of freedom which are not computed. The interface $\eta$, defined in space and scale, separates the region dominated by nonlinear interaction (red) from the region dominated by linear dissipation (blue). The horizontal green line corresponds to the Kolmogorov dissipation scale $\langle \eta \rangle$ is defined by the statistical mean (either ensemble or space average).}
\label{fig:cvs1}
\end{figure}

Direct numerical simulation (DNS) of turbulent flow has a large computational cost due to the huge
number of degrees of freedom to be taken into account.
The required spatial resolution thus becomes prohibitive, {\it e.g.}, scaling as $Re^{9/4}$ for hydrodynamics using Kolmogorov like arguments \cite{Pope00}.
The CVS method, introduced in \cite{FSK99,FS01}, proposes to reduce the computational cost by taking only into account the degrees of freedom that are nonlinearly active.
To this end, the coherent structure extraction method (presented in section~3) is combined with a deterministic integration of the Navier--Stokes equations. At each time step the CVE is applied to retain only the coherent degrees of freedom, typically a few percent of the coefficients. 
Then, a set of neighbor coefficients in space and scale, called `safety zone', is added to account for the advection of coherent vortices and the generation of small scales due to their interaction.
Afterwards the Navier--Stokes equations are advanced in time using this reduced set of a degrees of freedom.
Subsequently, the CVE is applied to reduce the number of degrees of freedom and the procedure is repeated for the next time step.
A graphical illustration, in wavelet coefficient space, of the degrees of freedom retained at a given time step, is given in Fig.~\ref{fig:cvs1}.
%
%
This procedure allows to track the flow evolution in space and scale selecting a reduced number of degrees of freedom in a dynamically adaptive   way.
With respect to simulations on a regular grid, much less grid points are used in CVS.

In \cite{YOKSF13} we extended CVS to compute 3D incompressible magnetohydrodynamic (MHD) turbulent flow
and developed a simulation method called coherent vorticity and current sheet  simulation (CVCS).
The idea is to track the time evolution of both coherent vorticity and coherent current density, {\it i.e.}, current sheets.
Both the vorticity and current density fields are, respectively, decomposed at each time step  into two orthogonal
components, corresponding to the coherent and incoherent contribution, using an orthogonal wavelet representation.
Each of the coherent fields is reconstructed from the wavelet coefficients whose modulus is
larger than a threshold, while their incoherent counterparts are obtained from the remaining
coefficients. The two threshold values depend on the instantaneous kinetic and magnetic
enstrophies. The induced coherent velocity and magnetic fields are computed from the coherent
vorticity and current density, respectively, using the Biot--Savart kernel. In order to compute the flow
evolution, one should retain not only the coherent wavelet coefficients but also their neighbors
in wavelet space, the safety zone. 
%
\begin{figure}[htbp!]
\begin{center}
\includegraphics[width= 0.8\linewidth]{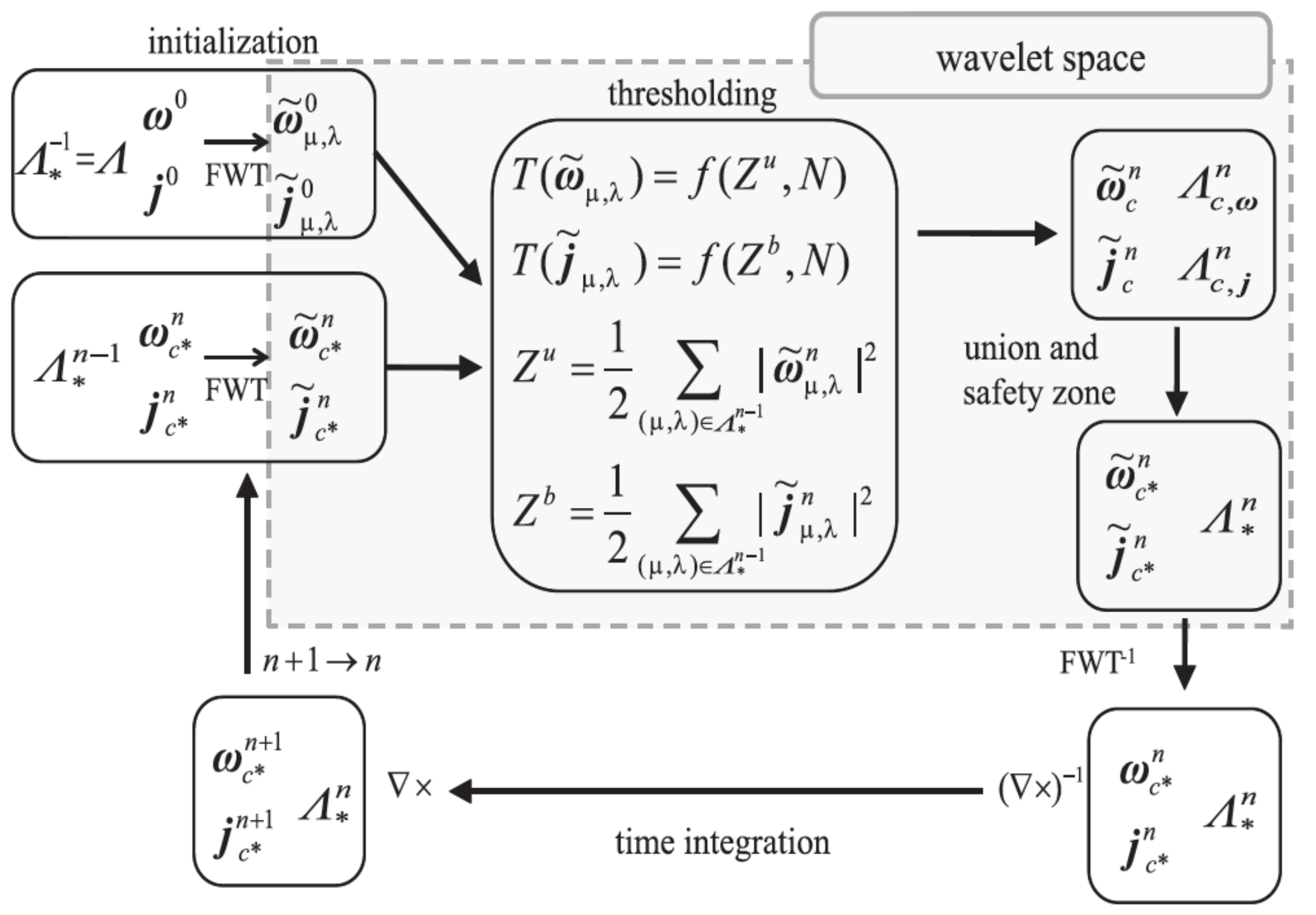}
\end{center}
\caption{Flowchart describing the principle of CVCS. The superscripts $n$ and $n+1$ denote time steps.
FWT and FWT$^{-1}$ denote the fast wavelet transform and its inverse. 
Operators performed in wavelet coefficient space are framed by the dashed rectangle.
From \cite{YOKSF13}.}
\label{fig:cvs2}
\end{figure}
%
%
\begin{figure}[htbp!]
\begin{center}
\includegraphics[width= 0.55\linewidth]{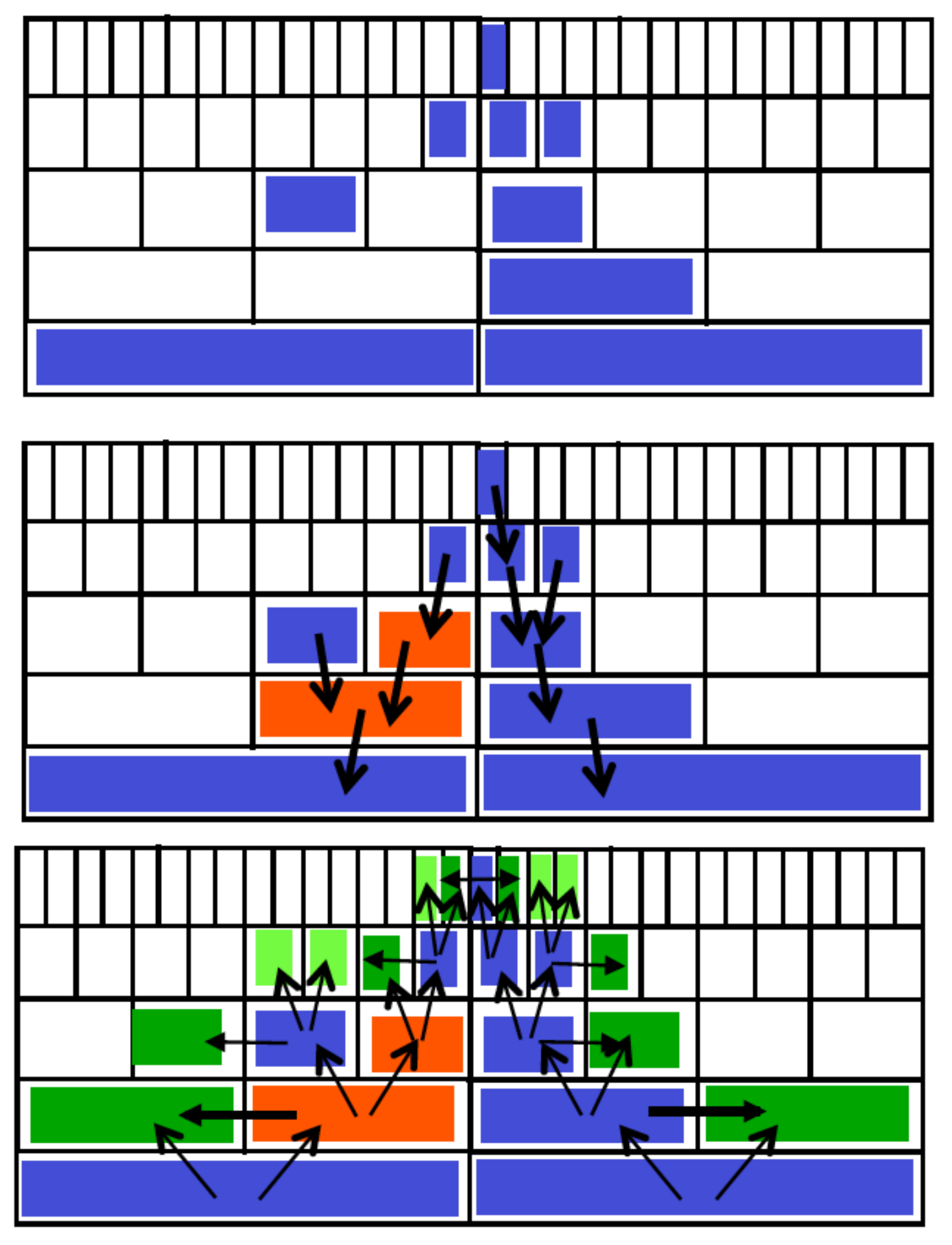}
\end{center}
\caption{Adaption strategy in wavelet coefficient space used in CVCS:
retained wavelet coefficients (blue), added wavelet coefficients to ensure a graded tree (red) and added wavelet coefficients
corresponding to the safety zone (green).}
\label{fig:cvs3}
\end{figure}
A flowchart summarizing the principle of CVCS is shown in Fig.~\ref{fig:cvs2}
and the adaption strategy in orthogonal wavelet coefficient space in Fig.~\ref{fig:cvs3}.

\begin{figure}[htbp]
\begin{center}
\includegraphics[width= 0.6\linewidth]{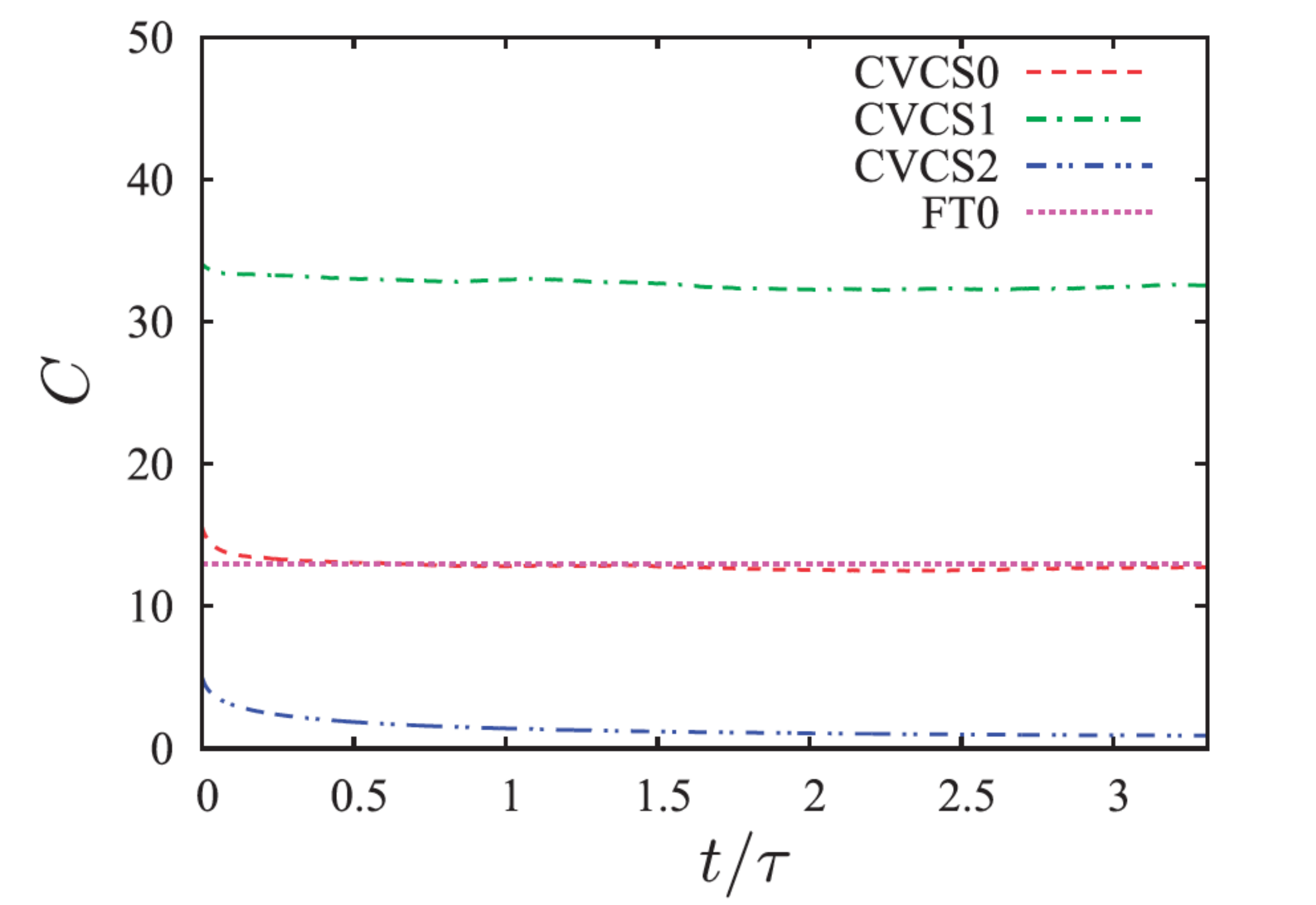}
\end{center} \vspace{-0.7cm}
\caption{Evolution of the percentage $C$ of retained wavelet coefficients for CVCS with three different adaption strategies 
in comparison with Fourier filtering (FT0) with a fixed cut-off wavenumber. 
From \cite{YOKSF13}.}
\label{fig:cvcs_comp}
\end{figure}

\begin{figure}[htbp]
\begin{center}
\includegraphics[width= \linewidth]{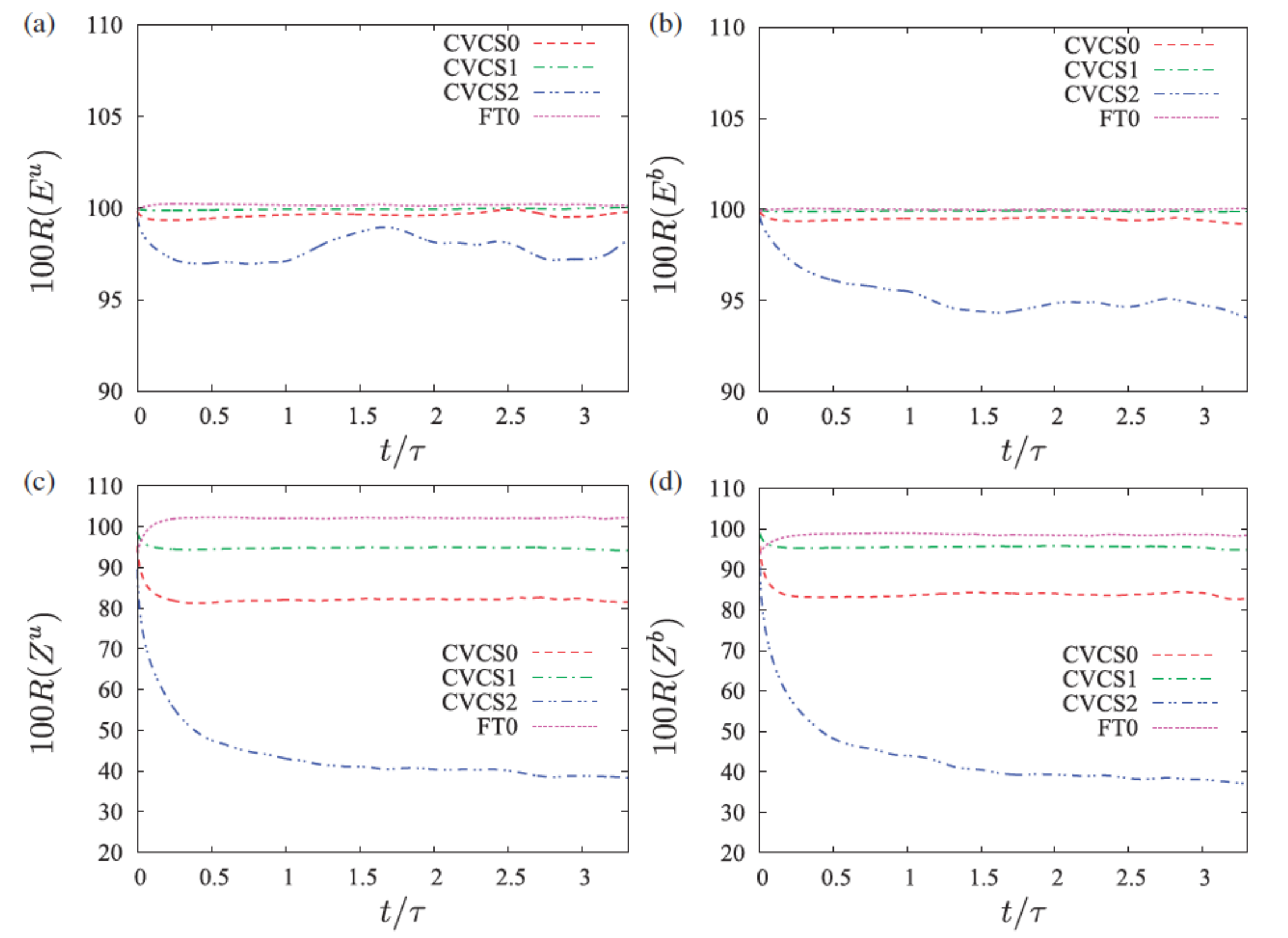}
\end{center}
\caption{Percentage of retained kinetic energy (a), magnetic energy (b), kinetic enstrophy (c) and
magnetic enstrophy (d) for the three different CVCS strategies in comparison with Fourier filtering (FT0).  
From \cite{YOKSF13}.}
\label{fig:cvcs_energy}
\end{figure}

In \cite{YOKSF13} CVCS was performed for 3D forced incompressible homogeneous MHD turbulence without mean
magnetic field, for a magnetic Prandtl number equal to unity.
The Navier--Stokes equations coupled with the induction equation were solved with a pseudospectral method 
using $256^3$ grid points and integrated in time with a Runge--Kutta scheme. 
Different adaption strategies to select the optimal saftey zone for CVCS  have been studied.
We tested the influence of the safety zone and of the threshold, as defined in section~3.1.3, by considering three cases: 
\begin{itemize}
\item CVCS0 with safety zone but without iterating the threshold $\epsilon_0$,
\item CVCS1 with safety zone but with iterating the threshold once $\epsilon_1$,
\item CVCS2  without safety zone but without iterating the threshold $\epsilon_0$,
\end{itemize}
details can be found in~\cite{YOKSF13}.
The quality of CVCS was then assessed by comparing the results with a direct numerical simulation. 
It is found that CVCS with the safety zone well preserves the statistical predictability of the
turbulent flow with a reduced number of degrees of freedom. 
CVCS was also compared with a Fourier truncated simulation using a spectral cutoff filter, where the number of retained Fourier
modes is similar to the number of the wavelet coefficients retained by CVCS0. 
Figure~\ref{fig:cvcs_comp} shows the percentage of retained wavelet coefficients for CVCS (with three different adaption strategies)  in comparison to Fourier filtering (FT0) with a fixed cut-off wavenumber.
%
%
The percentage of retained kinetic energy, magnetic energy, kinetic enstrophy and
magnetic enstrophy for the three different CVCS strategies in comparison with Fourier filtering (FT0) is plotted in Fig.~\ref{fig:cvcs_energy}. 
%

\begin{figure}[htbp]
\begin{center}
\includegraphics[width= \linewidth]{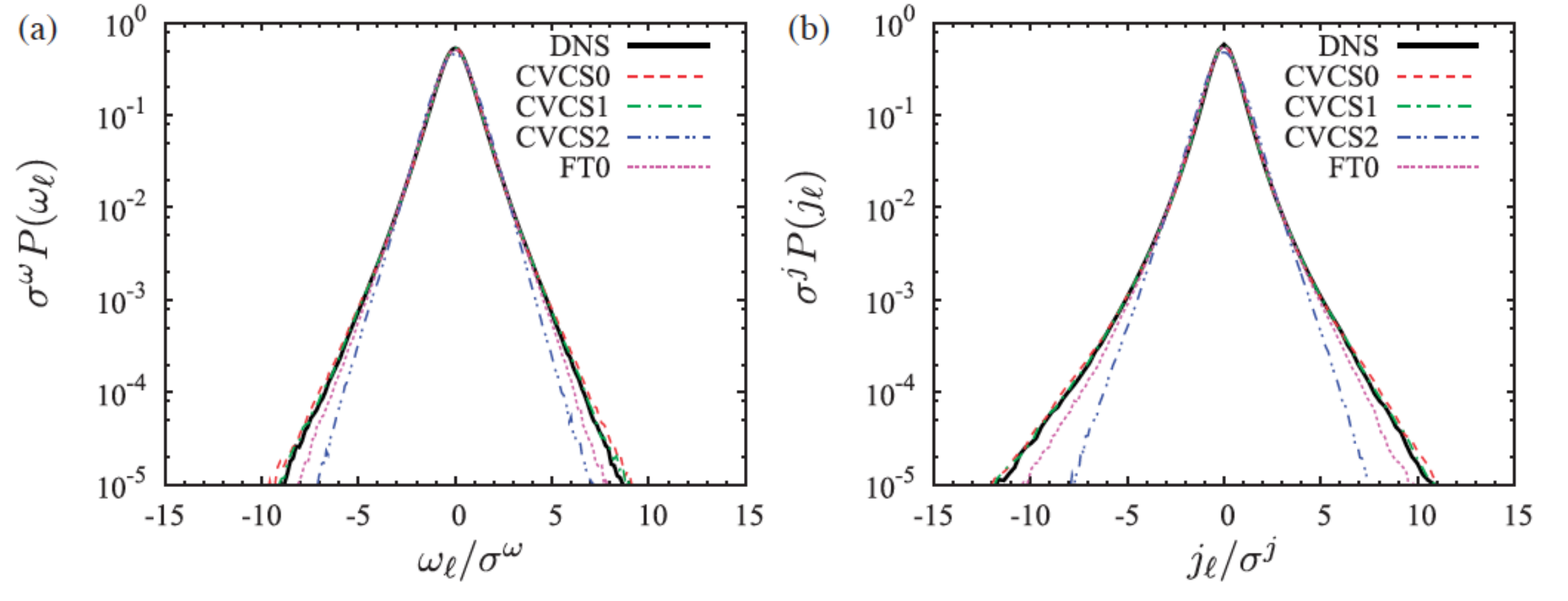}
\end{center}
\caption{PDFs of the $\ell$-th component of vorticity (a) and current density (b) normalized by the corresponding standard deviation.
From \cite{YOKSF13}.}
\label{fig:cvcs_pdf}
\end{figure}

\begin{figure}[htbp]
\begin{center}
\includegraphics[width= \linewidth]{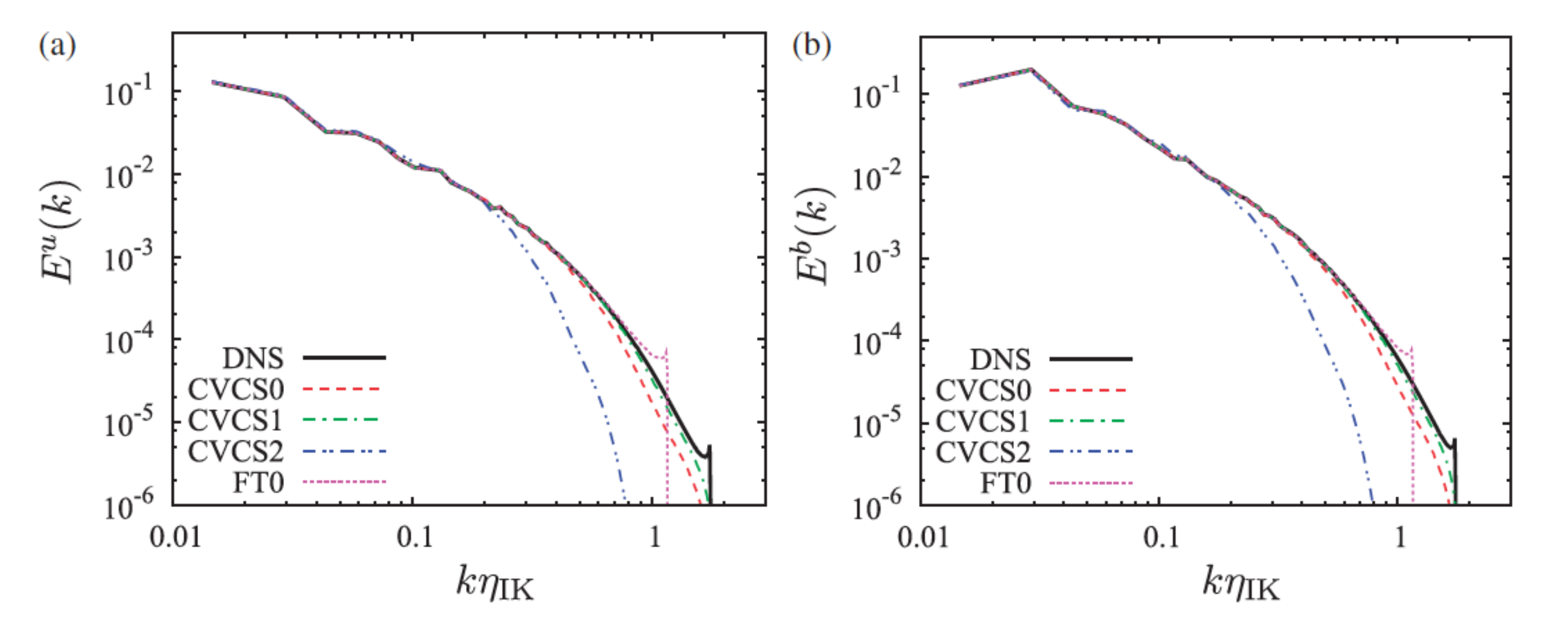}
\end{center}
\caption{ Kinetic (a) and magnetic energy spectra (b). The wavenumber is normalized with the Iroshnikov-Kraichnan scale.
From \cite{YOKSF13}.}
\label{fig:cvcs_spectra}
\end{figure}

Probability density functions of vorticity and current density, normalized by the corresponding standard deviation,
in Fig.~\ref{fig:cvcs_pdf} show that CVCS0 and CVCS1 capture well the high order statistics of the flow,
while in FT0 and in CVCS2 the tails of the PDFs are reduced with respect to the DNS results.
%
%
The energy spectra of kinetic and magnetic energy  in Fig.~\ref{fig:cvcs_spectra}
confirm that CVCS0 and CVCS1 reproduce perfectly the DNS results in the inertial range, where all nonlinear acticity takes place, and only differs in the dissipative range.
%

\begin{figure}[htbp]
\begin{center}
\includegraphics[width= 0.45\linewidth]{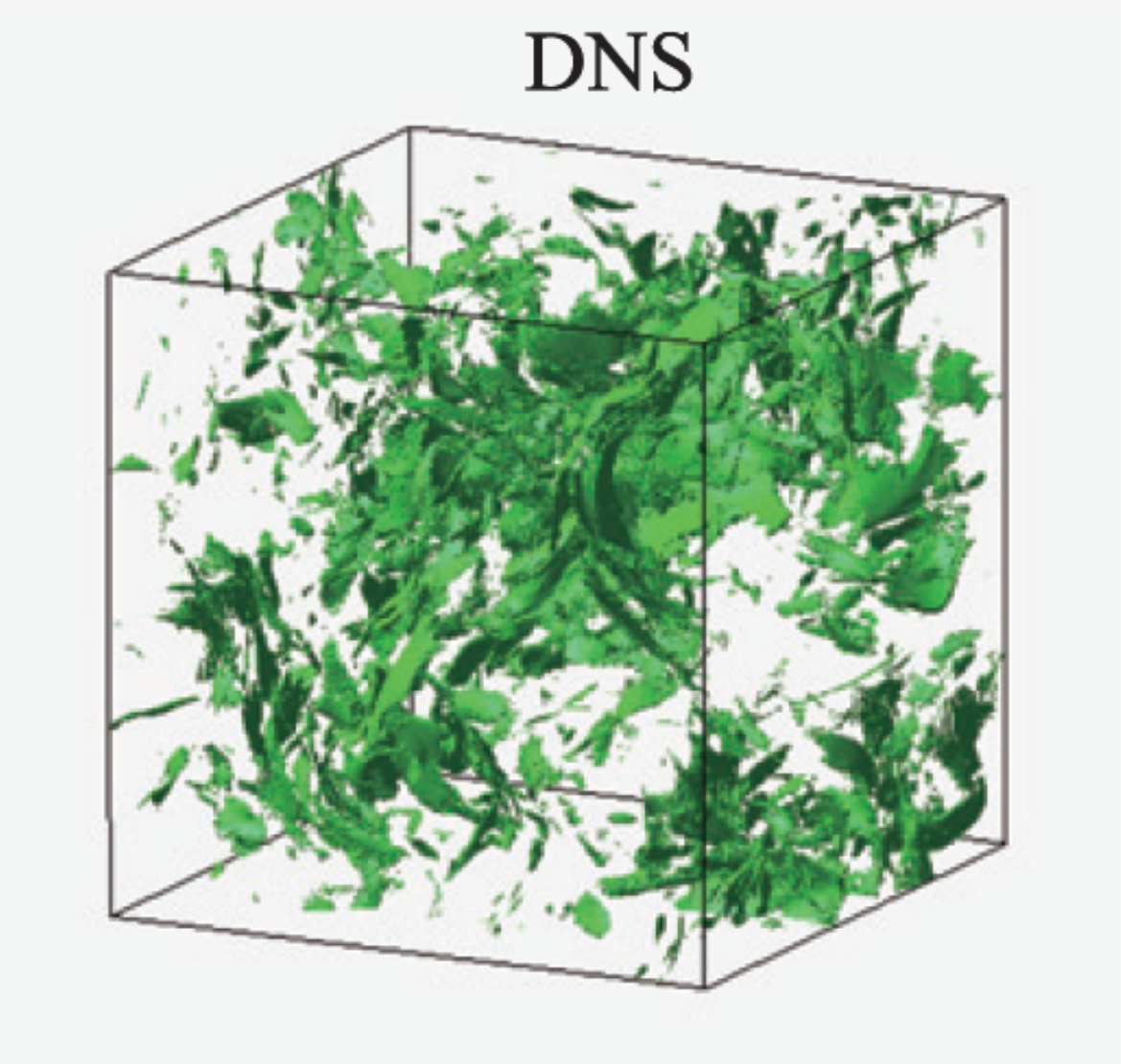}
\includegraphics[width= 0.45\linewidth]{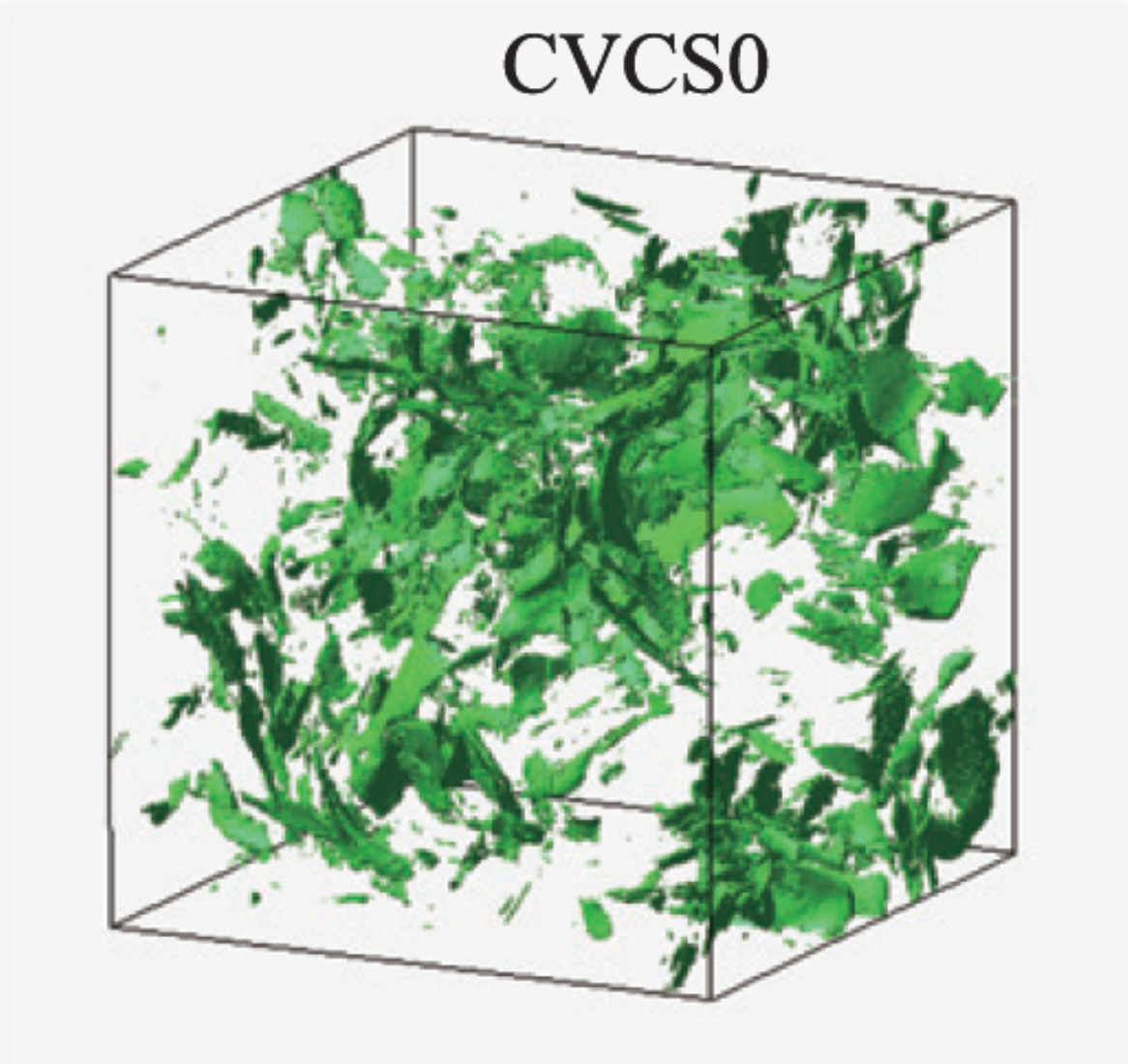}\\
\includegraphics[width= 0.45\linewidth]{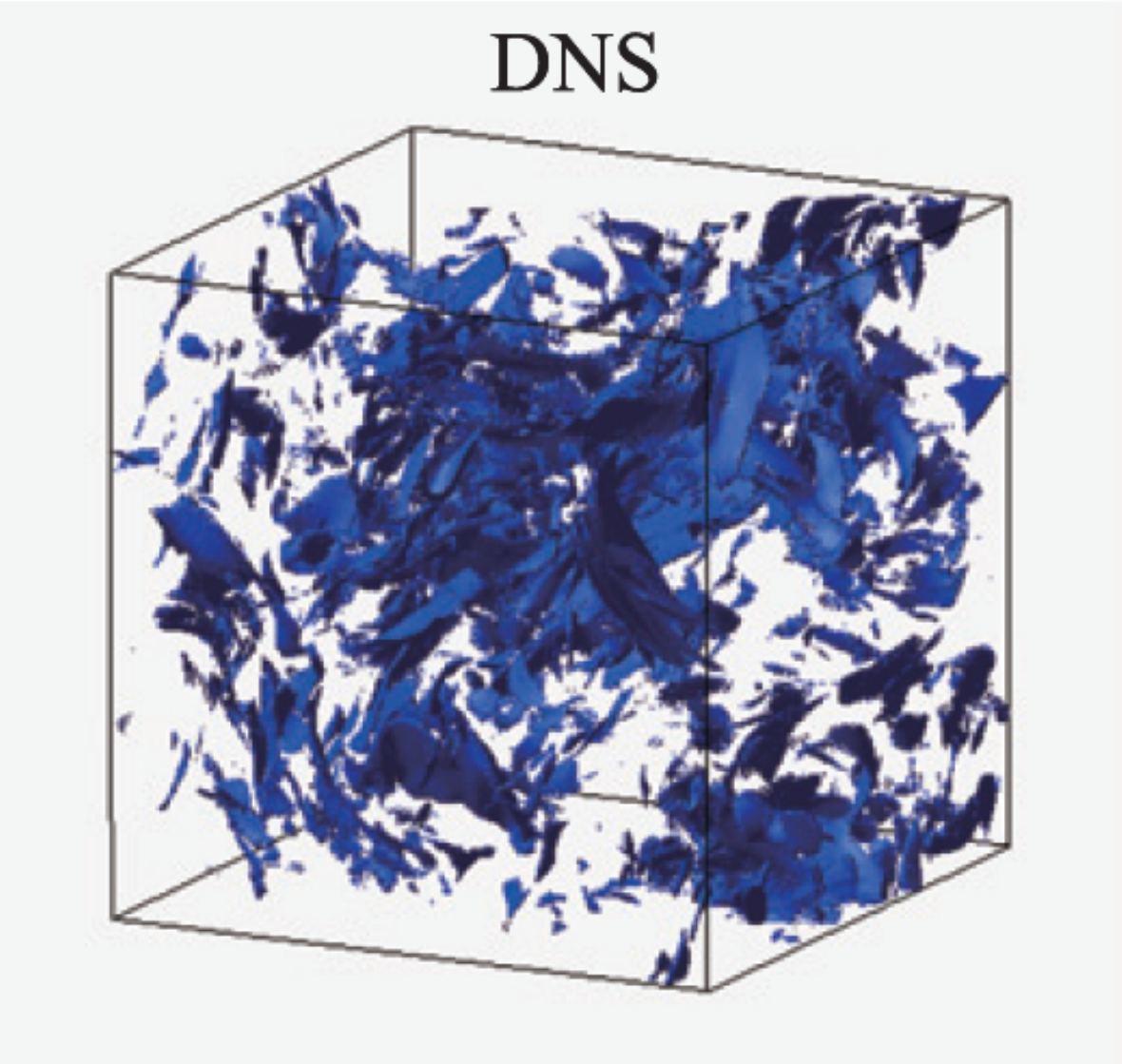}
\includegraphics[width= 0.45\linewidth]{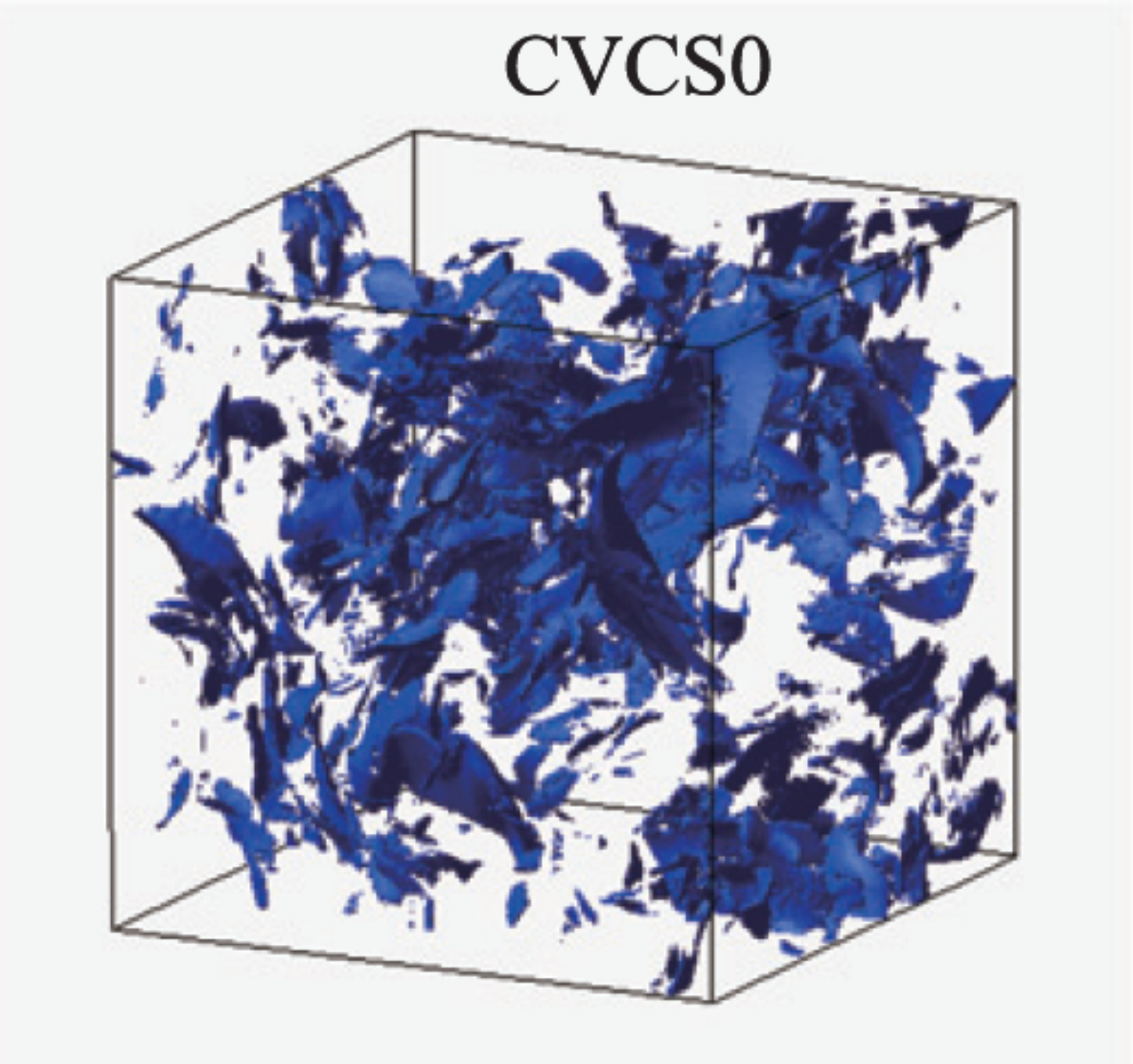}
\end{center}
\caption{Visualization of isosurfaces of modulus of vorticity (top) and modulus of current density (bottom) for DNS (left) and CVCS0 (right). 
From \cite{YOKSF13}.}
\label{fig:cvcs_vort_curr}
\end{figure}

The results thus show that the wavelet representation is more suitable than the Fourier representation, especially
concerning the probability density functions of vorticity and current density and that only about 13\% of the degrees of freedom (CVCS0)
compared to DNS are sufficient to represent the nonlinear dynamics of the flow.
A visualization comparing both the vorticity and current density field for DNS and CVCS0 is presented in Fig.~\ref{fig:cvcs_vort_curr}.
%

\section{Conclusion}

This paper reviewed different wavelet techniques and showed several of their applications to MHD and plasma turbulence.
Continuous and orthogonal wavelet transforms were presented and some wavelet-based statistical tools described, after selecting those most appropriate to study turbulence, such as scale-dependent second and higher order moments, intermittency measure, together with scale-dependent directional statistical measures.
Examples of applications to three-dimensional incompressible MHD turbulence, computed by DNS, illustrated how the flow intermittency can be quantified and how its anisotropy and helicity might vary with scale.
The wavelet-based coherent structure extraction algorithm was detailed and validated for a test signal.
Different applications to experimental and numerical turbulent plasma data, in one, two and three dimensions, were shown. 
The underlying methodology of a wavelet-based tomographic reconstruction algorithm for denoising images and movies obtained with fast cameras in tokamaks were explained and results were presented. 
Applications to an academic example and to fast camera data from Tore Supra  proved the efficiency of the algorithm to extract blobs and fronts while denoising the data.
Wavelet-based simulation schemes developed in the context of kinetic plasma equations were also described. 
Results computed with them showed how wavelet denoising accelerates the convergence of classical PIC schemes and how a particle-in wavelet (PIW) scheme solves the Vlasov-Poisson equation directly and efficiently in wavelet space. 
Concerning the fluid equations, in particular the resistive non-ideal MHD equations, the coherent vorticity and current sheet simulation (CVCS) methods were explained and examples illustrated the properties and insights the wavelet-based approach offers in the context of MHD and plasma turbulence.

\section*{Acknowledgements}

MF and KS are grateful to Sadri Benkadda for inviting them to give review lectures at the ITER International School 2014``High performance computing in fusion science" held in Aix-en-Provence, France. 
The manuscript is based on MF's lecture there, entitled `Wavelet transforms and their applications for ITER', given on August 26th, 2014.
The authors are also indebted to Wouter Bos, Romain Nguyen van yen, Naoya Okamoto and Katsunori Yoshimatsu with whom we published several papers on wavelet applications, from which the material of this review has been taken.
This work was supported by the French Research Federation for Fusion Studies carried out within the framework of the European Fusion Development Agreement (EFDA).


\end{document}